\documentclass[a4paper,fleqn,usenatbib]{mnras}

\usepackage{newtxtext,newtxmath}

\usepackage[T1]{fontenc}
\usepackage{ae,aecompl}

\usepackage{amsmath}
\usepackage{graphicx,longtable,mathrsfs,color,array}
\usepackage{epsfig,subfigure,placeins,float}
\usepackage[usenames,dvipsnames]{xcolor}
\usepackage{hyperref}
\usepackage{multirow}
\usepackage{todonotes}
\usepackage{caption}

\usepackage[normalem]{ulem}
\providecommand{\adsurl}[1]{\href{#1}{ADS}}

\newcommand*{\rom}[1]{\expandafter\@slowromancap\romannumeral #1@}

\newcommand{\ms}{M$_{\sun}$}

\newcommand{\mcrit}{$M_{\rm 500crit}$}

\newcommand{\mtot}{$M_{\rm tot}$}
\newcommand{\msim}{$M_{\rm 500,SIM}$}

\newcommand{\mspec}{$M_{\rm 500,SPEC}$}

\newcommand{\mgsim}{$M_{\rm 500c, SIM}^{\rm gas}$}
\newcommand{\mgspec}{$M_{\rm 500c, SPEC}^{\rm gas}$}

\newcommand{\ysz}{$Y_{\rm SZ}$}

\newcommand{\yszsim}{$Y_{\rm SZ, 500,SIM}$}
\newcommand{\yszspec}{$Y_{\rm SZ, 500,SPEC}$}

\newcommand{\yx}{$Y_{\rm X}$}
\newcommand{\yxsim}{$Y_{{\rm X, SIM}}$}
\newcommand{\yxspec}{$Y_{{\rm X, SPEC}}$}

\newcommand{\Lx}{$L_{\rm X}$}
\newcommand{\Lxce}{$L_{\rm X, ce}$}
\newcommand{\Lxsim}{$L_{{\rm X, SIM}}^{\rm 0.5 - 2 \, keV}$}
\newcommand{\Lxspec}{$L_{{\rm X, SPEC}}^{\rm 0.5 - 2 \, keV}$}
\newcommand{\Lxcesim}{$L_{{\rm X, ce, SIM}}^{\rm 0.5 - 2 \, keV}$}
\newcommand{\Lxcespec}{$L_{{\rm X, ce, SPEC}}^{\rm 0.5 - 2 \, keV}$}

\newcommand{\tx}{$T_{\rm X}$}
\newcommand{\txce}{$T_{\rm X, ce}$}
\newcommand{\txsim}{$T_{{\rm X, SIM}}$}
\newcommand{\txspec}{$T_{{\rm X, SPEC}}$}
\newcommand{\txcesim}{$T_{{\rm X, ce, SIM}}$}
\newcommand{\txcespec}{$T_{{\rm X, ce, SPEC}}$}

\newcommand{\rcrit}{$R_{\rm 500crit}$}
\newcommand{\rsim}{$R_{\rm 500,SIM}$}

\newcommand{\rspec}{$R_{\rm 500,SPEC}$}
\newcommand{\rsimc}{$R_{\rm 500crit, \, SIM}$}

\newcommand{\rspecc}{$R_{\rm 500crit, \, SPEC}$}

\newcommand{\xp}{$X_{\rm pivot}$}

\newcommand{\yxm}{\mbox{$Y_{\rm X}$ -- $M_{\rm tot}$}}
\newcommand{\yszm}{\mbox{$Y_{\rm SZ}$ -- $M_{\rm tot}$}}
\newcommand{\txm}{\mbox{$T_{\rm X}$ -- $M_{\rm tot}$}}
\newcommand{\txcem}{\mbox{$T_{\rm X, \, ce}$ -- $M_{\rm tot}$}}
\newcommand{\lxm}{\mbox{$L_{\rm X}$ -- $M_{\rm tot}$}}

\newcommand{\mgm}{\mbox{$M_{\rm gas}$ -- $M_{\rm tot}$}}

\newcommand{\tng}{\mbox{IllustrisTNG}}

\newcommand{\val}[3][0.000]{$#1^{+#2}_{-#3}$}

\makeatother

\title[X-ray and SZ Scaling Relations in IllustrisTNG]{Sunyaev-Zel'dovich effect and X-ray scaling relations of galaxies, groups and clusters in the IllustrisTNG simulations}

\author[A. R. Pop et al.]{Ana-Roxana Pop,$^{1}$\thanks{E-mail: ana-roxana.pop@cfa.harvard.edu}
Lars Hernquist,$^{1}$
Daisuke Nagai,$^{2}$
Rahul Kannan,$^{1}$ 
\newauthor
Rainer Weinberger,$^{3}$
Volker Springel,$^{4}$ 
Mark Vogelsberger,$^{5}$
Dylan Nelson,$^{6}$
\newauthor
R\"{u}diger Pakmor,$^{4}$
Annalisa Pillepich,$^{7}$ 
Paul Torrey$^{8}$
\\
\\
$^{1}$Center for Astrophysics $\rvert$ Harvard \& Smithsonian, 60 Garden Street, Cambridge, MA 02138, USA \\
$^{2}$Department of Physics, Yale University, New Haven, CT 06520, U.S.A. \\
$^{3}$Canadian Institute for Theoretical Astrophysics, 60 St. George Street, Toronto, ON M5S 3H8, Canada \\
$^{4}$Max-Planck-Institut f\"{u}r Astrophysik, Karl-Schwarzschild-Stra{\ss}e 1, D-85741 Garching bei M\"{u}nchen, Germany \\
$^{5}$Dept. of Physics, Kavli Institute for Astrophysics and Space Research, Massachusetts Institute of Technology, Cambridge, MA 02139, USA \\
$^{6}$Universit\"{a}t Heidelberg, Zentrum f\"{u}r Astronomie, Institut f\"{u}r theoretische Astrophysik, Albert-Ueberle-Str. 2, 69120 Heidelberg, Germany \\
$^{7}$Max-Planck-Institut f\"{u}r Astronomie, K\"{o}nigstuhl 17, D-69117 Heidelberg, Germany \\
$^{8}$Department of Physics, University of Florida, Gainesville, FL 32611, USA
}

\date{Accepted XXX. Received YYY; in original form ZZZ}

\pubyear{2022}

\begin{document}
\label{firstpage}
\pagerange{\pageref{firstpage}--\pageref{lastpage}}
\maketitle

\begin{abstract}
Observable thermodynamical properties of the intracluster medium (ICM) reflect the complex interplay between AGN feedback and the gravitational collapse of haloes. 
Using the large volume TNG300 simulation of the IllustrisTNG project we provide predictions for X-ray and Sunyaev-Zel'dovich (SZ) scaling relations for a sample of over 30,000 haloes that cover a wide mass range from galaxies to massive galaxy clusters \mbox{(\mcrit\, $\in [10^{12}$\,\ms\, -- $2\times 10^{15}$\,\ms]).}
 We produce mock X-ray observations of simulated haloes using methods that are consistent with observational techniques.  
 Thus, we investigate the scaling relations between the soft-band X-ray  luminosity, spectroscopic temperature, gas mass fraction, \yx\,, and \ysz\, as a function of halo mass, and we find broad agreement between IllustrisTNG and the observed relations. Our results highlight the scatter and bias introduced by estimated masses, and thus the importance of converting simulated ICM properties to the observable space when comparing simulations to current X-ray observations.
 The wide range of halo masses in our sample provides new insights into the shape of the X-ray and SZ scaling relations across three orders of magnitude in mass. Our findings show strong evidence for a break in $z=0$ scaling relations. 
 We introduce a smoothly broken power law model which robustly captures
 the location of this break, the width of the transition region around the break, as well as the slope dependence on halo mass. 
 Our results inform the next generation of subgrid black hole feedback models and provide predictions for ongoing and future observational surveys.

\end{abstract}


\section{Introduction}

Over the past decades, X-ray and microwave observations of the hot gaseous medium in galaxy clusters have provided unprecedented insights into cosmology \citep[e.g.,][]{Vikhlinin2009a,PlanckCollaboration2014b,Pratt2019} and cluster astrophysics \citep[e.g.,][]{Planck2011,McDonald2013,McDonald2017,Eckert2017}. Recent advances in X-ray and SZ surveys extended these measurements to smaller groups and galaxies \citep[e.g.,][]{Anderson2015,Amodeo2021,Bregman2022,Comparat2022,Chadayammuri2022}.

A plethora of X-ray and microwave surveys are underway to detect and characterize the properties of baryons in the form of warm-hot gas in galaxies, groups, and clusters. The latest observations in X-ray from \textit{eROSITA} \citep{Bulbul2021} and microwave from Simons Observatory \citep{Ade2019}, CMB-S4 \citep{CMB-S4}, and CMB-HD \citep{CMB-HD} provide remarkable constraints on cosmology and astrophysics \citep{Pillepich2018,Raghunathan2022}.

However, in order to harness the statistical power of these ongoing and upcoming surveys, it is critical to understand the X-ray and SZ observable-mass relations using both simulations \citep{Kravtsov2006,Nagai2006} and observations \citep{Vikhlinin2006,Pratt2009,Sun2009,Arnaud2010}. To date, these simulations have become an indispensable tool for understanding the impact of galaxy formation on X-ray and SZ observations \citep[e.g.,][]{Nagai2007b,Battaglia2012,Kay2012,Bahe2017} and characterizing X-ray and hydrostatic mass biases \citep[e.g.,][]{Rasia2006,Nagai2007a,Lau2009,Nelson2012,Nelson2014a,Biffi2016,Barnes2017a, Barnes2021}. The inclusion of feedback from active galactic nuclei (AGNs) in hydrodynamical simulations showed an increasingly large impact on X-ray and SZ scaling relations in galaxy groups \citep{Puchwein2008,Fabjan2011,Pike2014,Planelles2014,LeBrun2017,Truong2018,Henden2018,Henden2019,Lim2021,Yang2022}.

In this work, we investigate the X-ray and SZ scaling relations of galaxies, groups, and clusters, spanning over three orders of magnitude in mass (\mcrit\, $\in [10^{12} - 2\times 10^{15}]$\,\ms).  Taking full advantage of the large volume TNG300 simulation of the IllustrisTNG suite, we vastly expand on the number and mass range of simulated haloes that have been previously compared to X-ray and SZ observations. The purpose of this paper is threefold: (1) we develop a pipeline that produces high-fidelity mock X-ray observations, which closely mimics the methods utilized by observers, (2) we test the new subgrid model for AGN feedback implemented in \tng\, against X-ray and SZ observations, with the goal of understanding its strengths and weaknesses, and (3) we extend the studies of \yxm\ and \yszm\ relations down to $10^{12}$\,\ms\, in \tng, using a large sample of low mass groups and galaxies to predict baryonic effects on the X-ray and SZ scaling relations. Most notably, we discover the presence of a break in all X-ray and SZ scaling relations. We propose an analytic model for a smoothly broken power law that provides a robust method to find unbiased estimates of scaling relation slopes at very high and very low halo masses, as well as the location of the break, and the width of the transition region around the break. This model allows us to provide predictions for ongoing and future X-ray and SZ surveys, which will include many more galaxy groups and low mass clusters. 

The paper is organized as follows. In Section~\ref{sec:simulations}, we describe the \tng\, simulations, the sample of galaxies, groups, and clusters, and mock X-ray pipeline used to compute X-ray luminosities and spectroscopic temperatures. In Section~\ref{sec:scaling}, we describe the scaling relation models and fitting procedure, highlighting the introduction of a smoothly broken power law model that captures the break in the scaling relations and the transition region around the break. We present our results and discussions for X-ray scaling relations in Section~\ref{sec:xrayresults} and for the \yszm\, scaling in Section~\ref{sec:yszresults}. We conclude by summarizing our results in Section~\ref{sec:conclusions}.

\section{Methods}
\label{sec:simulations}

\subsection{Simulations and Sample Selection}
\label{subsec:sample}

The IllustrisTNG simulations model the formation and evolution of galaxies and clusters from cosmological initial conditions using the moving-mesh magnetohydrodynamics code Arepo \citep{Springel2010, Pakmor2013, Pakmor2016}.
The \tng\, suite of simulations includes three different boxes, with side lengths of $\sim$50\,Mpc, $\sim$100\,Mpc (same as Illustris), and $\sim$300\,Mpc, respectively \citep{Pillepich2018a,Nelson2018,Springel2018,Naiman2018,Marinacci2018,Pillepich2019,Nelson2019b,Nelson2019a}. The larger simulation box, with a volume of $\sim$(300\,Mpc)$^3$, provides us with significantly better statistics for high mass haloes compared to the previous generation of simulations ($>$2,500 haloes with \mcrit $\geq 10^{13}$\,\ms\, in TNG300, compared to $<$100 in Illustris). 
The IllustrisTNG model \citep{Weinberger2017b,Pillepich2018b} includes sub-resolution models for star formation, stellar feedback, massive black hole formation, growth via accretion and mergers, as well as AGN feedback with modifications to the earlier Illustris simulation \citep{Vogelsberger2014b,Vogelsberger2014a,Genel2014} that improve agreement with a variety of observations. The most important changes for X-ray and SZ properties of high mass gaseous haloes are the modeling of feedback effects from AGNs: IllustrisTNG employs a two-mode feedback model with a moderately efficient thermal mode at low black hole masses and high accretion rates (predominantly active in star-forming galaxies) and a highly efficient kinetic mode at high black hole masses and low accretion rates (active in most quiescent massive galaxies and galaxy clusters, \citealt{Weinberger2018}). The kinetic AGN feedback in IllustrisTNG is more efficient in quenching star formation, but less effective in removing gas from the halo than the corresponding AGN radio-mode feedback employed in Illustris \citep{Weinberger2017b}.

In this work, we select and analyze all central haloes in the $z=0$ snapshot with \mcrit \, $\geq 10^{12}$\,\ms\, from TNG300. Our final sample includes over 30,000 haloes in total, with more than 2,500 having \mcrit\,$\geq 10^{13}$\,\ms\,, and more than 150 clusters above $10^{14}$\,\ms. 
Haloes in IllustrisTNG are identified through a friends-of-friends (FoF) algorithm with linking length $b=0.2$ that is run on the dark matter particles. 
Other particle types (gas, stars, and black holes) are assigned to the same halo as their closest dark matter particle.
Galaxies (or ``subhaloes") are identified by running the \textsc{subfind} algorithm \citep{Springel2001, Dolag2009} to find gravitationally bound substructures that include all particle types.

In order to explore the impact of the dynamical state of a cluster on its estimated mass and X-ray properties, we classify each halo in our sample as relaxed or unrelaxed. Previous studies have identified numerous ways to distinguish relaxed from unrelaxed clusters \citep[e.g.,][]{Neto2007, Duffy2008, Klypin2011, Dutton2014, Klypin2016}. In this paper, we define relaxed clusters based on the fraction of kinetic to thermal energy inside \rcrit. In particular, we consider all haloes with:
\begin{equation}
    \frac{E_{{\rm kinetic, \, <R}_{\rm 500crit}}}{E_{{\rm thermal, \, <R}_{\rm 500crit}}} < 0.1,
\end{equation}
to be relaxed. In the equation above, $E_{{\rm kinetic, \, <R}_{\rm 500crit}}$ is the total kinetic energy from all gas cells within \rsim\, of the cluster center, while $E_{{\rm thermal, \, <R}_{\rm 500crit}}$ is the total thermal energy of those same gas cells.

\subsection{X-ray pipeline}
\label{subsec:xraypipeline}

Previous studies have shown that the observable properties of hot gas such as X-ray luminosities and temperatures can be biased due to multi-temperature structures \citep[e.g.,][]{Mazzotta2004, Rasia2005, Nagai2007a, Rasia2014} and gas clumping and inhomogeneities \citep[e.g.,][]{Nagai2011, Zhuravleva2013, Khedekar2013, Vazza2013, Avestruz2014}. 

Therefore, in order to assess the role of such systematics and to enable robust comparisons with observed data, we generate mock X-ray observations for all the haloes in our sample using a modified version of the \textsc{Mock-X} pipeline, which was introduced in \citet{Barnes2021}. Our approach mirrors numerous different methods that have been developed by other groups over the years \citep[e.g.,][]{Gardini2004, Nagai2007a, Rasia2008, Heinz2009, Biffi2012, ZuHone2014, LeBrun2014, Henden2018}.

First, we start by generating a rest-frame X-ray spectrum in the 0.5\,--\,10 keV band, using the density, temperature, and metallicity of every gas cell within a sphere of radius $1.5 $ \rcrit. This sphere is centered on the halo potential minimum, which is identified by the \textsc{subfind} algorithm as the location of the most bound particle in the halo. The X-ray spectrum is computed using the Astrophysical Plasma Emission Code \citep[\textsc{apec};][]{Smith2001} via the \textsc{pyatomdb} module with atomic data from \textsc{atomdb} v3.0.9 \citep{Foster2012}.
For each of the 11 chemical elements tracked in IllustrisTNG (H, He, C, N, O, Ne, Mg, Si, S, Ca, and Fe), we compute an individual spectrum. A particle's spectrum is the sum of the individual spectra for each chemical element, scaled by the particle's elemental abundance. We exclude any star-forming gas, as well as all cold gas cells with temperatures below \mbox{$10^5$\,K} or any gas cells that have a positive cooling rate. These gas cells are a small fraction ($<1\%$) of the total gas cells considered and they would bring a negligible contribution to the total X-ray surface brightness because most of the X-ray emission from these cells would fall outside the 0.5\,--\,10\,keV energy band. 
Moreover, this removes uncertainties due to the imprecise thermal properties of star-forming or cooling gas. Similar cuts have also been applied successfully in previous studies \citep[e.g.,][to name a few]{Nagai2007a, Henden2018, Barnes2021}. This cut is also complementary to X-ray observations, where compact sources with strong X-ray emission would either be unresolved or would be excised during the post-processing analysis. We further remove any substructures using the \textsc{subfind} algorithm to exclude gas cells bound to galaxies other than the central halo we are modeling. 

We produce mock X-ray spectra for each halo, and then model the X-ray luminosity and spectroscopic temperature of the given object within \rcrit. To do this, we start by binning all gas cells within 1.5\,\rcrit\, of the halo center in 25 linearly-spaced radial bins. 
We adopt an energy resolution for the X-ray spectrum of 150\,eV for energies between 0.5\,--\,10\,keV, and we convolve the resulting spectrum with the response function and effective area of the \textit{Chandra} ACIS-I detector. We adopt a relatively long exposure time of $10^6$ seconds, because we are interested in obtaining accurate predictions for the scaling relations all the way down to low mass groups and galaxies. This exposure time ensures that we are not limited by photon noise even at the low mass end of our sample.
Next, we fit a single temperature and metallicity \textsc{apec} model to the resulting spectrum for each radial bin and thus obtain an estimate of the density, temperature and metallicity of X-ray emitting gas as a function of radial distance. We derive X-ray temperatures from a single temperature fit to the total spectrum integrated within $r=(0-1)$\rcrit\, (for $T_{\rm X}$) and $r=(0.15-1)$\,\rcrit\, (for core-excised $T_{\rm X, ce}$).

\section{Scaling Relations}
\label{sec:scaling} 

\subsection{Self-Similar Scaling Relations}
\label{subsec:selfsimilar}

Under the assumption that the hot gas in the intracluster medium is in hydrostatic equilibrium, we expect the pressure $P_g$ and density $\rho_g$ of the gas inside the cluster to be related through:
\begin{equation}
    \frac{1}{\rho_g} \frac{{\rm d} P_g}{{\rm d} \, r} = - \frac{G M }{r^2}.\label{eqn:hydroeqn}
\end{equation}
The hydrostatic equilibrium equation above has the following solution for the total cluster mass as a function of the gas density $\rho_g(r)$ and temperature $T(r)$ profiles inside the cluster:
\begin{equation}
M (<r) = - \frac{k_B r \, T(r)}{\mu m_p G} \left( \frac{{\rm d \, ln}\, \rho_g(r)}{{\rm d \, ln}\, r} + \frac{{\rm d \, ln}\, T(r)}{{\rm d \, ln}\, r}\right).\label{eqn:masseqn}
\end{equation}

If one defines the total cluster mass ($M_\Delta$) to be the mass enclosed in a spherical region of radius $r_\Delta$ and mean overdensity $\Delta$ relative to the critical density of the Universe ($\rho_{\rm crit} = 3 H(z)^2 / (8 \pi G)$), then:
\begin{equation}\label{eqn:eqn_scaling_Mdelta}
M_\Delta = \frac{4\pi}{3} \Delta \rho_{\rm crit}(z) r_\Delta^3,
\end{equation}
where $H(z)=H_0 E(z)$ is the Hubble parameter and $E(z) \equiv \sqrt{\Omega_m(1+z)^3 + \Omega_\Lambda}$. 
Assuming a self-similar relation between clusters (where bigger clusters are just scaled up versions of smaller ones), equation~(\ref{eqn:eqn_scaling_Mdelta}) implies that the cluster size scales with the cluster mass as:
\begin{equation}\label{eqn:eqn_scaling_mass_radius}
M_\Delta \propto \Delta \rho_{\rm crit}(z) r_\Delta^3 \propto E^2(z) r_\Delta^3.
\end{equation}

As gas is streaming along intergalactic filaments and falling into the cluster potential ($\Phi$), it is slowed down and heated up via accretion shocks. Under the assumption of a collapsed isothermal sphere, the infalling gas will eventually reach a temperature approximately equal to the virial temperature of the cluster:
\begin{equation}
k_B T_\Delta \propto -\frac{1}{2}\Phi = \frac{G M_\Delta \mu \, m_p}{2 r_\Delta},
\end{equation}
where $k_B$ is the Boltzmann constant, $m_p$ is the proton mass, and $\mu$ is the mean molecular weight.

Thus, the self-similar relation between temperature and total cluster mass is:
\begin{equation}\label{eqn:eqn_scaling_temp_mass}
T_\Delta \propto \frac{M_\Delta}{r_\Delta} \propto M_\Delta^{2/3} E^{2/3}(z).
\end{equation}
The scaling relation between the X-ray luminosity and cluster mass can be derived in a similar fashion. Assuming that the main cooling channel for the ICM is through thermal bremsstrahlung, the bolometric X-ray luminosity can be written as:
\begin{equation}
L_{\rm X, \Delta}^{\rm bol} \propto \rho_g^2 \Lambda(T) r_\Delta^3 \propto \frac{M^2_{{\rm gas}, \Delta}}{r_\Delta^3} T^{1/2},
\end{equation}
where $\Lambda(T) \propto T^{1/2}$ is the cooling function for bolometric emission \citep{Sarazin1986}. 

Under the self-similar model \citep{Kaiser1986}, the density profile of gas, $\rho_{\rm gas}(r)$, scales self-similarly and the ratio between the total gas mass ($M_{{\rm gas}, \Delta}$) and the total cluster mass ($M_\Delta$) is independent of halo mass, i.e.  $M_{{\rm gas}, \Delta} \propto M_\Delta $. Together with equations (\ref{eqn:eqn_scaling_mass_radius}) and (\ref{eqn:eqn_scaling_temp_mass}), this allows us to derive the self-similar prediction for the bolometric X-ray luminosity - total mass relation:
\begin{equation}
L_{\rm X, \Delta}^{\rm bol} \propto \frac{M_\Delta^2}{r_\Delta^3} T_\Delta^{1/2} \propto  M_\Delta^{4/3} E(z)^{7/3}.
\end{equation}
Similarly, the X-ray luminosity - temperature relation is given by:
\begin{equation}
L_{\rm X, \Delta}^{\rm bol} \propto  T_\Delta^{2} \, E(z).
\end{equation}

Finally, in the case of the thermal Sunyaev-Zel'dovich effect, the integrated Compton y-parameter is a measure of the total thermal energy of the ICM and it is proportional to $Y_{\rm SZ} \propto \int n_e T_e dV \propto M_{\rm gas} T_m$, where $n_e$ and $T_e$ are the electron number density and temperature, and $T_m$ is the mass-weighted mean temperature. The baryonic content of clusters is dominated by the hot ICM, and thus the gas fraction inside clusters is roughly equal to the mass fraction of baryons $f_g \approx \frac{M_{\rm baryons}}{M_{\rm total}}$. This fraction is similar to the ratio between $ \frac{\Omega_b}{\Omega_m}$ in the Universe, and thus it is not expected to vary significantly from one halo to another. Thus, we can relate both the integrated SZ signal, $Y_{\rm SZ}$, and its X-ray analogue, $Y_{\rm X} \equiv M_{{\rm gas}} T$, to the product between mass and temperature: $Y_{{\rm SZ}, \Delta} \propto Y_{{\rm X}, \Delta} \propto M_{{\rm gas}, \Delta} T_{\Delta} \propto M_{ \Delta} T_{\Delta} \propto M_\Delta^{5/3} E^{2/3}(z)$.
As a result, the self-similar relations for $Y_{{\rm SZ}, \Delta}$ and $Y_{\rm X, \Delta}$ follow from equation~(\ref{eqn:eqn_scaling_temp_mass}):
\begin{equation}
    Y_{{\rm SZ}, \Delta} \propto Y_{{\rm X}, \Delta}  \propto M_\Delta^{5/3} E^{2/3}(z).
\end{equation}

\subsection{Fitting Scaling Relations} 
\label{subsec:fitting} 

Our goal is to accurately determine the slopes characterizing the underlying power-law distributions for a variety of X-ray and SZ scaling relations. In this work, we consider the following four forms (described in Sections~\ref{subsec:SPL}\,--\,\ref{subsec:SBPL}) of scaling relation when fitting the data.  

\subsubsection{Simple Power Law}
\label{subsec:SPL}
Under the assumption of self-similarity \citep{Kaiser1986}, the scaling relations characterizing galaxy clusters are expected to follow a simple power-law distribution. Thus, the first model we consider is characterized by
a power-law model with two free parameters ($A$, $\alpha$):
\begin{equation}
    Y = 10^A \left( \frac{X}{X_{\rm norm}}\right)^\alpha \label{eqn:eqnSPL}.
\end{equation}

\subsubsection{Broken Power Law with Fixed Pivot}
Since lower mass haloes have shallower potentials than galaxy clusters, feedback processes such as AGN jets have a strong impact on the distribution of hot gas at the cores of groups and galaxies \citep[e.g.,][]{LeBrun2014}. This can result in a steepening of the slope in X-ray and SZ scaling relations for low mass clusters and groups, leading to a break in the self-similar power law. 
One way this break has been modeled in previous studies \citep[e.g.,][]{LeBrun2017} is by fitting a broken power-law, with the break in the scaling being fixed. In this study, we will test two models for broken power-law scaling relations. 

First, we will perform a broken power-law fit with fixed pivot at $X_{\rm piv}^0=10^{14}$ \ms\, and three free parameters ($A_1$, $\alpha_1$, $\alpha_2$):
\begin{equation}
Y =
    \begin{cases}
    10^{A_1} \left( \frac{X}{X_{\rm norm}}\right)^{\alpha_1}, \;\; X < X_{\rm piv}^0\\
    10^{A_2} \left( \frac{X}{X_{\rm norm}}\right)^{\alpha_2}, \;\; X \geq X_{\rm piv}^0
    \end{cases}\label{eqn:eqnBPLfixed}
\end{equation}

\subsubsection{Broken Power Law with Free Pivot}

We will also perform a broken power-law fit with free pivot $X_{\rm piv}$ and thus four free parameters ($A_1$, $\alpha_1$, $\alpha_2$, $X_{\rm piv}$):
\begin{equation}
 Y =
    \begin{cases}
    10^{A_1} \left( \frac{X}{X_{\rm norm}}\right)^{\alpha_1}\text{, } X < X_{\rm piv} \\
    10^{A_2} \left( \frac{X}{X_{\rm norm}}\right)^{\alpha_2} \text{, } X \geq X_{\rm piv}
    \end{cases}\label{eqn:eqnBPLfree}
\end{equation}
For the last two models describing a broken power-law, requiring the two power-laws to match at the breaking point, $X_{\rm piv}$, will uniquely constrain one of the parameters. In our case, we only allow $A$, $\alpha_1$, $\alpha_2$ (and $X_{\rm piv}$ for the last model) to vary freely, whilst $A_2$ can be derived from the relation: 
\begin{equation}
    A_2 = A_1 + (\alpha_1 - \alpha_2) \log_{10}\left(\frac{X_{\rm piv}}{X_{\rm norm}} \right).
\end{equation}
For each scaling relation, we choose the normalization $X_{\rm norm}$ close to the median of the sample, and perform the least square best-fits in log-space, thus avoiding numerical errors and ensuring that the residuals are normally distributed.
For the broken power-law model with a fixed pivot, we choose $X_{\rm piv}^0=10^{14}$\ms\, in order to ease comparisons to previous studies \citep[e.g.,][]{LeBrun2017}.

\subsubsection{Smoothly Broken Power Law}
\label{subsec:SBPL}
Still, from a physical standpoint, there is no reason to expect a sharp transition for haloes of mass \mcrit\, $\simeq 10^{14}$ \ms\,. Instead, the deviation from hydrostatic equilibrium is expected to occur smoothly as we transition from high mass clusters to smaller groups and galaxies. As a result, the final model we include in our study is a smoothly varying broken power law (SBPL) which asymptotically reproduces the behaviour of a simple power-law for the highest mass clusters. 
The local slope of the smoothly broken power law has a built-in mass (or more generally, X) dependence using a hyperbolic tangent term that ensures the model reduces to simple power laws for X-quantities very far away from the break in the power law. The slope at a given X location is defined as:
\begin{equation}
\alpha_{\rm SBPL} = \frac{{\rm d} \log Y}{{\rm d} \log X} = \frac{\alpha_2 - \alpha_1}{2} \tanh \left[ \frac{1}{\delta} {\rm log}_{10} \left(\frac{X}{X_{\rm p}} \right) \right] + \frac{\alpha_2 + \alpha_1}{2}.\label{eqn:eqnSBPLslope}
\end{equation}
In order to characterize the full SBPL model, we integrate the equation above with respect to X, and we thus obtain the last model considered in this study.
We then fit a smoothly broken power law fit with free pivot ($X_{\rm piv}$) and a total of five free parameters ($A_1$, $\alpha_1$, $\alpha_2$, $\delta$, $X_{\rm pivot}$):
\begin{equation}
\frac{Y(X)}{Y(X_{\rm norm})} =  \left( \frac{X}{X_{\rm norm}} \right)^{ \left(\frac{\alpha_2 + \alpha_1}{2}\right)} \left[ \frac{{\rm cosh} \left(\frac{1}{\delta} {\rm log}_{10} \left( \frac{X}{X_{\rm p}}\right) \right)  }{ {\rm cosh} \left( \frac{1}{\delta} {\rm log}_{10} \left(\frac{X_{\rm norm}}{X_{\rm p}}\right)\right) } \right]^{\left( \frac{\alpha_2 - \alpha_1}{2}\right) \, \delta \, {\rm ln} 10}.\label{eqn:eqnSBPL}
\end{equation}
Here, $\delta$ is a measure of the width of the transition region between the two slopes ($\alpha_1$ for $X \ll X_{\rm pivot}$ and $\alpha_2$ for $X \gg X_{\rm pivot}$, respectively). In the limit of $\delta \rightarrow 0$, the smoothly broken power law model is identical to the broken power law model with free pivot (eqn.~\ref{eqn:eqnBPLfree}). In order to maintain the variable names consistent with the previous models, we define the normalization parameter $A$ through the relation $10^A \equiv Y (X_{\rm norm})$, or in other words, $A$ is the logarithm of the scaling relation evaluated at the normalization point $X_{\rm norm}$.

\subsubsection{Fitting Procedure}
We fit the four models described above for three different samples (the full sample, relaxed and unrelaxed), as well as two different apertures: the true simulation aperture, \rsim\,, and the spectroscopic aperture, \rspec\,. Quantities measured within the halo radius extracted directly from the simulation using the \textsc{subfind} algorithm are denoted with the subscript {\footnotesize SIM}. In order to provide a better match to observations, we also compute X-ray and SZ observables integrated inside the spectroscopic aperture (\rspec) and denote all such quantities with the subscript {\footnotesize SPEC}. To estimate the spectroscopic aperture, \rspec, we first employ the thermodynamic profiles extracted from the synthetic X-ray pipeline detailed in Section \ref{subsec:xraypipeline} in order to compute the cumulative total mass profile of each halo assuming hydrostatic equilibrium (equation \ref{eqn:masseqn}). Then, \rspec\, is defined as the point where the X-ray-derived total density profile of a given halo intersects the value $500 \times \rho_{\mathrm{crit}}$.

Since our sample is not uniformly distributed in mass, we perform the fits by binning our sample by mass rather than fitting each individual object. Due to the shape of the mass function, low mass galaxies are much more common than massive clusters; thus, a naive allocation of equal weights for all galaxies would result in a fit that is dominated by the lowest mass objects in the sample. 
Given the log-normal distribution of halo masses in our sample, we use logarithmically spaced bins. For halo masses \mcrit\, between $10^{12}$ and $10^{14}$\,\ms\, we bin our data in 20 bins, each \mbox{0.1\,dex} apart. In order to maintain at least 5 objects in each mass bin, we split our haloes between $10^{14}$ and $10^{14.6}$\,\ms\, into 3 log-spaced bins \mbox{0.2\,dex} apart, and we include all remaining haloes above $10^{14.6}$\,\ms\, into one last bin.

In choosing the best summary statistics for binned data, we particularly want to avoid introducing any biases that would alter the true value of the underlying slope. The most common choice adopted in previous studies is to use the median of each bin and then perform a single power law fit using the new reduced data set ($x_{\rm median}$, $y_{\rm median}$). However, when fitting a line in log-log space, the relative norm of quantities along the $x$ and $y$ dimensions is irrelevant. Particularly for the highest mass bins which contain very few clusters, using the medians of those bins can artificially skew the slope.
Instead, we aim to use a transformation from the original full sample ($x_i$, $y_i$) to the binned reduced data sample ($\bar{x}$, $\bar{y}$) that preserves the original slope in logarithmic space. Thus, the natural choice is to use the geometric mean of each mass bin, since in log-space the logarithm of the geometric mean is the arithmetic mean of $\log(x)$:
\begin{equation}
\log \left( \sqrt[n]{\Pi_{i=1}^n x_i}\right) = \frac{1}{n} \Sigma_{i=1}^n \log(x_i).
\end{equation}
It can be nicely shown geometrically that, by construction:
\begin{eqnarray}
(x_i, y_i) \rightarrow (\bar{x}_{\rm geom \; mean}, \bar{y}_{\rm geom\; mean}) = (\sqrt[n]{\Pi_{i=1}^n x_i}, \sqrt[n]{\Pi_{i=1}^n y_i}).
\end{eqnarray}
preserves the original slope of the data in log-log space. This property holds as long as geometric means are the summary statistics used for \textit{both} y-axis and x-axis quantities.

The fits are performed using a weighted least-squares fitting routine that assigns different weights based on the level of scatter in each mass bin. For estimating the scatter around the geometric mean values $\bar{x}_{\rm geom \; mean}$ and $\bar{y}_{\rm geom \; mean}$ in each mass bin $i$, we use geometric standard deviations, defined by:
\begin{equation}
{\rm ln}\, \sigma_{\rm geom, \, i} = \sqrt{\frac{\sum_{j=1}^{N_i}( {\rm ln}\,  Y_j - {\rm ln}\, \mu_{\rm geom, i})^2} {N_i}},
\end{equation}
where  $Y_j$ is the value of the y-axis quantity for each object $j$ in mass bin $i$, and $\mu_{\rm geom, i} \equiv \sqrt[N_i]{\Pi_{j=1}^{N_i}  Y_j}$ is the geometric mean of all $Y_j$ values for objects in mass bin $i$, which contains $N_i$ objects in total.
In order to derive the uncertainties of our best-fit parameters, we perform the same analysis on $10^4$ bootstrap samples with replacement. For each resample, we only include in the fit those mass bins that contain at least five clusters.

For all models that we fit, we compute the normalization $A$ of the relations at the points $X_{\rm norm}$, such that $10^A \equiv Y(X_{\rm norm})$. On one hand, the location of the pivot points, $X_{\rm pivot}$, contains rich physical information, as it marks the break from the power law model governing large clusters to the power law model that best characterizes galaxy groups. Yet unlike $X_{\rm pivot}$, the exact location of the normalization point $X_{\rm norm}$ does not greatly impact our model interpretations. In particular, the only constraint we should consider in choosing $X_{\rm norm}$ is that it is located within the same scale range as the $X$ quantity considered in the fit. This has the benefit of reducing numerical errors and improving the fit precision. By construction, there is a clear degeneracy between the choice of $X_{\rm norm}$ and the best-fit normalization parameter: $A = {\rm log}_{10} Y(X_{\rm norm})$. In order to facilitate easier comparisons between our best-fit parameter $A$ and those found by previous studies, we fix $X_{\rm norm} = 10^{14}$\,\ms \, for all our fits. Whilst this places $X_{\rm norm}$ towards the more massive range of our full sample (\mcrit \, $\in [10^{12} - 2\times10^{15}]$\,\ms), it still maintains good numerical accuracy. Thus, our normalization point $X_{\rm norm} = 10^{14}$\,\ms\, is the same as that used in \cite{LeBrun2017} and it is very close to the normalizations used in \cite{Truong2018} (\mcrit\,$\approx 1.5 \times 10^{14}$\,\ms) and \textsc{fable} \citep[\mcrit\,\(= 2 \times 10^{14}\)\,\ms,][]{Henden2019}. In contrast, the normalization point chosen in \textsc{macsis} and \textsc{c-eagle} is somewhat higher \citep[\mcrit\,\(= 4 \times 10^{14}\)\,\ms,][]{Barnes2017a, Barnes2017b}, which in turn will bias their normalization parameters $A$ to higher values than ours.

\subsection{Application to the TNG simulation}
\label{subsec:application}

\begin{figure*}
\centering
\includegraphics[width=0.95\textwidth]{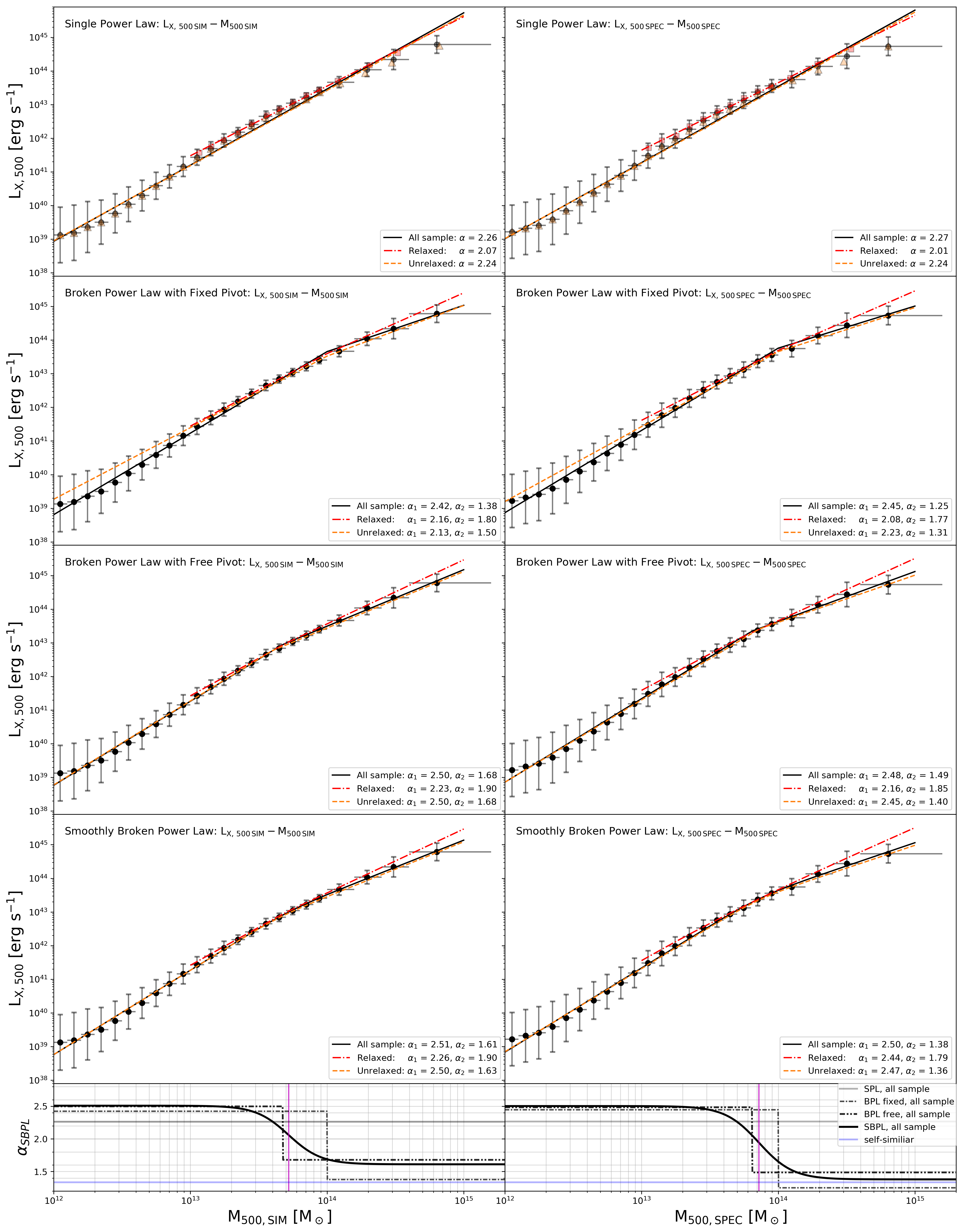}
\caption{Best-fits for the \lxm\, scaling relation. 
From top to bottom, the fits to the data become increasingly better and they correspond
 to a simple power law (SPL, eqn.~\ref{eqn:eqnSPL}), a broken power law with a pivot fixed at \mcrit\, $=10^{14}$\,\ms\, (BPL fixed, eqn.~\ref{eqn:eqnBPLfixed}), a broken power law with free pivot (BPL free, eqn.~\ref{eqn:eqnBPLfree}), and finally, our new model for a smoothly broken power law (SBPL, eqn.~\ref{eqn:eqnSBPL}). The bottom panels show the mass dependence  of the slopes for each model. The SPL model predicts a constant slope (horizontal light grey line). The BPL model with a pivot fixed at $10^{14}$\,\ms\, predicts slopes that are too shallow compared to the data for high mass clusters. 
The BPL model with a free pivot comes closest to the predicted slopes and  pivot location for the SBPL model, but even small changes in pivot location generate significant differences in the slopes predicted for the highest mass clusters. The vertical purple line marks the location of the break in the scaling relation, as predicted by the SBPL model. 
}
\label{fig:Lx8panel}
\end{figure*}

In Figure~\ref{fig:Lx8panel}, we use the \lxm\, scaling relation as a case study to show how the smoothly broken power law model compares to the other models presented in this section. As discussed in Section~\ref{subsec:fitting}, we fit our models to the geometric means of each mass bin (marked with black circles), with error bars corresponding to geometric standard deviations.  The top four rows in Figure~\ref{fig:Lx8panel} show the best-fitting lines obtained from four different models. From top to bottom, these models correspond to a simple power law (SPL, eqn.~\ref{eqn:eqnSPL}), a broken power law with a pivot fixed at \mcrit\,$=10^{14}$\,\ms\, (BPL fixed, eqn.~\ref{eqn:eqnBPLfixed}), a broken power law with free pivot (BPL free, eqn.~\ref{eqn:eqnBPLfree}), and finally,  a smoothly broken power law  model (SBPL, eqn.~\ref{eqn:eqnSBPL}). 
The left column shows quantities measured inside the true \rsim\, aperture from the simulation, while the right column uses the spectroscopic aperture \rspec. We perform the fits to the full sample of haloes (black line), as well as a sample of unrelaxed haloes (orange dashed lines) and a sample of relaxed haloes (red dash-dotted line). In all these scenarios, the SBPL provides robust predictions for the slope dependence on halo mass and it correctly identifies the break in the scalings.
The legends in the bottom-right corners of each panel include the best-fitting slope for the SPL, and the preferred slope at the lowest masses ($\alpha_1$) and at the highest masses ($\alpha_2$) for the other three models.
Due to the large range of scales shown in these figures, the differences between the four models may seem relatively small. Upon closer inspection of the best-fitting slopes, our results show that the predicted slopes, especially for high-mass clusters, are significantly different among the four models we consider. The SPL model strongly overpredicts the slope at the high-mass end, as the lower mass bins force a steep tilt in the model. The BPL model with a fixed pivot provides a better fit but it underestimates the slope at the highest mass end of the sample due to the constraint that \xp\,$=10^{14}$\,\ms. In IllustrisTNG, the  break in \lxm\, occurs at \mcrit\,$<10^{14}$\,\ms, as suggested by both the BPL model with a free pivot and the SBPL model. 
The bottom panels present the slopes for each model as a function of halo mass. The SPL model predicts a constant slope (horizontal light grey line). The BPL model with a pivot fixed at $10^{14}$\ms\, predicts slopes that are too shallow compared to the data for high mass clusters. The vertical purple line marks the location of the break in the scaling relation, as predicted by the SBPL model. 
While the BPL model with a free pivot comes closest to the predicted slopes and the pivot location for the SBPL model, small changes in the pivot location can translate to significant biases in the predicted slopes. In particular, the self-similar prediction, $L_{\rm X} \propto M^{4/3}$ is shown as a blue horizontal line in the bottom panels of Fig.~\ref{fig:Lx8panel}. The BPL model with a free pivot predicts a break that is only mildly shifted to smaller masses than the break in the SBPL model. However, as seen in the top panels, the \lxm\, scaling relation does not exhibit a sharp break. Instead, the slope becomes gradually steeper with decreasing halo mass, from clusters down to groups and galaxies. The SBPL model correctly captures this behaviour and as a result, it also offers more accurate predictions of the slope at cluster scales. For quantities measured inside the spectroscopic aperture (right column of Fig.~\ref{fig:Lx8panel}), the BPL model with a free pivot predicts a slope that is steeper than self-similar. However, the SBPL model predicts a slope consistent with the self-similar prediction, $\alpha = 4/3$, for all clusters above \mcrit\,$\simeq 4 \times 10^{14}$\,\ms.

\begin{figure}
\centering
\includegraphics[width=0.5\textwidth]{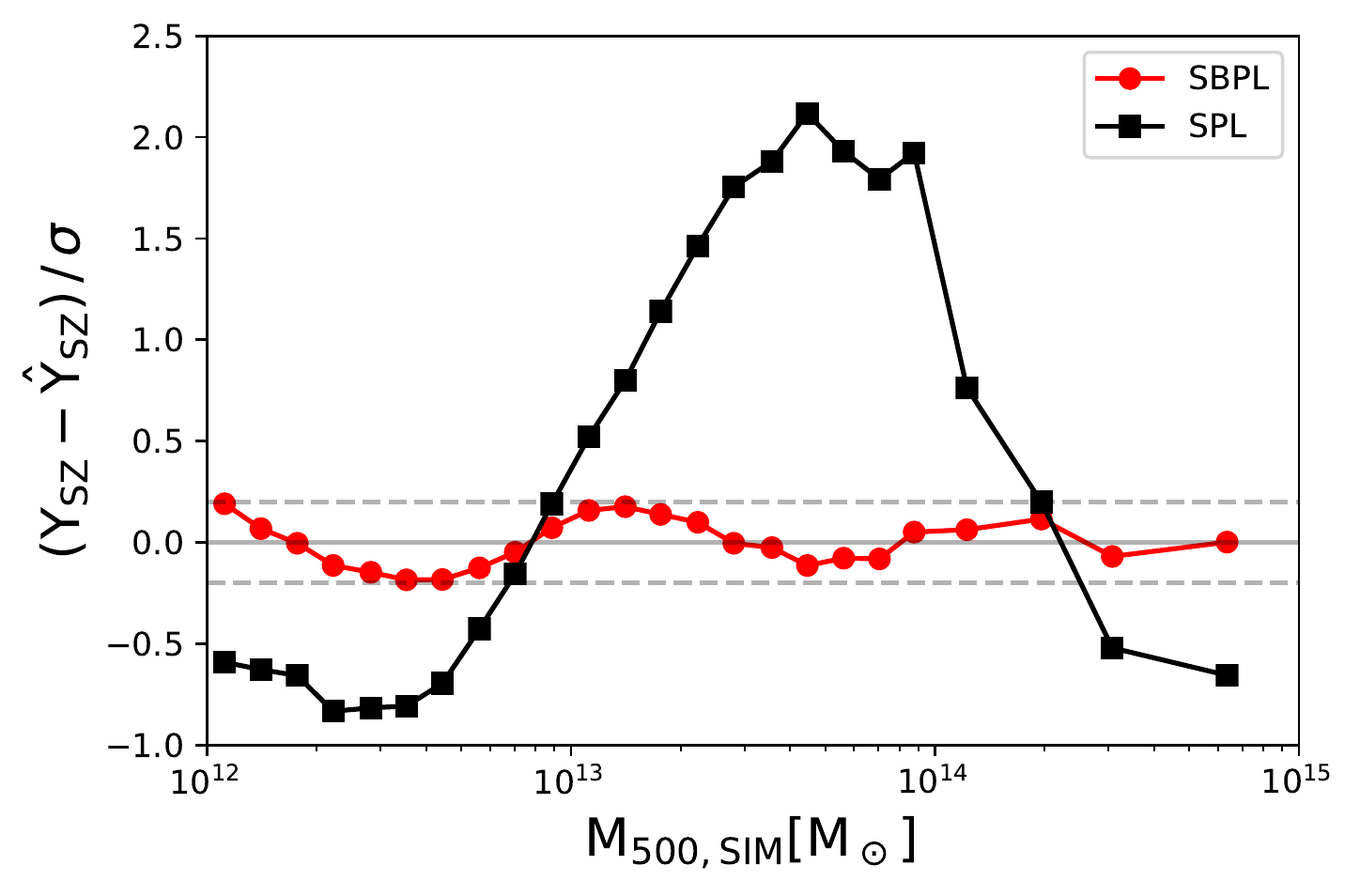}
\caption{Comparison between the mass dependence of residuals from the SPL and SBPL models. The residuals are computed as the difference between the geometric mean of \ysz\,  for haloes in a given bin and the expected \ysz\, from the SPL or SBPL model, normalized to the variance of \ysz\, values in that mass bin. Using the SBPL model, we find residuals that are randomly distributed around zero, with deviations $\lesssim \pm 0.2$. On the other hand, a simple SPL fit suffers from extremely large residuals (more than 10 times larger than SBPL residuals in the galaxy group mass range ($10^{13}-10^{14}$\,\ms).}
\label{fig:Yszresiduals}
\end{figure}

Upon closer inspection, we can see that the SPL model tends to over-predict the luminosities of galaxies and clusters, while under-predicting the luminosities of intermediate-mass galaxy groups. 
In Figure~\ref{fig:Yszresiduals}, we compare the residuals between the data and the predictions of the SPL and SBPL models, respectively. This time, we use the \yszm\, scaling relation as our case study, although we checked that the trends are similar among all scaling relations considered in this paper. A robust model should not only have small residuals, but those residuals should also be randomly distributed around the null value. We find that the SBPL model fulfills both of these requirements -- the residuals oscillate around $Y_{\rm SZ}$ -- $\hat{Y}_{\rm SZ} \simeq 0$ without any clear mass dependence. Overall, residuals stay below $\pm 0.2$ for the entire mass range. This is an impressive result for the SBPL model, considering that it only has 5 free parameters and yet it is able to capture the evolution of the scaling relations over three orders of magnitude in halo mass. In comparison, the SPL model suffers from extremely large residuals, which are more than $10\times$ larger than those of the SBPL model in the galaxy group mass range ($10^{13} - 10^{14}$\,\ms). 
In Sections~\ref{sec:xrayresults} and \ref{sec:yszresults}, we compare the best-fit parameters obtained from fitting X-ray and SZ scaling relation using the simple power law model (employed commonly in previous studies) to the results obtained using the smoothly broken power law model introduced in this section.

\section{X-ray Scaling Relations}
\label{sec:xrayresults}

\subsection{Gas mass - total mass scaling relation}
\label{subsec:scalingsmgas}

\begin{figure*}
\centering
\includegraphics[width=0.99\textwidth]{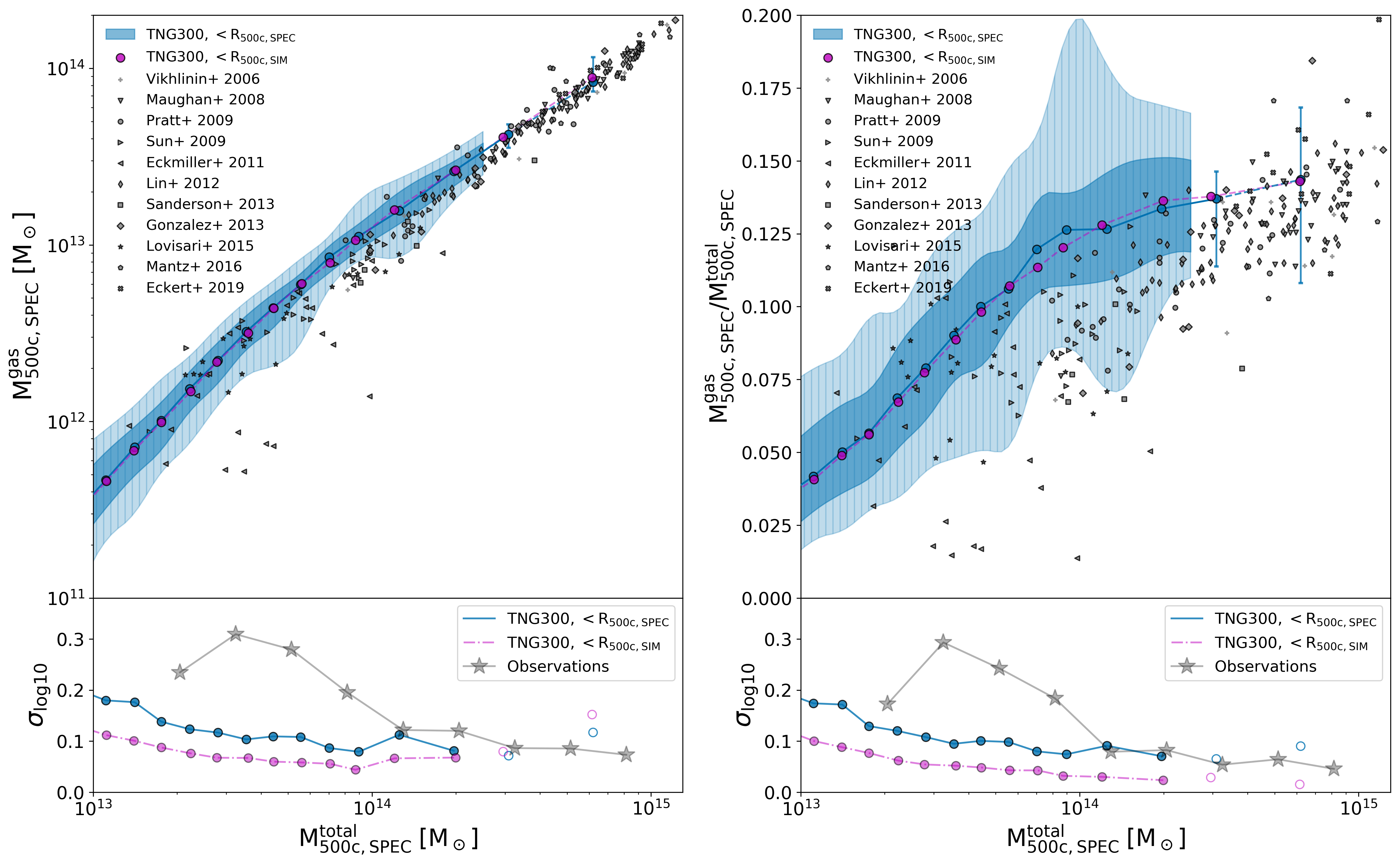}
\caption{\textit{Left panel:} Gas mass within \rcrit\, as a function of the total halo mass at $z=0$ in TNG300. Blue dots mark the median values for \mgspec\, as a function of the total mass measured inside the corresponding spectroscopic aperture, \mspec. Dark (light) blue contours mark $1 \sigma$ (and $2 \sigma$) scatter around the median values for bins containing 10 haloes or more, with contour limits smoothed using quadratic interpolation. The light purple dots show the median values of \mgsim\, as a function of the total mass measured inside the corresponding simulation aperture, \msim. 
\textit{Right panel:} Gas mass fraction within \rcrit\, as a function of the total halo mass at $z=0$ in TNG300. Blue dots mark the median values for \mgspec / \mspec\, as a function of the total mass measured inside the corresponding spectroscopic aperture, \mspec. Dark (light) blue contours mark $1 \sigma$ (and $2 \sigma$) scatter around the median values for bins containing 10 haloes or more, with contour limits smoothed using quadratic interpolation. The light purple dots show the median values of \mgsim/\msim\, as a function of the total mass measured inside the corresponding simulation aperture, \msim.
For comparison, we show in both panels grey symbols representing observations from \citet{Vikhlinin2006, Maughan2008, Pratt2009, Sun2009, Eckmiller2011, Lin2012, Sanderson2003, Gonzalez2013,  Lovisari2015, Mantz2016b, Eckert2019} .
The bottom panels show the level of intrinsic scatter $\sigma_{{\rm log} 10}$ for the gas mass (left) and gas mass fraction (right), measured inside \rspec \, (blue dots + line) and \rsim \, (purple dots + dashed line). We compare these results to grey stars indicating the average intrinsic scatter in observations, computed over the combined sample of all observational data points shown in the top panels.
}
\label{fig:Mgascontour}
\end{figure*}

\begin{figure*}
\centering
\includegraphics[width=0.99\textwidth]{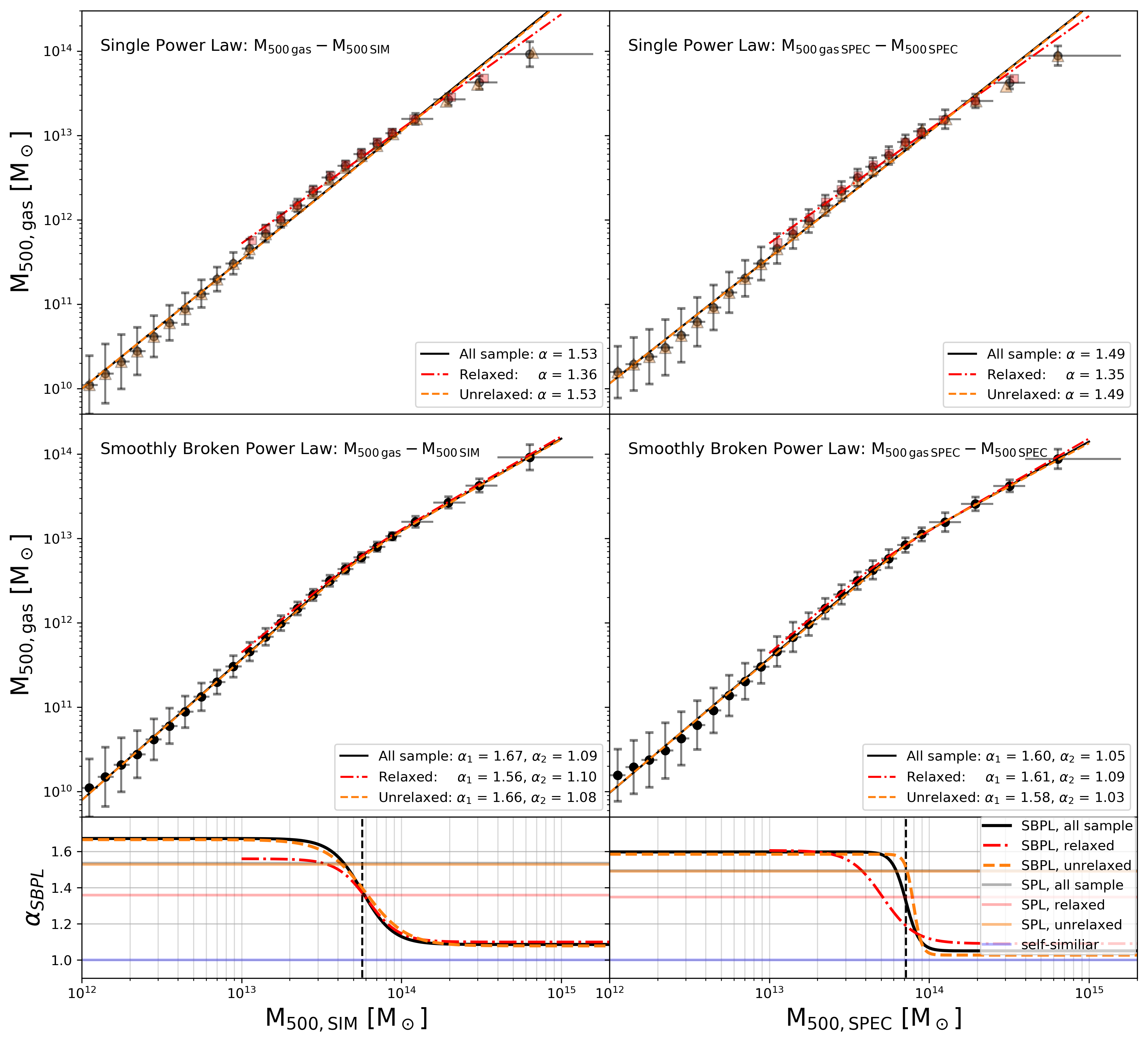}
\caption{Best-fits for the \mgm\, scaling relation at $z=0$ in TNG300. Left column shows quantities measured inside the true \rsim\, aperture from the simulation, while the right column uses the spectroscopic aperture \rspec. Black circles mark the geometric mean in mass bins {0.1\,dex} apart from $10^{12}$ to $10^{14}$\,\ms\, and \mbox{0.2\,dex} apart for higher masses. Black error bars indicate the geometric standard deviation in each mass bin. In the top row, we show the best-fitting simple power law (SPL, eqn.~\ref{eqn:eqnSPL}) fit for the relaxed (red dashed-dotted line), unrelaxed (orange dashed line), as well as the full sample of galaxies (black solid line). The second row panels show the best-fitting smoothly broken power law models (SBPL, eqn.~\ref{eqn:eqnSBPL}), which provide a better fit to the data across all mass scales. The bottom panels present the slope $\alpha_{\rm SBPL}$ (eqn.~\ref{eqn:eqnSBPLslope}) as a function of halo mass. The location of the best-fitting pivot which marks the break in the scaling relations is shown with a vertical dashed black line. Horizontal thin lines correspond to the self-similar prediction of $\alpha = 1$ (blue line) and the best-fitting simple power law fits for the full sample, as well as the relaxed and unrelaxed haloes.}
\label{fig:Mgas4panel}
\end{figure*}

In Figure~\ref{fig:Mgascontour}, we show the gas mass and gas fraction of IllustrisTNG haloes as a function of halo mass. Since we are interested in gas mass estimates derived from X-ray observations, the masses we present refer only to the hot gas component. We remove all cold gas ($<10^{5}$\,K), as well as any star-forming gas or cells that have a net positive cooling rate. This cut only removes less than one percent of the total number of gas cells in our haloes and is motivated by the fact that emission from gas with temperatures $<10^{5}$\,K is below the lower detection limit of typical X-ray surveys. In mimicking X-ray measurements, which remove small-scale X-ray clumps, we also remove any mass associated with self-gravitating structures by excluding gas cells bound to  subhaloes from our $M_{\rm gas}$ and $M_{\rm tot}$ estimates. 

We present in Figure~\ref{fig:Mgascontour} the medians of the gas masses and gas fractions computed for each mass bin, with quantities measured both within the \textsc{subfind}-estimated aperture, \rsimc\, (light purple), as well as the spectroscopic aperture, \rspecc\, (light blue). Additionally, for masses measured within \rspec, we also show 1 and 2 $\sigma$ scatter (dark and light blue contours, respectively) for those mass bins containing at least ten objects. We compare our results to a range of observational data \citep{Vikhlinin2006, Maughan2008, Pratt2009, Sun2009, Eckmiller2011, Lin2012, Sanderson2003, Gonzalez2013,  Lovisari2015, Mantz2016b, Eckert2019} represented by different grey symbols in Figure~\ref{fig:Mgascontour}. We find that IllustrisTNG clusters (\msim\,$\gtrsim 2\times 10^{14}$\,\ms) are in relatively good agreement with observations. However, galaxy groups between $\approx 5\times 10^{13}$\,\ms\,-- $2\times 10^{14}$\,\ms\, lie on the upper end of the scatter in observations. At even smaller halo masses, we find better agreement with observations, but there is large scatter in current X-ray estimates in this mass range. A few of the higher mass galaxy groups and small clusters in IllustrisTNG scatter above the universal baryon fraction, $f_{\rm b} = \Omega_{\rm b} / \Omega_{\rm m} = 0.157$, as can been seen from the $2\sigma$ light-blue contours. This happens only for measurements within \rspec, which suffer from a higher level of scatter and because the X-ray mass bias leads to \mspec\, being biased low compared to \msim.

We do not find a significant difference in the median gas fractions based on the choice of aperture (\rsimc\, or \rspecc). This can be explained by the relatively flat cumulative $f_{\rm gas}$ profile at a radius $\simeq$\rsim\, away from the cluster center. However, X-ray estimates for $R_{\rm 500crit}$ introduce a significantly larger scatter in gas mass measurements. In the bottom panels of Figure~\ref{fig:Mgascontour}, we compare the intrinsic scatter for $M_{\rm gas}$
and $f_{\rm gas}$ in each mass bin, for quantities evaluated inside \rsim\, (blue line) and \rspec\, (purple dashed line), respectively. For IllustrisTNG groups and clusters, we find that spectroscopic mass estimates can almost double the $\sigma_{\rm log10}$ scatter in the $M_{\rm gas}$ -- $M_{\rm tot}$ and $f_{\rm gas}$ -- $M_{\rm tot}$ relations. We also compare these estimates to the average intrinsic scatter in observations (shown with grey stars in Figure~\ref{fig:Mgascontour}), computed over the combined sample of all observational data points from the top panels. We find that $M_{\rm gas}$
and $f_{\rm gas}$ measured within \rspec\, produce a similar level of scatter as that seen in observations of galaxy clusters. Below $\sim 10^{14}$\,\ms, there is a large jump in the intrinsic scatter measured from observational samples. However, this is likely driven by uncertainties in estimating gas masses for lower mass galaxy groups, as well as discrepancies between the calibrations adopted by different surveys. In IllustrisTNG, the intrinsic scatter in $M_{\rm gas}$ - $M_{\rm tot}$ shows a mild mass dependence, with $\sigma_{\rm log10}$ decreasing with halo mass, from $\sigma_{\rm log10} \simeq 0.19$ at \mspec\,$=10^{13}$\,\ms\, to $\sigma_{\rm log10} \simeq 0.08$ at \mspec\,$=10^{14}$\,\ms.

The self-similar expectation for the $M_{\rm gas}$--$M_{\rm tot}$ scaling relation is $\alpha=1$ (see Section~\ref{subsec:selfsimilar}), which corresponds to a constant gas fraction within \rcrit. 
However, radiative cooling and star formation play an important role in converting a fraction of the hot gas reservoir into stars. As shown by previous numerical studies comparing radiative and non-radiative simulations, the higher star formation efficiency in groups relative to clusters can cause a tilt to $\alpha > 1$ for the $M_{\rm gas}$--$M_{\rm tot}$ relation \citep[e.g.,][]{Stanek2010, Battaglia2013, Planelles2014, LeBrun2017}. There is, nonetheless, a potentially even more important player in determining the best-fit slope. AGN feedback can more efficiently eject the hot gas at the centers of galaxy groups beyond \rcrit, due to the lower potential wells of groups compared to clusters \citep[see, e.g.,][]{Puchwein2008, Fabjan2011, McCarthy2011, Planelles2013, Gaspari2014,LeBrun2014}.

In Figure~\ref{fig:Mgas4panel}, we show the geometric means and standard deviations for the scaling relation \mgm. Top panels include best-fitting lines corresponding to the simple broken power law (SPL) model in equation~(\ref{eqn:eqnSPL}), while middle panels show our results for a smoothly broken power law (SBPL) model, as defined in 
equation~(\ref{eqn:eqnSBPL}). 
We perform the fits both for quantities measured inside \rsim\, (left column), as well as inside \rspec\, (right column). All the models are fit to the geometric means of each mass bin and they take into account the error equal to the corresponding geometric standard deviations for each bin. We also investigate the best-fits, not just for the full sample (shown in black), but also for the relaxed clusters (shown in red) and unrelaxed haloes (shown in orange) in our simulation. 
The simple power law model predicts very steep slopes: $\alpha_{\rm SPL}^{\rm SIM} = 1.534^{+0.004}_{-0.004}$ and $\alpha_{\rm SPL}^{\rm SPEC} = 1.493^{+0.030}_{-0.008}$ for the full sample. These slopes are not only much higher than the self-similar prediction, $\alpha=1$, but they are also much steeper than the slopes predicted by previous simulations, such as \textsc{fable}: $1.25^{+0.04}_{-0.04}$ \citep{Henden2019} or \textsc{macsis}: $1.25^{+0.01}_{-0.03}$ \citep{Barnes2017a}. These overly steep slopes are entirely the result of our sample extending to much lower masses than the samples considered in previous studies. The lower mass galaxy groups in our sample force the SPL model to become significantly more steep. This effect is evident from inspecting the top panels of Figure~\ref{fig:Mgas4panel}, where the SPL best-fits over-predict the amount of gas in the most massive bins. Several previous studies have chosen to perform mass cuts in their sample, throwing away data from galaxy groups, in an effort to reduce the bias in the best-fitting slopes \citep[e.g.,][]{Henden2019}. However, with the smoothly broken power law model we introduce in this paper, we can now provide a much better fit for samples that include both massive clusters and low mass galaxy groups.
The best-fitting parameters for all our fits are summarized in Table~\ref{tab:tabmg}.
For the \rsim\, aperture, the SBPL model prefers a slope of $\alpha_1=1.671^{+0.002}_{-0.002}$ for low mass groups and a slope of $\alpha_2 = 1.086^{+0.014}_{-0.013}$ at the high mass end. This confirms that the very steep slope preferred by the SPL model was caused by the lower mass haloes in our sample. For massive clusters in IllustrisTNG, the SBPL model predicts a slope that is only mildly steeper than self-similar. We also find that the slopes at the high mass end are consistent with each other irrespective of the chosen aperture ($\alpha_2^{\rm SIM} = 1.086^{+0.014}_{-0.013}$ \, and $\alpha_2^{\rm SPEC} = 1.051^{+0.037}_{-0.042}$). For \rspec, the best-fitting slope for the highest mass clusters is only $\sim1\sigma$ above the self-similar expectation. 

\begin{figure*}
\centering
\includegraphics[width=0.7\textwidth]{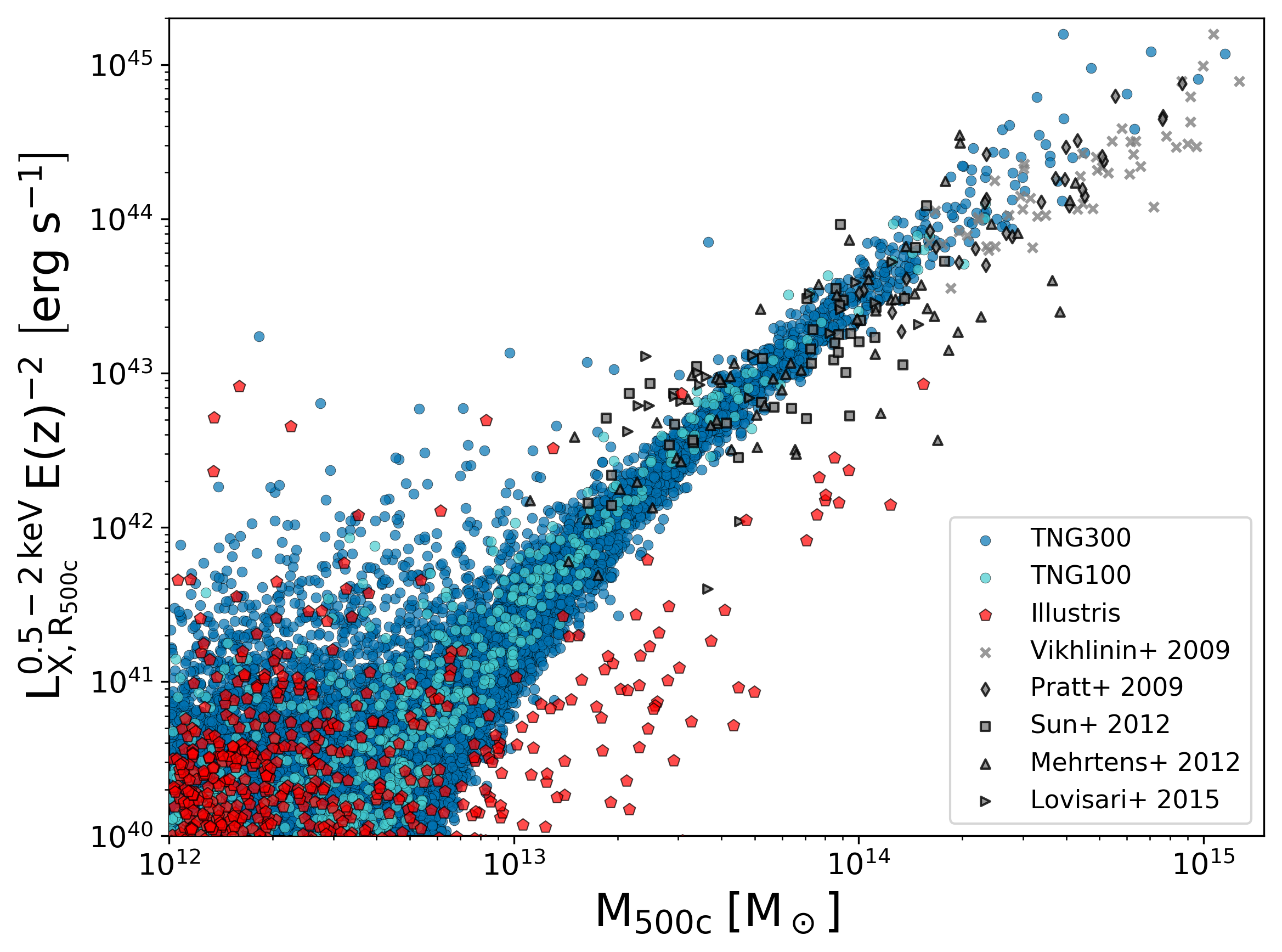}
\caption{Scatter points showing the soft-band (0.5 - 2 keV) X-ray luminosity as a function of halo mass at $z=0$. Colored data points correspond to simulated haloes from Illustris (red pentagons), TNG100 (cyan circles), and TNG300 (blue circles). We compare to observational data (grey symbols) from \citet{Vikhlinin2009b, Pratt2009, Sun2012, Mehrtens2012, Lovisari2015}. The subgrid model for AGN feedback in Illustris was too strong and evacuated most of the hot gas within \rsim\, of the cluster center. This led to unrealistically low X-ray luminosities. On the other hand, the new AGN feedback model in IllustrisTNG shows a great improvement compared to the previous generation of simulations. IllustrisTNG groups and clusters reproduce the observed trend with total mass, and they lie on the upper end of the scatter in observations.  Note also the agreement between TNG100 and TNG300, indicating numerical convergence.}
\label{fig:Lxillustris}
\end{figure*}

As expected, the SPL slope predictions have values in-between the $\alpha_1$ and $\alpha_2$ values predicted by the SBPL model. This is because the SPL model is forced to find a compromise between the slope preferred by low and high mass haloes. In the bottom panels of Figure~\ref{fig:Mgas4panel}, we include the mass dependence of the best-fitting slope, $\alpha_{\rm SBPL}$ (eqn.~\ref{eqn:eqnSBPLslope}), for all three samples we are considering. Horizontal thinner lines mark the self-similar slope (in blue), as well as the SPL best-fitting slopes from the top panels. For quantities measured inside \rsim, the SBPL model predicts slopes steeper than self-similar at all masses. For the spectroscopic aperture, the SBPL predictions are overall shifted to shallower slopes, and $\alpha_{\rm SBPL}$ asymptotically converges to a value only mildly steeper than self-similar above $10^{14}$\,\ms.

The fact that the SBPL model is providing a much better fit than the SPL model provides evidence for a break in the \mgm\, scaling relation.
The location of the best-fitting pivot, \xp, is shown with a vertical black dashed line in the bottom panels of Figure~\ref{fig:Mgas4panel}. For the full sample, we find \xp\,$\simeq 5.7 \times 10^{13}$\,\ms\, and \xp\,$\simeq 7.2 \times 10^{13}$\,\ms\, for the simulation and spectroscopic apertures, respectively. Moreover, the break appears to be smoother for quantities measured inside \rsim: $\delta^{\rm SIM} = 0.19 \pm 0.02$ versus $\delta^{\rm SPEC} =$ \,\val[0.08]{0.04}{0.03}.

If we focus only on the sample of relaxed groups and clusters, the SPL best-fit slope ($\alpha_{\rm SPL} = 1.358^{+0.031}_{-0.009}$) is considerably steeper than the prediction ($\alpha_{\rm SPL} = 1.534^{+0.004}_{-0.004}$) for the full sample. We note that we only perform the fit down to \msim $=10^{13}$\,\ms\, for the relaxed sample, because below this mass, most of our mass bins do not contain sufficient (if any) relaxed objects. Thus, given the overall higher median masses in the relaxed sample, it is not surprising to notice that the SPL slope for relaxed clusters shifts closer to the SBPL prediction for high masses. Nevertheless, our sample of relaxed haloes still contains a significant amount of contamination from group scales which drives the overall slope to be higher than those reported for the \textsc{macsis} relaxed sample \citep[\(1.05^{+0.04}_{-0.04}\)\,,][]{Barnes2017a} or the \textsc{c-eagle} relaxed clusters \citep[\(1.09^{+0.03}_{-0.08}\)\,,][]{Barnes2017b}. Once we switch to the SBPL fit, we find the best-fit slopes at cluster scales, $\alpha_2^{\rm SIM} = 1.100^{+0.021}_{-0.027}$ and $\alpha_2^{\rm SPEC} = 1.091^{+0.026}_{-0.040}$, to be in good agreement with previous results for relaxed clusters  \cite{Barnes2017a, Barnes2017b}. Our SBPL results are also consistent with observations of relaxed clusters: $1.04\pm 0.05$ \citep{Mantz2016a}.

Previous numerical studies of scaling relations analyzed samples composed almost exclusively of galaxy clusters, and sometimes high mass galaxy groups. For example, \citet{Henden2019} apply a mass cut of \mcrit $> 3\times 10^{13}$\ms\, to their \textsc{fable} sample and \textsc{macsis} only includes clusters with  \mcrit $> 10^{14}$\,\ms\, \citep{Barnes2017a}. 
Our results are in broad agreement with previous simulations. \citet{Truong2018} find $\alpha = 1.08 \pm 0.01$, which is a very good match to the slope predicted by our SBPL model for clusters ($\alpha_2 = 1.086^{+0.014}_{-0.013}$). Both \textsc{fable} and \textsc{macsis} predict slightly steeper slopes: $1.25 \pm 0.04$ and \val[1.25]{0.01}{0.03}, respectively. This can be explained by differences in sample selection, because as we have seen in the bottom panel of Figure~\ref{fig:Mgas4panel}, simple power laws tend to predict significantly steeper slopes as soon as low mass clusters and groups are included in the sample. To further strengthen this hypothesis, \citet{Henden2019} report a significantly shallower slope of $1.16 \pm 0.05$ when they shift their mass cut to \mcrit\, $>10^{14}$\ms. \citet{LeBrun2017} fit both a simple power law ($\alpha = 1.32 \pm 0.02$) as well as a broken power law with a pivot fixed at \xp\,$=10^{14}$\,\ms, for which they find a slope of $1.18 \pm 0.02$ at the high mass end. 

The slopes preferred by IllustrisTNG clusters are in very good agreement with several observational studies that predict \mgm\, slopes of $1.04\pm 0.10$ \citep{Mahdavi2013}, $1.04 \pm 0.05$ \citep{Mantz2016a}, $1.13 \pm 0.03$ \citep{Lin2012}, and \val[1.21]{0.11}{0.10} \citep{Eckert2019}. On the other hand, a few other observational samples predict slightly steeper slopes, such as $1.25 \pm 0.06$ \citep{Arnaud2007}, $1.26 \pm 0.03$ \citep{Gonzalez2013}, and $1.32 \pm 0.07$ \citep{Chiu2018}. Overall, our results from IllustrisTNG agree with the consensus from both observations and simulations that \mgm\, slopes are steeper than self-similar. 

\subsection{X-ray luminosity - total mass scaling relation}
\label{subsec:xraymass}

In this subsection, we present the dependence of the soft-band (0.5 - 2 keV) X-ray luminosity, $L_{{\rm X, R}_{\rm 500c}}^{\rm 0.5 - 2\, keV}$, on the total halo mass. 
The X-ray luminosity in clusters and groups is highly sensitive to the thermal structure of the intracluster medium and the density profile of hot gas. Due to this, the \lxm\, scaling relation is one of the best probes for testing different AGN feedback models and their role in shaping the distribution of X-ray emitting gas. 
To exemplify how different choices of subgrid models for AGN feedback can drastically alter \Lx\, measurements, in Figure~\ref{fig:Lxillustris} we compare the X-ray luminosities of IllustrisTNG haloes against those of the previous generation Illustris simulation. IllutrisTNG clusters from both the TNG300 (dark blue circles) and TNG100 (cyan circles) simulations show relatively good agreement with observations. 
On the other hand, the AGN feedback model implemented in Illustris (red pentagons) was too extreme in its evacuation of significant amounts of the hot gas inside massive galaxies \citep{Genel2014}, leading to a strong suppression of X-ray luminosities at all mass scales. 
We further note that the good agreement between TNG100 and the lower resolution TNG300, as indicated by Figure~\ref{fig:Lxillustris}, implies that the results are reasonably well-converged numerically.

\begin{figure*}
\centering
\includegraphics[width=0.99\textwidth]{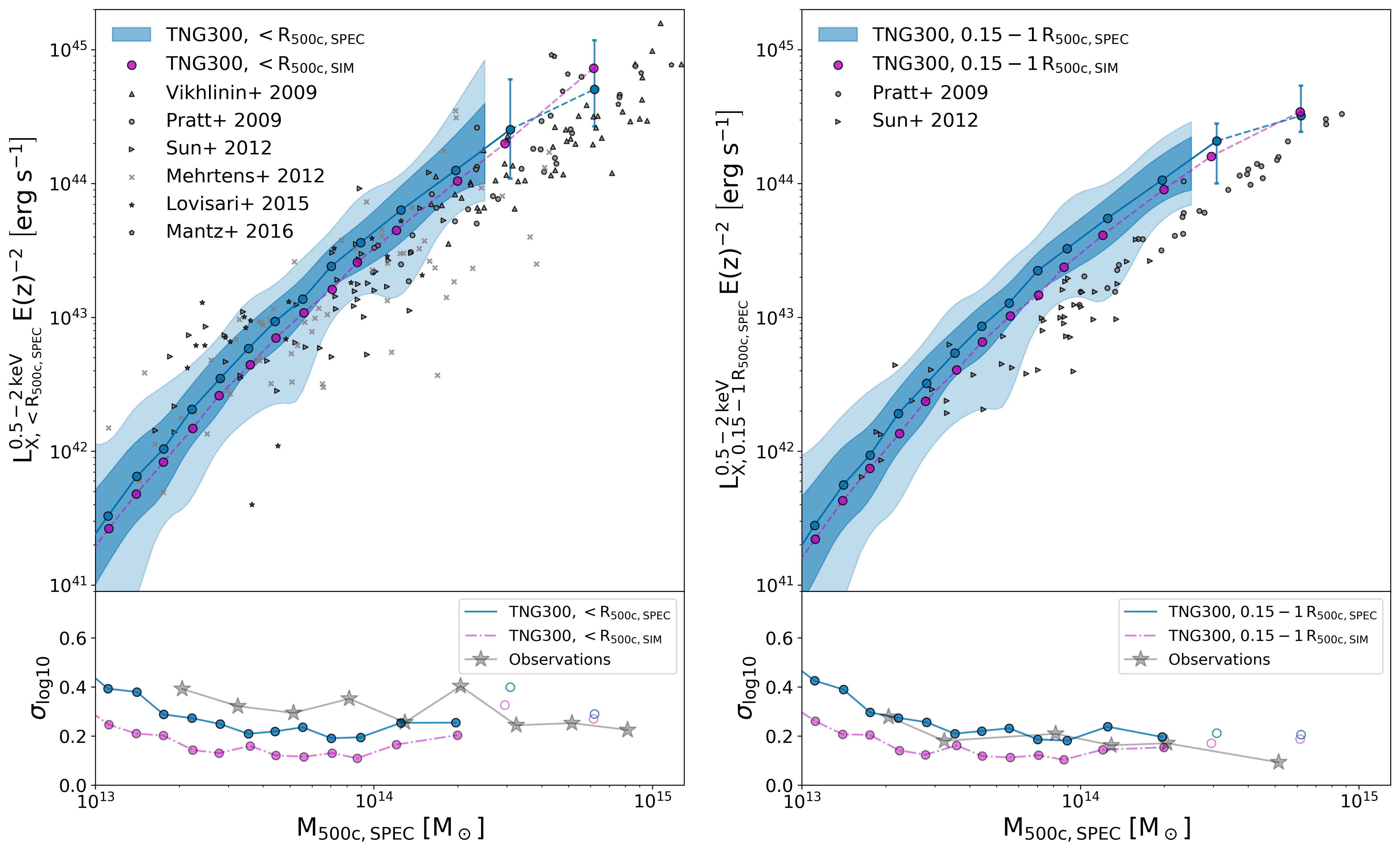}
\caption{\textit{Left panel:} Soft band X-ray luminosity within \rcrit\, as a function of the total halo mass at $z=0$ in TNG300. Blue dots mark the median values for \Lxspec\, as a function of the total mass measured inside the corresponding spectroscopic aperture, \mspec. Dark (light) blue contours mark $1 \sigma$ (and $2 \sigma$) scatter around the median values for bins containing 10 haloes or more, with contour limits smoothed using quadratic interpolation. The light purple dots show the median values of \Lxsim\, as a function of the total mass measured inside the corresponding simulation aperture, \msim. For comparison, we show grey symbols representing observations from \citet{Vikhlinin2009b, Pratt2009, Sun2012, Mehrtens2012, Lovisari2015, Mantz2016b}.
\textit{Right panel:} Core-excised soft band X-ray luminosity as a function of the total halo mass at $z=0$ in TNG300. Blue dots mark the median values for \Lxcespec\, as a function of the total mass measured inside the corresponding spectroscopic aperture, \mspec. Dark (light) blue contours mark $1 \sigma$ (and $2 \sigma$) scatter around the median values for bins containing 10 haloes or more, with contour limits smoothed using quadratic interpolation. The light purple dots show the median values of \Lxcesim\, as a function of the total mass measured inside the corresponding simulation aperture, \msim.
For comparison, we show grey symbols representing observations for core-excised X-ray luminosities from \citet{Pratt2009} and \citet{Sun2012} .
The bottom panels show the level of intrinsic scatter $\sigma_{{\rm log} 10}$ for the soft band X-ray luminosity, measured inside \rspec \, (blue dots + line) and \rsim \, (purple dots + dashed line). We compare these results to grey stars indicating the average intrinsic scatter in observations, computed over the combined sample of all observational data points shown in the top panels.}
\label{fig:Lxcontour}
\end{figure*}


As seen in Figure~\ref{fig:Lxcontour}, IllustrisTNG groups and clusters are in overall good agreement with observations. 
Quantities measured within the simulation aperture, \Lxsim\, and \msim, are shown in light purple, while quantities measured within the spectroscopic aperture, \Lxspec\, and \mspec, are shown in light blue. We also include the $1$ and $2 \sigma$ scatter for \Lxspec\, -- \mspec, across all bins that contain at least ten objects. 
In the left panel of Figure~\ref{fig:Lxcontour}, we present median values of the soft-band X-ray luminosity measured from $r=(0-1)$\rcrit, $L_{\rm X, R_{\rm 500c}}^{\rm 0.5-2 keV}$. We compare these results to relevant observational data from \citet{Vikhlinin2009b, Pratt2009, Sun2012, Mehrtens2012, Lovisari2015, Mantz2016b}. IllustrisTNG data reproduces the observed trend in \Lx\, with increasing mass, and it also shows excellent agreement with observations of intermediate and low mass groups. For more massive objects, IllustrisTNG medians lie on the upper end of the scatter in observations. 

In the right panel of Figure~\ref{fig:Lxcontour}, we present median values for the core-excised soft-band X-ray luminosity measured within $r=(0.15 - 1)$\rcrit, $L_{{\rm X, (0.15-1)\, R}_{\rm 500c}}^{\rm 0.5-2 keV}$. As we would expect, our data indicates that removing the core region ($r<0.15$\rcrit) leads to a suppression in the amount of scatter measured at cluster scales. Moreover, the highest mass bins show $L_{{\rm X, (0.15-1)\, R}_{\rm 500c}}^{\rm 0.5-2 keV}$ biased low compared to $L_{{\rm X, R}_{\rm 500c}}^{\rm 0.5-2 keV}$ measurements. These trends are also visible in observational data  \citep{Pratt2009, Sun2012}.

For a more detailed analysis of the intrinsic scatter in the \lxm\, scaling, we include in the bottom panels of Figure~\ref{fig:Lxcontour} results for $\sigma_{\rm log10}$ computed for simulation and spectroscopic apertures. Just like in the case of the \mgm\, scaling relation, spectroscopic apertures lead to a significant increase in the level of intrinsic scatter. We find quite a bit of variation from one mass bin to another, but on average, $\sigma_{\rm log10}$ decreases with increasing halo mass. This effect is even more pronounced for core-excised luminosities. Compared to the intrinsic scatter measured in the combined sample of all observational data points included in the top panels, $\sigma_{\rm log10}$ in $L_{{\rm X, R}_{\rm 500c}}^{\rm 0.5-2 keV}$ is about 50\% too low. Estimates that take into account the X-ray mass bias (\Lxspec\, -- \mspec) come closer to reproducing the level of scatter seen in observations. For core-excised luminosities, measurements inside the spectroscopic apertures are in excellent agreement with observations, reproducing both the normalization and trend with mass from $\sim 2\times 10^{13}$\ms\, all the way to $\sim 2 \times 10^{14}$\ms. For the highest mass bins we only have data from \citet{Pratt2009}, which shows a very small level of scatter, below both our \rsim\, and \rspec\, estimates. 

Compared to observations, IllustrisTNG simulations predict core-excised luminosities that lie at the upper end of the scatter in observational data. Accounting for the bias from X-ray estimated masses further increases the normalization of the predicated \lxm\, scaling. The fact that galaxy groups between $\approx 5\times 10^{13}$\ms\,-- $2\times 10^{14}$\ms\, appear to be too gas rich in IllustrisTNG (see Figure~\ref{fig:Mgascontour}) is likely the biggest culprit. X-ray luminosities are highly sensitive to the total amount of X-ray emitting hot gas inside \rsim\,, as well as to the density distribution of the gas. Thus, it is unsurprising that the general trends observed in \mgm\, also translate to the \lxm\, scaling. While gas fractions in IllustrisTNG seem to be in overall agreement with observations for massive clusters, X-ray luminosities continue to be biased high even in this mass range. This hints that second order effects such as the shape of the gas density profile, the temperature profile of the hot gas, or the presence of high-density X-ray-bright gas clumps could further bias \Lx\, measurements to too high values at the cluster scale.

\begin{figure*}
\centering
\includegraphics[width=0.9\textwidth]{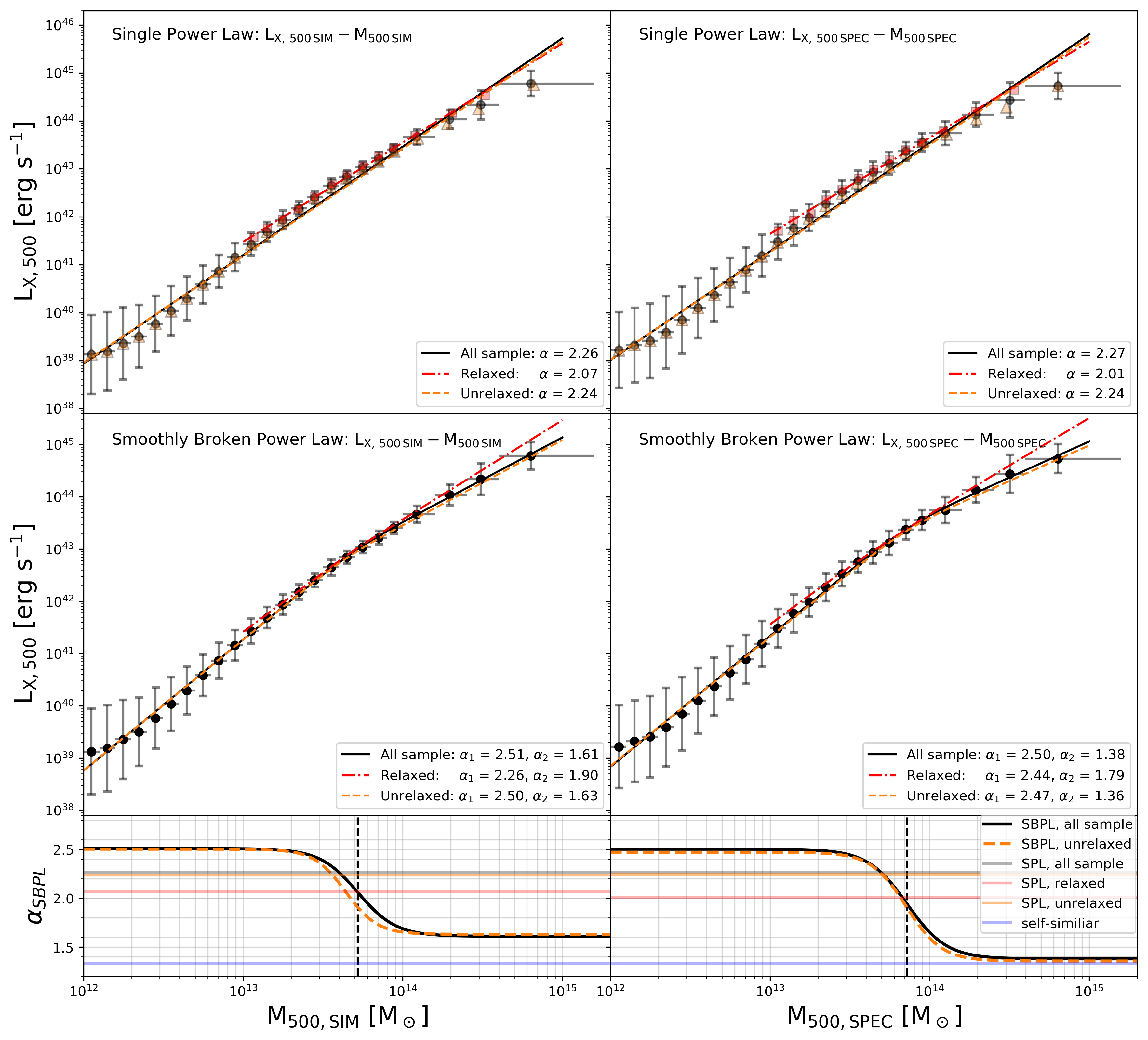}
\caption{Best-fits for the \lxm\, scaling relation at $z=0$ in TNG300. See Figure~\ref{fig:Mgas4panel} for the description of the panels and lines, noting that $\alpha = 4/3$ for the self-similar prediction of this relation.}
\label{fig:Lx4panel}
\end{figure*}

Several numerical and observational studies suggest that the \lxm\, scaling relation has a slope significantly steeper than the self-similar expectation, $\alpha = 4/3$ \citep[e.g.,][]{LeBrun2017, Barnes2017a, Barnes2017b, Henden2019, Truong2018, Pratt2009, Giles2016, Short2010, Stanek2010}. To investigate the best-fitting model for IllustrisTNG soft-band X-ray luminosities, we present in Table~\ref{tab:tablxall} the results from applying four different analytic models: SPL, BPL with fixed pivot at \xp\, =$10^{14}$\,\ms, BPL with free pivot, and SBPL (as defined in Section~\ref{subsec:fitting}). Simple power law fits result in very steep slope predictions: $\alpha^{\rm SIM}$ = \val[2.264]{0.016}{0.013} and $\alpha^{\rm SPEC}$ = \val[2.266]{0.072}{0.022}. However, as shown in Figure~\ref{fig:Lx8panel}, a broken power law with fixed pivot and a broken power law with free pivot provide increasingly better fits for \lxm\, compared to the best-fitting simple power law. In order to accommodate the steep slope preferred by galaxies and groups, the SPL model overestimates cluster luminosities and thus leads to a slope biased high. A broken power law with a pivot fixed at \xp\, =$10^{14}$\,\ms\, predicts a high mass slope that actually agrees with the self-similar model: $\alpha_2^{\rm SIM} =$ \val[1.377]{0.076}{0.068} and $\alpha_2^{\rm SPEC} =$ \val[1.250]{0.114}{0.116}. Nonetheless, our data prefers a pivot shifted to lower masses, with the best-fitting pivot from a BPL fit giving \xp $\simeq 4.7 \times 10^{13}$\ms. When we account for this lower mass break in the scaling, the high mass slope returns to steeper than self-similar values: $\alpha_2^{\rm SIM} =$ \val[1.679]{0.084}{0.076}. We find that the SBPL model provides the best fit for the \lxm\, scaling relation among all 4 models we considered. The model prefers a break that happens over a considerable mass range ($\sim 2 \times 10^{13} - 10^{14}$\ms), and it provides predictions for $\alpha_1$ and $\alpha_2$ with higher accuracy. The power of the SBPL model is further showcased by the fact that the added flexibility in the model allows us to find an interesting difference between \Lxsim\, and \Lxspec\, predictions. For simulation estimates of $L_{{\rm X, R}_{\rm 500c}}^{\rm 0.5-2 keV}$, the SBPL model predicts a high mass slope clearly steeper than self-similar $\alpha_{\rm SBPL, 2}^{\rm SIM}$ = \val[1.610]{0.125}{0.163}. However, when we account for the contribution of X-ray mass bias, the slope at cluster scales becomes significantly shallower ($\alpha_{\rm SBPL, 2}^{\rm SPEC}$ = \val[1.379]{0.271}{0.272}) and it is in good agreement with the self-similar expectation.

Most simulations that fit a simple power law model predict slopes that are steeper than self-similar: \val[1.97]{0.10}{0.08} in \textsc{fable} \citep{Henden2019}, \val[1.88]{0.03}{0.05} in \textsc{macsis} \citep{Barnes2017a}, $1.812 \pm 0.019$ in cosmo-\textsc{owls} \citep{LeBrun2017}, and $1.70 \pm 0.06$ \citep{Truong2018}. Some of the trends we noticed in the \mgm\, comparison with numerical studies are also evident here. Simulations such as \textsc{fable} and \textsc{macsis} find SPL slopes that are shallower than our SPL estimates (since we include more low mass objects) but steeper than our SBPL predictions at high cluster masses (likely because low mass clusters are biasing their slope estimates). On the other hand, the \citet{Truong2018} predictions agree well with our estimates inside \rsim. 
Other numerical studies find a range of slopes, from $1.87 \pm 0.01$ in \citet{Stanek2010}  to $1.77 \pm 0.03$ in the Millenium Gas runs of \citet{Short2010}, and the shallower slope of $1.45 \pm 0.05$ predicted by the \textsc{music} simulations \citep[][]{Biffi2014}.
Most observations predict relatively steep slopes: $1.96 \pm 0.10$ from \citet{Maughan2007}, $2.08 \pm 0.13$ from \citet{Pratt2009}, $2.22 \pm 0.24$ from \citet{Giles2017}. Some of the steeper slopes can be explained by differences in scalings predicted through bolometric or soft-band luminosity measurements. For example, \citet{Mantz2010b} find a significant decrease in the best-fit slope from $1.59 \pm 0.09$ for $L_{\rm bol}$ -- \mtot\, to $1.29 \pm 0.07$ for $L_{\rm X}$ -- \mtot. Moreover, the results in \citet{Mantz2010b} agree with our prediction for the  $L_{\rm X, SPEC}$ -- \mspec\, scaling at high masses.

In Figure~\ref{fig:Lxce4panel} and Table~\ref{tab:tablxce}, we provide a summary of the best-fitting SPL and SBPL models for the core-excised X-ray luminosity as a function of mass. 
Several previous studies found shallower slopes when excising the inner core region of clusters \citep[e.g.,][]{Truong2018, Henden2019}. Our results agree with these findings. The slope predicted by the SBPL model at high mass scales drops from \val[1.610]{0.125}{0.163} to \val[1.381]{0.084}{0.068}, which is now  consistent with the self-similar prediction. The median value of $\alpha_{\rm SBPL, 2}$ derived from bootstrap resampling is also lower inside a spectroscopic aperture, where excising the core reduces the slope from \val[1.379]{0.271}{0.272} to \val[1.233]{0.204}{0.265}. We note that core-excised luminosities have slopes consistent with self-similar ($\alpha = 4/3$) irrespective of the choice of aperture in our sample. Observations also report shallower slopes for core-excised luminosities, such as \citet{Pratt2009} who measure $\alpha = 1.80 \pm 0.05$ from the REXCESS sample.
Lastly, observations from \citet{Mantz2016b} also point to shallower slopes for core-excised X-ray luminosities, compared to \Lx\, measurements that include the core.

Finally, our results indicate that relaxed clusters have steeper slopes than unrelaxed clusters. This trend is more pronounced when we include the effect of estimated X-ray apertures. For \lxm\,, the relaxed sample predicts $\alpha_{\rm SBPL,2} = $\val[1.793]{0.106}{0.255} and for core-excised luminosities, we find $\alpha_{\rm SBPL,2} = $\val[1.571]{0.101}{0.178}. The steepening of the slope for relaxed samples leads us to conclude that discrepancies between some of the predictions from observations and our simulations can be attributed to a combination of differences in the (1) mass distributions and (2) fractions of relaxed objects between our respective samples.

\subsection{X-ray temperature - total mass scaling relation}
\label{subsec:scalingstx}

\begin{figure*}
\centering
\includegraphics[width=0.99\textwidth]{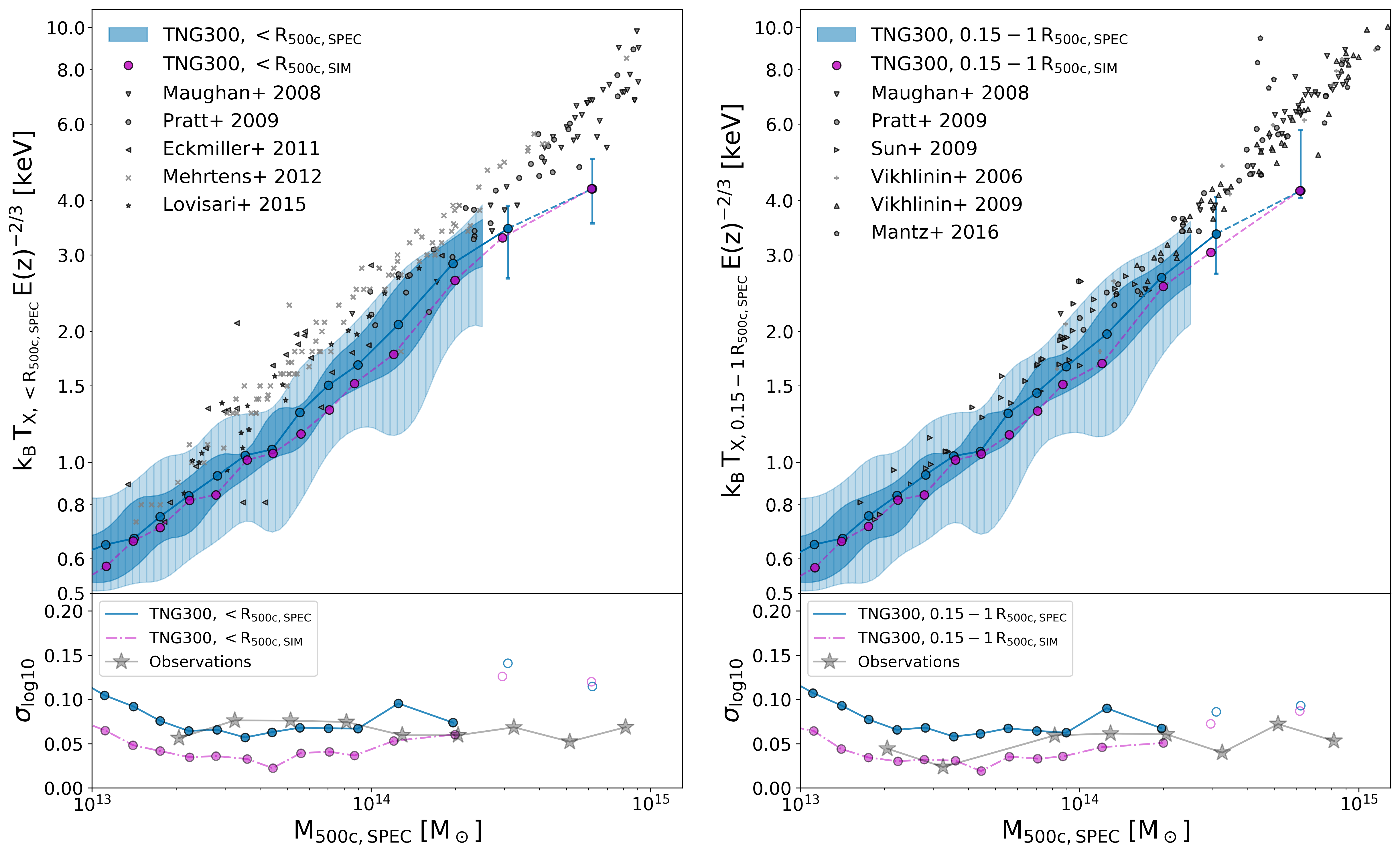}
\caption{\textit{Left panel:} Spectroscopic temperature measured within \rcrit\, as a function of the total halo mass at $z=0$ in TNG300. Blue dots mark the median values for \txspec\, as a function of the total mass measured inside the corresponding spectroscopic aperture, \mspec. Dark (light) blue contours mark $1 \sigma$ (and $2 \sigma$) scatter around the median values for bins containing 10 haloes or more, with contour limits smoothed using quadratic interpolation. The light purple dots show the median values of \txsim\, as a function of the total mass measured inside the corresponding simulation aperture, \msim. For comparison, we show grey symbols representing observations from \citet{Maughan2008, Pratt2009, Eckmiller2011, Mehrtens2012, Lovisari2015} .
\textit{Right panel:} Core-excised spectroscopic temperature as a function of the total halo mass at $z=0$ in TNG300. Blue dots mark the median values for \txcespec\, as a function of the total mass measured inside the corresponding spectroscopic aperture, \mspec. Dark (light) blue contours mark $1 \sigma$ (and $2 \sigma$) scatter around the median values for bins containing 10 haloes or more, with contour limits smoothed using quadratic interpolation. The light purple dots show the median values of \txcesim\, as a function of the total mass measured inside the corresponding simulation aperture, \msim.
For comparison, we show grey symbols representing observations for core-excised temperatures from \citet{Maughan2008, Pratt2009, Sun2009, Vikhlinin2006, Vikhlinin2009b, Mantz2016b} .
The bottom panels show the level of intrinsic scatter $\sigma_{{\rm log} 10}$ for the spectroscopic temperatures, measured inside \rspec \, (blue dots + line) and \rsim \, (purple dots + dashed line). We compare these results to grey stars indicating the average intrinsic scatter in observations, computed over the combined sample of all observational data points shown in the top panels.}
\label{fig:Txcontour}
\end{figure*}

\begin{figure*}
\centering
\includegraphics[width=0.9\textwidth]{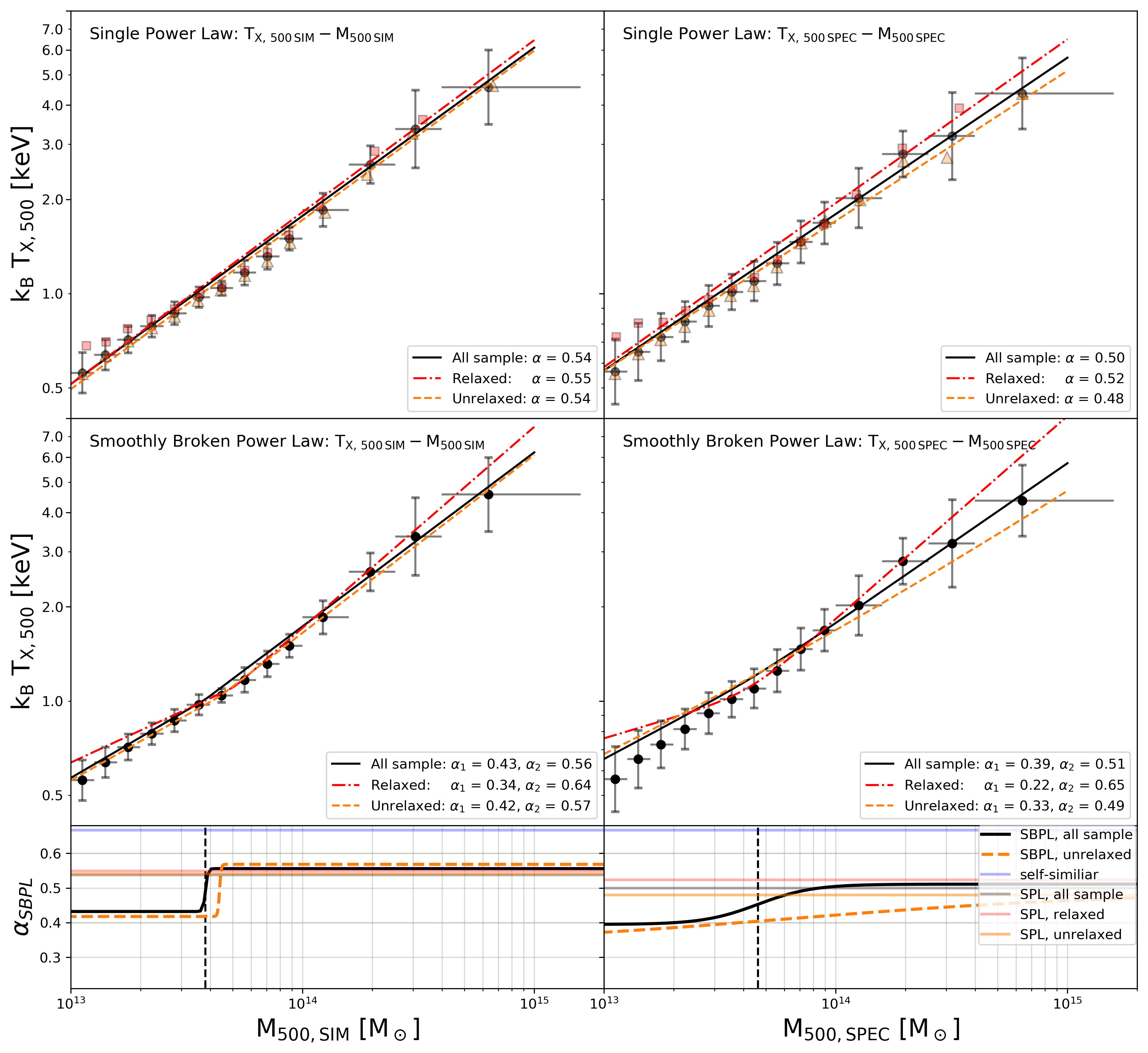}
\caption{Best-fits for the \txm\, scaling relation at $z=0$ in TNG300. See Figure~\ref{fig:Mgas4panel} for the description of the panels and lines, noting that $\alpha = 2/3$ for the self-similar prediction of this relation.}
\label{fig:Tx4panel}
\end{figure*}

In Figure~\ref{fig:Txcontour}, we investigate IllustrisTNG spectroscopic temperatures ($T_{\rm X}$) as a function of halo mass. 
For spectroscopic temperatures that include the core region, IllustrisTNG appears to systematically underestimate $T_{\rm X}$ compared to observations (shown with grey symbols in Figure~\ref{fig:Txcontour}) from \citet{Maughan2008, Pratt2009, Eckmiller2011, Mehrtens2012, Lovisari2015}. 
Accounting for the bias introduced by X-ray estimates for the \rcrit\, aperture  brings IllustrisTNG group temperatures closer to those measured in observations.

In the right panel of Figure~\ref{fig:Txcontour}, we show core-excised spectroscopic temperatures, which we compare to observational samples from \citet{Maughan2008, Pratt2009, Sun2009, Vikhlinin2006,  Vikhlinin2009b, Mantz2016b}. We find that IllustrisTNG groups show better agreement with observations of core-excised temperatures, especially when we account for the X-ray mass bias. However, the highest mass clusters in IllustrisTNG appear to be consistently too cool compared to observations. 

In the bottom panels of Figure~\ref{fig:Txcontour}, we show the intrinsic scatter $\sigma_{\rm log10}$ for quantities measured both inside the true simulation aperture, \rsim, and the spectroscopic aperture, \rspec. Consistent with the results from previous subsections, spectroscopic apertures also lead to higher levels of scatter in the \txm\, scaling relation. We compare to the level of scatter in the combined sample composed of all observations that are included in the top panels. For $T_{{\rm X, <R}_{\rm 500c}}$, the overall intrinsic scatter in observations is well matched by the values for $\sigma_{\rm log10}$ measured $<$\rspec\, in \tng. For core-excised temperatures,  
$T_{{\rm X, 0.15 - 1\,R}_{\rm 500c}}$, the intrinsic scatter in observations lies in between the values we obtain within $(0.15 - 1)$\rsim\, and $(0.15 - 1)$\rspec. In \tng, the intrinsic scatter in \txm\, is minimized at group scales, with the lower mass groups having increasingly higher $\sigma_{\rm log 10}$. At the other mass end, there is some evidence that \tng\, clusters show too much variation in \tx\, estimates from one cluster to another. We have too few massive clusters in our sample to draw a definitive conclusion. However, given the suppression of scatter in \tng\, clusters once we excise the core region, we are led to believe that additional physics might be needed in order to correctly model cluster cores. Some of the missing physical processes that could impact the temperatures of inner cluster regions include cosmic-rays \citep{Pfrommer2017}, anisotropic thermal conduction \citep{Kannan2016, Kannan2017, Barnes2019}, and a better modelling of the interaction between AGN jets and the ICM \citep[e.g.,][]{Weinberger2017}.

Previous numerical studies mostly predict slopes that are shallower than self-similar: $0.59$ \citep{Truong2018}, $0.64$ \citep{Henden2019}, $0.58$ \citep{Barnes2017b, LeBrun2017}, $0.55 \pm 0.01$ \citep{Short2010}, $0.576 \pm 0.002$ \citep{Stanek2010}, $0.54 \pm 0.01$ \citep{Planelles2014}, $0.56 \pm 0.03$ \citep{Biffi2014}, and $0.60 \pm 0.01$ \citep{Pike2014} . There is some disagreement among observations, with some studies predicting slopes shallower than self-similar: $0.58 \pm 0.03$ in \citep{Arnaud2007}, 
$0.57$  \citep{Reichert2011}, $0.60$ \citep{Lieu2016}, $0.61$  \citep{Sun2009}, $0.60$  \citep{Eckmiller2011}, and $0.61$  \citep{Lovisari2015}. Nonetheless, several groups find slopes that are consistent with the self-similar prediction, such as $0.66$  \citep{Kettula2015}, $0.66 \pm 0.05$ \citep{Mantz2010a} and $0.65 \pm 0.04$ \citep{Vikhlinin2009a}.

We find best-fitting slopes that are consistently shallower than self-similar ($\alpha = 2/3$) for all models and at all mass scales in IllustrisTNG. The only exception relates to relaxed clusters, which are discussed at the end of this subsection. 
For spectroscopic temperatures evaluated on spectra computed between $0-1$\rcrit, we find SPL slopes $\alpha^{\rm SIM} = $ \val[0.537]{0.024}{0.037} and $\alpha^{\rm SPEC} = $ \val[0.500]{0.040}{0.043}. We restrict our best-fit modelling only to haloes with $T_{\rm X} > 0.5$ keV which corresponds to a mass cut of \mcrit $\simeq 10^{13}$\ms. 
Spectra are convolved with the \textit{Chandra} response function (as detailed in Section~\ref{subsec:xraypipeline}) and spectroscopic temperatures of objects below this $T_{\rm X}$ cut are prone to uncertainties due to the single temperature fit. 
As a result of our fits being limited to higher masses for the \txm\, scaling relation, SPL slopes are driven closer to the values predicted by SBPL models for the highest mass clusters ($\alpha_{\rm SBPL, 2}^{\rm SIM} =$\val[0.556]{0.043}{0.042} and $\alpha_{\rm SBPL, 2}^{\rm SPEC} =$\val[0.511]{0.080}{0.031}). We observe a significant difference in the break of the scaling relation depending on the chosen aperture. For quantities within \rsim\,, \xp $\simeq 3.8 \times 10^{13}$\ms\, and the model prefers a break that is extremely sharp ($\delta \simeq 0.01$). On the other hand, for spectroscopic apertures, \xp $\simeq 4.6 \times 10^{13}$\ms, and the transition happens over a much larger mass range ($\delta \simeq 0.24$).

The best-fitting models for core-excised spectroscopic temperatures (see Figure~\ref{fig:Lxce4panel} and Table~\ref{tab:tablxce}) yield consistent slopes to those we found for temperatures that include the core region. Inside \rsim, the SBPL prediction at high masses changes from \val[0.556]{0.043}{0.042} for $T_{\rm X}$ to \val[0.599]{0.273}{0.032} for $T_{\rm X, ce}$. For spectroscopic apertures, SBPL slopes from clusters suffer a small adjustment: \val[0.511]{0.080}{0.031} for $T_{\rm X}$ to \val[0.549]{0.092}{0.106} for $T_{\rm X, ce}$.
Core-excised X-ray temperatures have lower normalizations ($A = $ \val[0.206]{0.008}{0.004} for $T_{\rm X, ce}$ versus $A$ = \val[0.238]{0.023}{0.017} for $T_{\rm X}$). This is to be expected, as removing the cluster cores has the net effect of lowering the spectroscopic temperatures. 
Core-excised temperatures also seem to prefer break locations that are significantly shifted to higher masses, from $3.8 \times 10^{13}$\ms\, for  $T_{\rm X}$ to $8.3 \times 10^{13}$\ms\, for  $T_{\rm X, ce}$, and the breaks are smoother, i.e., the change in slope happens over a wider mass range.

While the full sample in \tng\, predicts slopes for \txm\, that are shallower than the self-similar prediction ($\alpha = 2/3$), our results indicate that the mass dependence of relaxed clusters is consistent with self-similarity. Simple power law fits completely wash out this effect because they concomitantly try to fit $T_{\rm X}$ measurements for groups and clusters. However, once we apply the SBPL model, we uncover that relaxed clusters prefer slopes that are consistent with $\alpha_{\rm self-similar} = 0.67$: $\alpha_{\rm SBPL, 2}^{\rm SIM} = $\val[0.642]{0.058}{0.019} and $\alpha_{\rm SBPL, 2}^{\rm SPEC} = $\val[0.648]{0.108}{0.044}. 
\subsection{\yx\, -- \mtot\, scaling relation}
\label{subsec:yxmscaling}

We finally consider the scaling relation between \yx\, and \mtot. The quantity \yx\, is defined as:
\begin{equation}
    Y_{\rm X} = T_{\rm X, \, ce} \times M_{\rm gas}
\end{equation}
and it serves as an X-ray analogue to the integrated SZ effect. It combines the core-excised spectroscopic temperature ($T_{\rm X, \, ce})$ and the hot gas mass  estimated from X-ray measurements within \rcrit\, of the cluster center.
Introduced by \citet{Kravtsov2006}, this X-ray proxy of the total cluster mass has been shown to exhibit remarkably low scatter at a fixed halo mass. 

\begin{figure}
\centering
\includegraphics[width=0.5\textwidth]{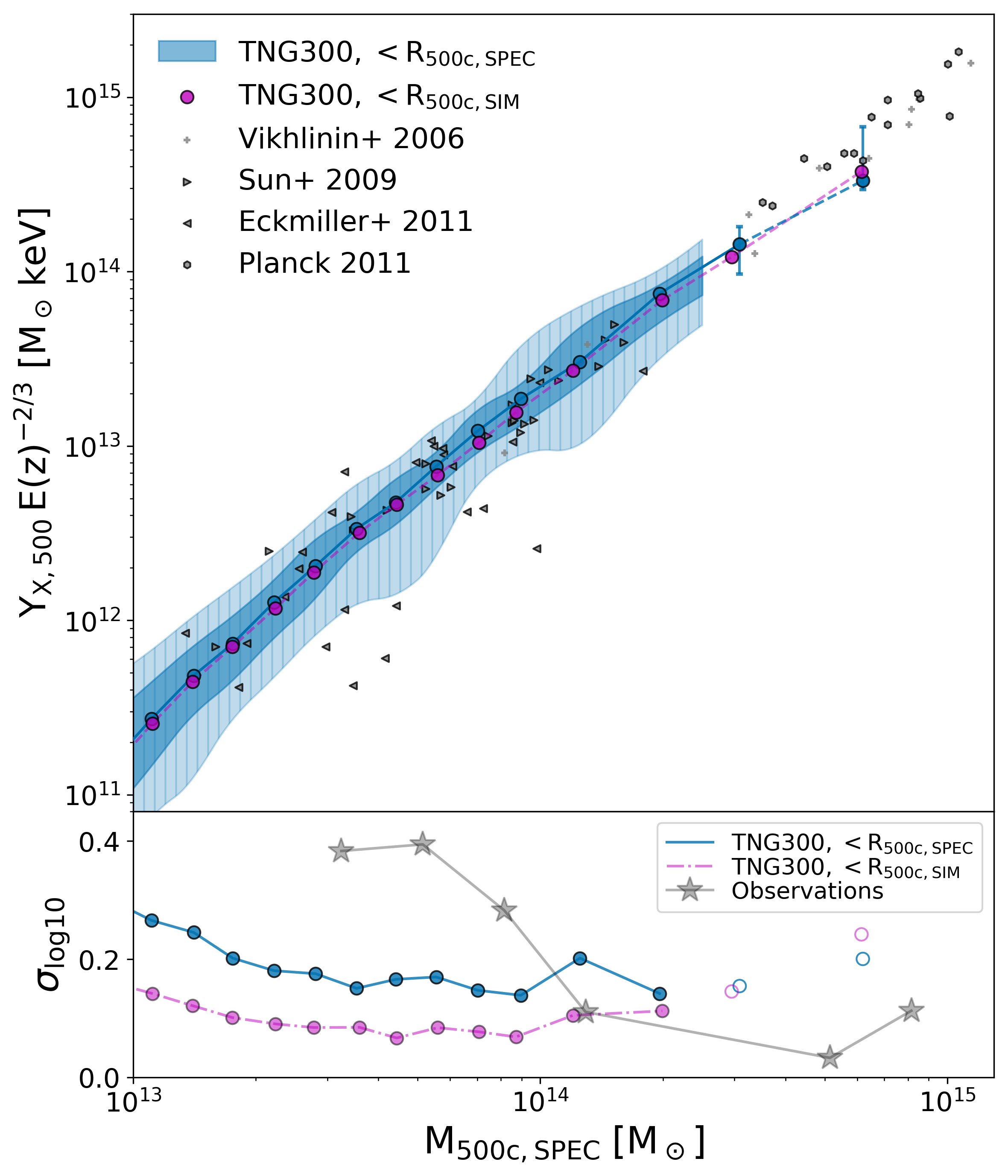}
\caption{\yx \, measured within \rcrit\, as a function of the total halo mass at $z=0$ in TNG300. Blue dots mark the median values for \yxspec\, as a function of the total mass measured inside the corresponding spectroscopic aperture, \mspec. Dark (light) blue contours mark $1 \sigma$ (and $2 \sigma$) scatter around the median values for bins containing 10 haloes or more, with contour limits smoothed using quadratic interpolation. The light purple dots show the median values of \yxsim\, as a function of the total mass measured inside the corresponding simulation aperture, \msim. For comparison, we show grey symbols representing observations from \citet{Vikhlinin2006, Sun2009, Eckmiller2011, PlanckCollaboration2011} .
The bottom panel shows the level of intrinsic scatter $\sigma_{{\rm log} 10}$ for \yx, measured inside \rspec \, (blue dots + line) and \rsim \, (purple dots + dashed line). We compare these results to grey stars indicating the average intrinsic scatter in observations, computed over the combined sample of all observational data points shown in the top panels.}
\label{fig:Yxcontour}
\end{figure}

In Figure~\ref{fig:Yxcontour}, we compare IllustrisTNG \yx measurements against observations from \citet{Vikhlinin2006, Sun2009, Eckmiller2011, PlanckCollaboration2011}. We find excellent agreement between \tng\, groups and clusters and observed data for the \yxm\, scaling relation. 
In Section~\ref{subsec:scalingsmgas}, we showed that \tng\, groups are too gas rich compared with observations. Nonetheless, the underestimated X-ray temperatures (see Section~\ref{subsec:scalingstx}) perfectly compensate for the too high gas fractions, resulting in \yx\, values that reproduce both the observed trend with mass and the normalization of \yxm. 

In the bottom panel of Figure~\ref{fig:Yxcontour}, we show the dependence of the intrinsic scatter in the \yxm\, relation on the total halo mass, \mtot. Our results indicate a relatively flat level of scatter in \yx\,, with $\sigma_{\log10}$ being minimized at group scales. 
The highest mass bins in \tng\, appear to have too much scatter in \yx\,, but we have insufficient statistics to draw conclusions about $\sigma_{\log10}$ above \mbox{$\approx 3\times 10^{14}$\,\ms}.

\begin{figure*}
\centering
\includegraphics[width=0.9\textwidth]{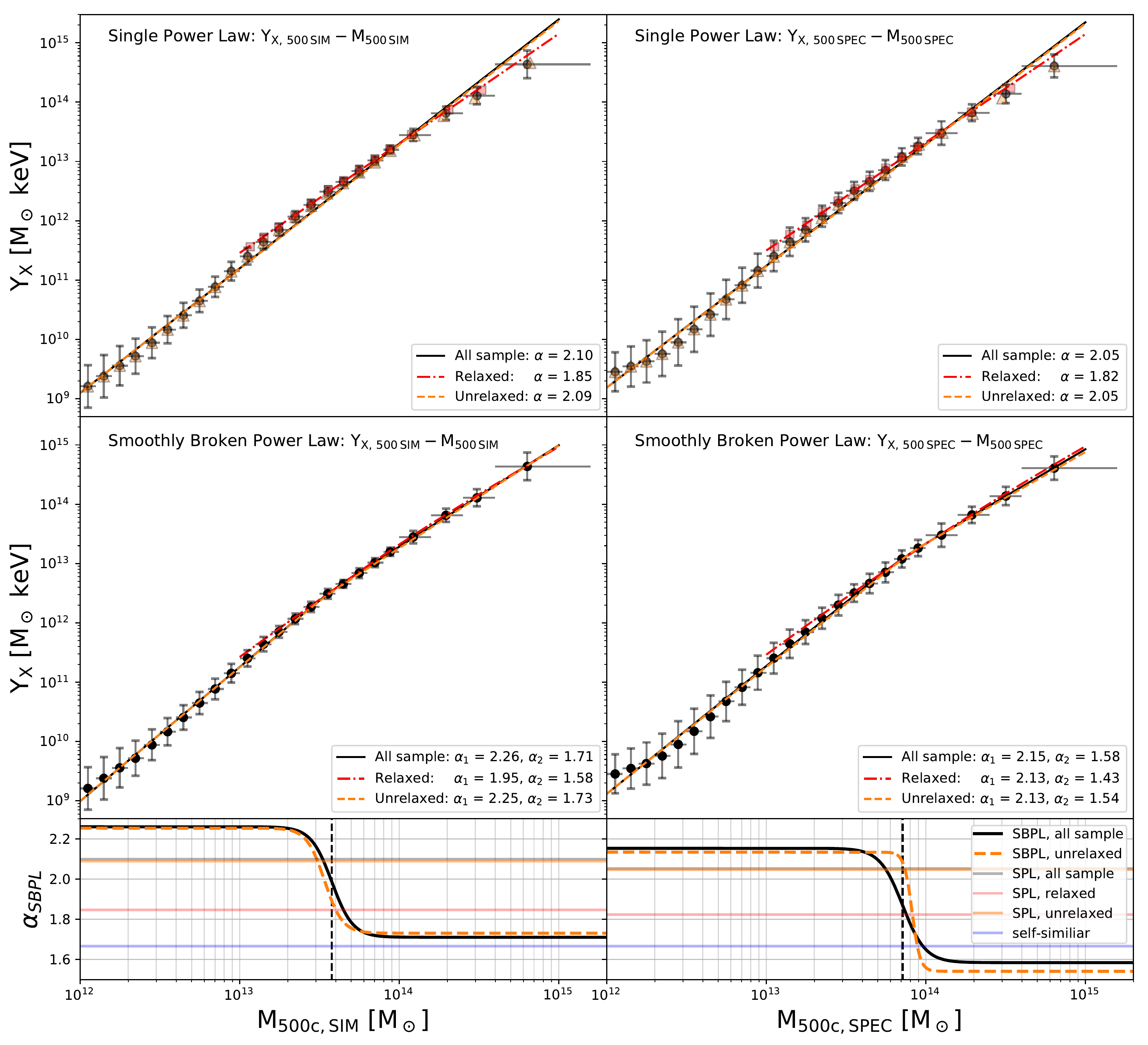}
\caption{Best-fits for the \yxm\, scaling relation at $z=0$ in TNG300. See Figure~\ref{fig:Mgas4panel} for the description of the panels and lines, noting that $\alpha = 5/3$ for the self-similar prediction of this relation. }
\label{fig:Yx4panel}
\end{figure*}

Since \yx\,  is the product of $T_{\rm X, \, ce}$ and $M_{\rm gas}$, the slope of the \yxm\, relation is expected to be equal to the sum of the slopes in $T_{\rm X, \, ce}$\,--\,\mtot\, and $M_{\rm gas}$\,--\,\mtot, i.e., $\alpha_{Y_{\rm X}} \simeq \alpha_{T_{\rm X, ce}} + \alpha_{M_{\rm gas}}$. We present the results of our best-fitting models for the \yxm\, scaling relation in Figure~\ref{fig:Yx4panel} and Table~\ref{tab:tabyx}. We indeed find best-fitting slopes that are consistent with $\alpha_{Y_{\rm X}} \simeq \alpha_{T_{\rm X, ce}} + \alpha_{M_{\rm gas}}$. When fitting a SPL model, we find slopes that are much steeper than the self-similar expectation: $\alpha^{\rm SIM}=2.098\pm0.004$ and $\alpha^{\rm SPEC}=$\val[2.052]{0.028}{0.012}. Since our sample includes a large fraction of low mass haloes, the SPL slope is biased to steeper values than predictions from other simulations: \val[1.88]{0.04}{0.05} in \textsc{fable} \citep{Henden2019}, \val[1.84]{0.02}{0.05} in \textsc{macsis} \citep{Barnes2017a}, and $1.888\pm 0.018$ in cosmo-\textsc{owls} \citep{LeBrun2017}. These results agree with observational measurements from \citet{Arnaud2007} ($1.82 \pm 0.09$), \citet{Eckmiller2011} ($1.82 \pm 0.07$) and \citet{Mahdavi2013} ($1.79 \pm 0.22$). 

The SBPL model allows us to estimate the slope ($\alpha_{\rm SBPL,2}$) that characterizes the \yxm\, relation for the highest mass clusters. As shown in Figure~\ref{fig:Yx4panel} and Table~\ref{tab:tabyx}, \tng\, data predicts slopes that are consistent with the self-similar expectation at the high mass end of our sample. We find $\alpha_{\rm SBPL, 2}^{\rm SIM} = $ \val[1.710]{0.021}{0.018} and $\alpha_{\rm SBPL, 2}^{\rm SPEC} = $ \val[1.584]{0.081}{0.132}. Through a closer inspection of the bottom panel of Figure~\ref{fig:Yx4panel}, we can see that our results are in broad agreement with the other studies mentioned above. For example, \cite{Henden2019} have a mass cut of $3\times 10^{13}$\ms\, and their slope estimate (\val[1.88]{0.04}{0.05}) is very close to the predicted value of $\alpha_{\rm SBPL}$ evaluated at \msim $\simeq 3\times 10^{13}$\ms\, for quantities measured inside the simulation apertures (as it is the case in their study). We also warn against making direct comparisons to the trends in $\alpha_{\rm SBPL}$ as a function of mass unless the data was binned. Moreover, if the lower mass cut of the sample is at or above the location of the break in the scaling relation, this could make it harder to capture the correct behaviour around the pivot location.

Lastly, we mention that several simulations also found a slope that is consistent with our results and with the self-similar expectation: \citet{Truong2018} ($1.66 \pm 0.02)$, \citet{Fabjan2011} and \citet{Biffi2014}. On the observational side, several groups also report results consistent with the self-similar prediction for \yxm: $1.67 \pm 0.08$ in \citet{Biffi2014} and $1.63 \pm 0.04$ in \citet{Mantz2016b}. 

Relaxed clusters also have slopes consistent with self-similarity according to our SBPL model. However, the best-fit values for $\alpha_{\rm SBPL, 2}$ are biased low and the agreement with the self-similar expectation is partly due to the larger uncertainity in these slopes, as estimated through bootstrapping with replacement. For quantities measured inside \rspec, there is good agreement between the pivot location for \mgm, \txm, and \yxm. Similarly to the scaling relations for gas mass and spectroscopic temperatures, the \yxm\, scaling relation in \tng\, predicts a break around \xp\,$\simeq 7.2 \times 10^{13}$\,\ms.  
\section{The Sunyaev-Zel'dovich Effect}
\label{sec:yszresults}

\begin{figure}
\centering
\includegraphics[width=0.5\textwidth]{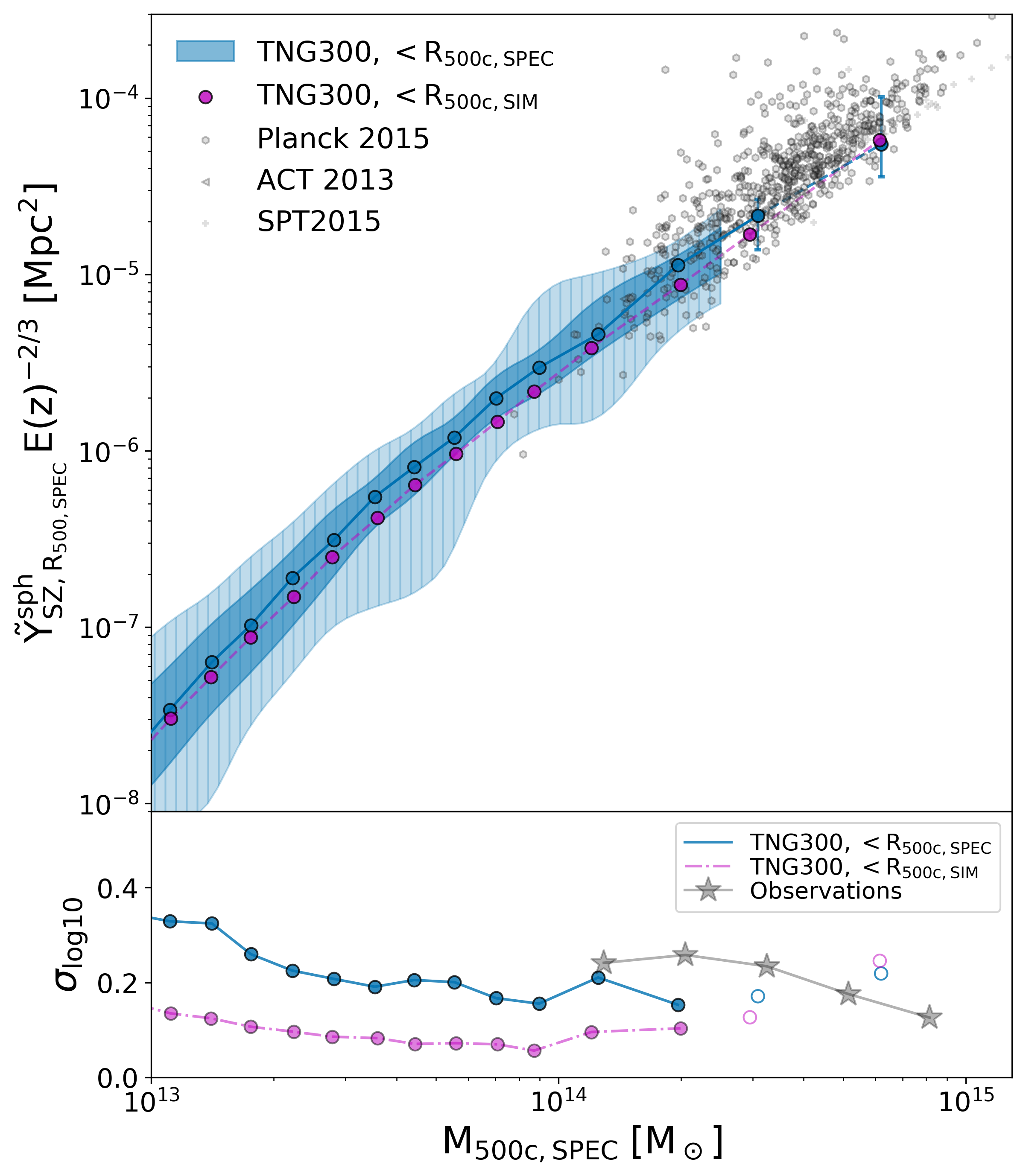}
\caption{Integrated SZ signal, \ysz \,, measured within \rcrit\, as a function of the total halo mass at $z=0$ in TNG300. Blue dots mark the median values for \yszspec\, as a function of the total mass measured inside the corresponding spectroscopic aperture, \mspec. Dark (light) blue contours mark $1 \sigma$ (and $2 \sigma$) scatter around the median values for bins containing 10 haloes or more, with contour limits smoothed using quadratic interpolation. The light purple dots show the median values of \yszsim\, as a function of the total mass measured inside the corresponding simulation aperture, \msim. For comparison, we show grey symbols representing observations from \citet{PlanckCollaboration2016b, Hasselfield2013, Bleem2015} .
The bottom panel shows the level of intrinsic scatter $\sigma_{{\rm log} 10}$ for \ysz, measured inside \rspec \, (blue dots + line) and \rsim \, (purple dots + dashed line). We compare these results to grey stars indicating the average intrinsic scatter in observations, computed over the combined sample of all observational data points shown in the top panels.}
\label{fig:yszcontour}
\end{figure}

Finally, we compare the integrated Sunyaev-Zel'dovich effect inside \rsim\, and \rspec\, spherical apertures \citep{Motl2005,Nagai2006} measured from \tng\, haloes to observations from \citet{PlanckCollaboration2016b, Hasselfield2013, Bleem2015}. \tng\, clusters show good agreement with observations of \yszm\,, both in reproducing the overall trend with mass, as well as in matching the observed normalization of the scaling. In the bottom panel of Figure~\ref{fig:yszcontour}, we show the level of intrinsic scatter for quantities measured inside \rsim\, and \rspec\,, respectively. For measurements inside true apertures, $\sigma_{\rm log10}$ shows no dependence on mass, except for the highest mass bins where we do not have good statistics. Overall, the intrinsic scatter from the combined sample of Planck, SPT and ACT observations appears to be higher than the \tng\, estimates.

\begin{figure*}
\centering
\includegraphics[width=0.9\textwidth]{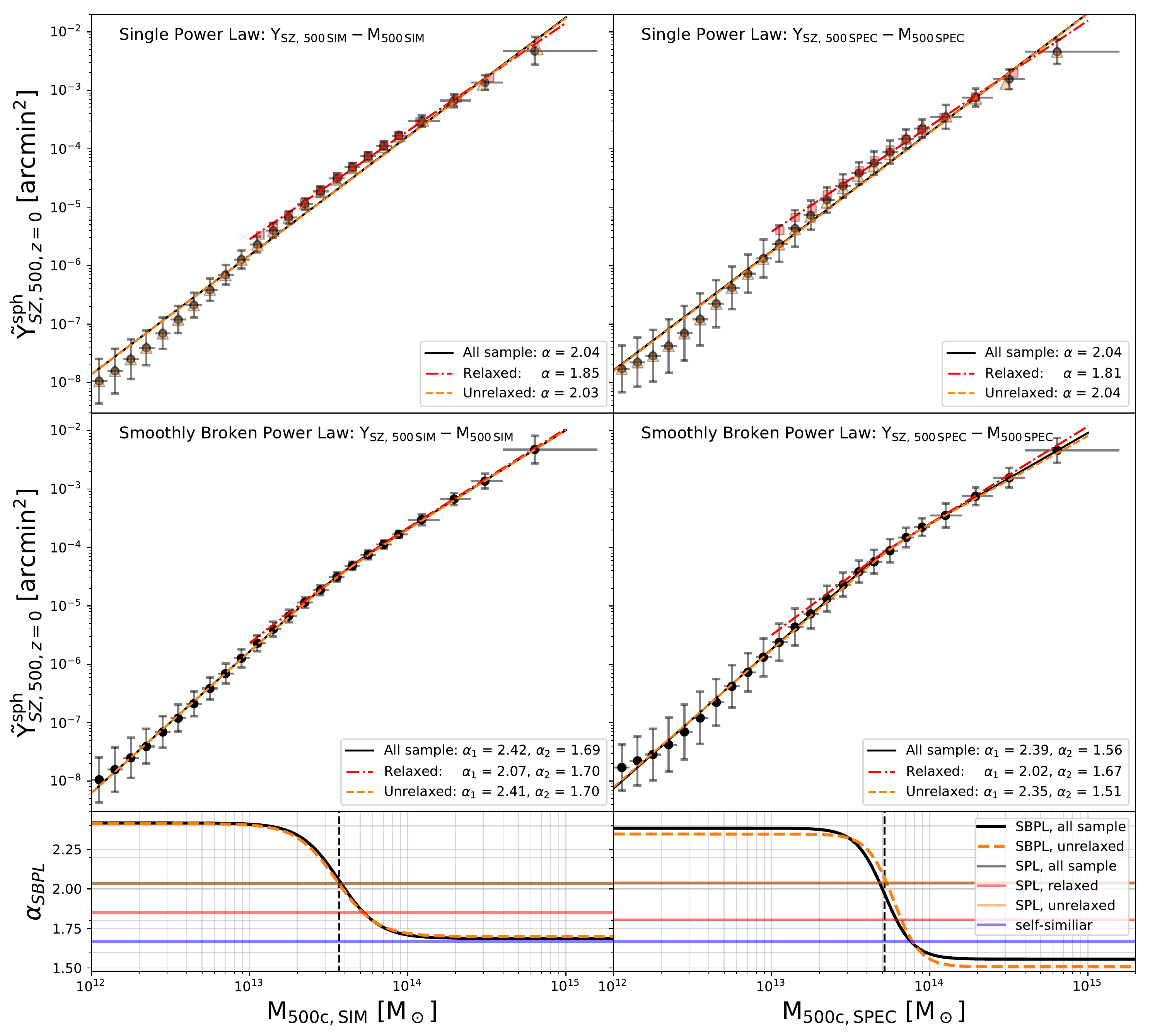}
\caption{Best-fits for the \yszm\, scaling relation at $z=0$ in TNG300. See Figure~\ref{fig:Mgas4panel} for the description of the panels and lines, noting that $\alpha = 5/3$ for the self-similar prediction of this relation.}
\label{fig:Ysz4panel}
\end{figure*}

In Figure~\ref{fig:Ysz4panel} and Table~\ref{tab:tabysz} we present \tng\, results from fitting the SPL and SBPL models to the \yszm\, scaling relation. We find very similar results to those found for \yxm. This is not surprising since \ysz\, is a measure of the total thermal energy in the ICM and thus, it should provide similar results to \yx. Since \ysz\, is computed using an integral over the cluster volume, the integrated SZ effect is an indicator of the mass-weighted temperature of the cluster, whereas \yx\, is directly proportional to the core-excised spectroscopic temperature. Thus, for the \yszm\, relation we find $\alpha_{\rm SPL} = $ \val[2.036]{0.021}{0.016}, which is significantly steeper than self-similar due to the presence of low mass haloes in our sample. The SBPL model predicts $\alpha_{\rm SBPL,2}^{\rm SIM} = $ \val[1.687]{0.015}{0.015}, which is consistent with the self-similar expectation of $5/3$. Relaxed clusters also prefer a self-similar slope at the highest mass end of our sample: $\alpha_{\rm SBPL,2}^{\rm SIM} = $ \val[1.696]{0.026}{0.052}. 

The pivot locations in \yxm\, and \yszm\, are in excellent agreement for quantities measured inside the true simulation aperture, \rsim: \xp\,$\simeq3.8\pm0.02\times10^{13}$\,\ms\, for \yxm\, compared to \xp\,$\simeq3.7\pm 0.01\times 10^{13}$\,\ms\, for \yszm\;. The break is slightly sharper in \yxm\, ($\delta = $ \val[0.13]{0.03}{0.04}) whereas \yszm\, exhibits a smoother break ($\delta = $ \val[0.25]{0.02}{0.02}). 

\section{Conclusions}
\label{sec:conclusions}
In this work, we have studied the X-ray and Sunyaev-Zel'dovich (SZ) scaling relations with halo mass using a sample spanning over three orders of magnitude in mass (\mcrit\,$\in [10^{12}$\,--\,$2\times10^{15}]$\ms), including over 30,000 galaxies, galaxy groups, and clusters from the \tng\, cosmological simulation.
Taking full advantage of the large volume TNG300 simulation of the IllustrisTNG project, we vastly expand on the number and mass range of simulated haloes 
that have been previously 
compared to X-ray and SZ observations. The purpose of this paper was threefold:

(1) We developed a pipeline that produces high-fidelity mock X-ray observations, which closely mimics the methods utilized by observers. Our method is designed to be computationally efficient and to yield reliable results for very large samples of haloes, which probe vastly different regimes in X-ray temperatures and emission lines that contribute to their respective soft-band spectrum. 
We also account for the scatter and bias introduced by X-ray estimated masses. Thus, our work highlights the importance of converting ICM properties of simulated haloes to the observable space.

(2) We tested the subgrid model for AGN feedback implemented in \tng\, against X-ray and SZ observations, with the goal of understanding its strengths and weaknesses. Our results also provide insights for the next generation of feedback models that will be implemented in cosmological simulations.

(3) By extending our study to include all haloes above $10^{12}$\,\ms\, in \tng\,, our sample includes sufficiently low mass groups and galaxies to  confidently predict the presence of a break in all X-ray and SZ scaling relations included in this study. We propose an analytic model for a smoothly broken power law (eqn.~\ref{eqn:eqnSBPL}) that provides a robust method to find unbiased estimates of scaling relation slopes at very high and very low halo masses, as well as the location of the break, and the width of the transition region around the break. This model allows us to provide predictions for ongoing and future X-ray and SZ surveys, which will include many more galaxy groups and low mass clusters. 

We summarize our main results below.

(i) The $z=0$ X-ray and Sunyaev-Zel'dovich properties of \tng\, groups and clusters show a major improvement compared to Illustris, and they are in good overall agreement with observations. This improvement is largely due to changes in the AGN feedback model, where at low accretion rates, a new kinetic feedback model imparts momentum to the surrounding gas in a stochastic manner.

(ii)  \tng\, reproduces the observed trends in \yxm\, and \yszm, showing excellent agreement both with the observed normalizations and trends with mass. The predicted gas mass fractions and X-ray luminosities follow the observed trends with halo mass, but they seem to overestimate the normalizations of $f_{\rm gas}$ and \Lx, especially for galaxy groups with intermediate masses (\mcrit\, $\in [5\times10^{13}$\,\ms -- $2\times 10^{14}$\,\ms]). In contrast, \tng\, temperatures lie on the lower end of the scatter in observations, and some of the discrepancy is removed by accounting for the X-ray mass bias and by removing the inner core region ($<0.15$\,\rcrit).

(iii) Due to the shape of the halo mass function, cosmological simulations and volume-limited surveys contain many more low-mass haloes than massive ones.
It is often prohibitively expensive for cosmological simulations to model full-physics, high-resolution volumes that are large enough to include good statistics for very massive galaxy clusters. 
Nonetheless, clusters are unique astrophysical laboratories and the slopes predicted by X-ray and SZ scaling relations at cluster scales hold important clues for cosmology. 
In this work, we show that simple power law fits to individual data points will be highly biased by the lowest mass objects in the sample, and they will therefore provide poor predictions for the highest mass haloes, such as clusters. Even mass cuts as high as \mcrit\, $\simeq 5\times 10^{14}$\,\ms\, can still lead to biased slope estimates.

In order to obtain reliable predictions for samples that include a range of halo masses, it is necessary to bin the data. Commonly employed choices for the summary statistics in these bins include medians and means (i.e., simple averages). Nonetheless, these estimates can bias the resulting slope away from the underlying true slope reflected by the data, especially for bins that contain few objects.
In \S\ref{subsec:fitting}, we propose using geometric means as the summary statistics for binning the sample and show that it naturally preserves the original slope of the data in logarithmic space.

(iv) \textit{A model for a smoothly broken power law}: Simple power laws provide poor slope predictions for samples that cover a wide range of halo masses, even when correctly binning the data.  
We test three other models: a broken power law with a fixed pivot (eqn.~\ref{eqn:eqnBPLfixed}), a broken power law with a free pivot (eqn.~\ref{eqn:eqnBPLfree}), and finally a smoothly broken power law (SBPL, eqn.~\ref{eqn:eqnSBPL}) which is new to this study. We find that the SBPL model provides the best match to \tng\, haloes spanning over 3 orders of magnitude in \mcrit. This model also has sufficient functional flexibility to model all X-ray and SZ scaling relations considered in this paper, for both simulated and spectroscopic apertures, as well as for relaxed and unrelaxed samples. 

(v.1) \textit{\mgm\, scaling relation}: Using the SBPL model, we find that the best-fitting slope at high masses for the  \mgm\, scaling relation is lower than values suggested by previous simulations.  Although our values still indicate a preferred slope that is steeper than self-similar, we show that for the highest mass clusters, the slope only deviates mildly from the self-similar expectation.
Based on our findings, even intermediate and low mass clusters  can bias simple power law models to predict slopes significantly steeper than $\alpha_{\rm self-similar} = 1$.

(v.2) \textit{\lxm\, scaling relation}:  For the scaling relation between \Lx\, and \mtot, the SBPL model predicts a high-mass slope significantly steeper than self-similar if the X-ray mass bias is neglected. However, for X-ray luminosities measured inside spectroscopic apertures, we find that the slope for the highest mass clusters is consistent with the self-similar expectation. Moreover, the slope for core-excised X-ray luminosities vs. mass is consistent with self-similar irrespective of the chosen aperture. Relaxed clusters, on the other hand, prefer significantly steeper slopes in \lxm.

(v.3) \textit{\txcem\, scaling relation}:  The SBPL best-fits for spectroscopic temperatures as a function of mass in \tng\, agree with previous numerical and observational studies. We find \txm\, slopes that are shallower than the self-similar expectation, both for $T_{\rm X}$ and core-excised $T_{\rm X}$, as well as for simulation and spectroscopic apertures. If we restrict our fit only to relaxed clusters, we find steeper slopes which are consistent with the self-similar expectation \tx $\propto M^{2/3}$.

(v.4) \textit{\yxm\, scaling relation}: For the X-ray analogue of the integrated SZ effect, \yx, the scaling relation with \mtot\, for clusters is very slightly steeper than self-similar for quantities measured from the simulation, but it is consistent with self-similar once we account for the X-ray mass bias.

(v.5) \textit{\yszm\, scaling relation}: In \tng\,, the \yszm\, scaling relation  shows good agreement with the self-similar expectation for the highest mass clusters. The SBPL slope predicted at the high mass end is slightly biased low when we account for X-ray mass bias. 

(vi) \textit{Break in scaling relations}: We characterize the location of the break in each of the scaling relations included in this study. We find that all breaks occur between \mcrit\, $\sim 3 \times 10^{13}$\,\ms\, and $2 \times 10^{14}$\,\ms. However, our results indicate significant variability in the location of the break depending on the assumptions used to estimate masses and the particular scaling relation that is being considered. Unsurprisingly, the breaks in X-ray and SZ scaling relations are not sharp. Instead, the slope changes gradually between the high mass cluster regime down to (usually steeper) slopes that better characterize the trends in groups and galaxies. We provide estimates for the SBPL slope as a function of mass for all scalings and our results highlight that even samples with relatively high mass cuts (>$10^{14}$\ms) will suffer from significant biases when their slopes are estimated through simple power law models. 

Ongoing and future surveys such as 
the Dark Energy Survey \citep{DarkEnergySurveyCollaboration2016}, 
\textit{Euclid} \citep{Laureijs2011},
the Large Synoptic Survey Telescope \citep{LSSTDarkEnergyScienceCollaboration2012},
\textit{eROSITA} \citep{Merloni2012, Pillepich2012, Pillepich2018, Bulbul2021},
\textit{Athena} X-ray observatory \citep{Nandra2013},
SPT-3G \citep{Benson2014, Bender2018},
CMB-S4 \citep{Abazajian2016}, CMB-HD \citep{CMB-HD},
and Advanced ACTpol \citep{Henderson2016}
will expand current observational samples by several orders of magnitude in the number of clusters and galaxy groups being observed. These will be very useful for contrasting against current and future cosmological hydrodynamical simulations.

In this study, we combined mock X-ray measurements of a large sample of simulated haloes with a quantitative model for the dependence of X-ray and SZ scaling 
relations on halo mass.
\tng\, data indicates the presence of a break in all observable-mass relations considered in this study and this paper provides predictions for the mass dependence of the slope in the scaling relations across three orders of magnitude in halo mass.

\section*{Acknowledgments}
The authors would like to thank David Barnes for his contributions to \textsc{Mock-X} and helpful discussions prior to a career move. 
The work in this paper was supported by the NASA Earth and Space Science Fellowship (NESSF 80NSSC18K1111) awarded to Ana-Roxana Pop. LH acknowledges support from NSF grant AST-1815978. RW is supported by the Natural Sciences and Engineering Research Council of Canada (NSERC), funding reference CITA 490888-16. MV acknowledges support through NASA ATP 19-ATP19-0019, 19-ATP19-0020, 19-ATP19-0167, and NSF grants AST-1814053, AST-1814259, AST-1909831, AST-2007355 and AST-2107724. D.\,Nelson acknowledges funding from the Deutsche Forschungsgemeinschaft (DFG) through an Emmy Noether Research Group (grant number NE 2441/1-1). PT acknowledges support from NSF grant AST-1909933, AST-2008490, and NASA ATP Grant 80NSSC20K0502. The computations were run on the Odyssey cluster supported by the FAS Division of Science, Research Computing Group at Harvard University. This research made use of several software packages, including \verb+NumPy+ \citep{Harris2020}, \verb+SciPy+ \citep{Virtanen2020}, and all figures were generated using \verb+matplotlib+ \citep{Hunter2007}.

\section*{Data Availability}
Data from the Illustris and IllustrisTNG simulations used in this work are publicly available at the websites \href{https://www.illustris-project.org}{https://www.illustris-project.org} and \href{https://www.tng-project.org}{https://www.tng-project.org}, respectively \citep{Nelson2015, Nelson2019b}.

\bibliographystyle{mnras}
\bibliography{thesis}

\begin{thebibliography}{}
\makeatletter
\relax
\def\mn@urlcharsother{\let\do\@makeother \do\$\do\&\do\#\do\^\do\_\do\%\do\~}
\def\mn@doi{\begingroup\mn@urlcharsother \@ifnextchar [ {\mn@doi@}
  {\mn@doi@[]}}
\def\mn@doi@[#1]#2{\def\@tempa{#1}\ifx\@tempa\@empty \href
  {http://dx.doi.org/#2} {doi:#2}\else \href {http://dx.doi.org/#2} {#1}\fi
  \endgroup}
\def\mn@eprint#1#2{\mn@eprint@#1:#2::\@nil}
\def\mn@eprint@arXiv#1{\href {http://arxiv.org/abs/#1} {{\tt arXiv:#1}}}
\def\mn@eprint@dblp#1{\href {http://dblp.uni-trier.de/rec/bibtex/#1.xml}
  {dblp:#1}}
\def\mn@eprint@#1:#2:#3:#4\@nil{\def\@tempa {#1}\def\@tempb {#2}\def\@tempc
  {#3}\ifx \@tempc \@empty \let \@tempc \@tempb \let \@tempb \@tempa \fi \ifx
  \@tempb \@empty \def\@tempb {arXiv}\fi \@ifundefined
  {mn@eprint@\@tempb}{\@tempb:\@tempc}{\expandafter \expandafter \csname
  mn@eprint@\@tempb\endcsname \expandafter{\@tempc}}}

\bibitem[\protect\citeauthoryear{{Abazajian} et~al.,}{{Abazajian}
  et~al.}{2016}]{Abazajian2016}
{Abazajian} K.~N.,  et~al., 2016, arXiv e-prints, \href
  {https://ui.adsabs.harvard.edu/abs/2016arXiv161002743A} {p. arXiv:1610.02743}

\bibitem[\protect\citeauthoryear{{Abazajian} et~al.,}{{Abazajian}
  et~al.}{2019}]{CMB-S4}
{Abazajian} K.,  et~al., 2019, arXiv e-prints, \href
  {https://ui.adsabs.harvard.edu/abs/2019arXiv190704473A} {p. arXiv:1907.04473}

\bibitem[\protect\citeauthoryear{{Ade} et~al.,}{{Ade} et~al.}{2019}]{Ade2019}
{Ade} P.,  et~al., 2019, \mn@doi [\jcap] {10.1088/1475-7516/2019/02/056}, \href
  {https://ui.adsabs.harvard.edu/abs/2019JCAP...02..056A} {2019, 056}

\bibitem[\protect\citeauthoryear{{Amodeo} et~al.,}{{Amodeo}
  et~al.}{2021}]{Amodeo2021}
{Amodeo} S.,  et~al., 2021, \mn@doi [\prd] {10.1103/PhysRevD.103.063514}, \href
  {https://ui.adsabs.harvard.edu/abs/2021PhRvD.103f3514A} {103, 063514}

\bibitem[\protect\citeauthoryear{{Anderson}, {Gaspari}, {White}, {Wang}  \&
  {Dai}}{{Anderson} et~al.}{2015}]{Anderson2015}
{Anderson} M.~E.,  {Gaspari} M.,  {White} S. D.~M.,  {Wang} W.,   {Dai} X.,
  2015, \mn@doi [\mnras] {10.1093/mnras/stv437}, \href
  {https://ui.adsabs.harvard.edu/abs/2015MNRAS.449.3806A} {449, 3806}

\bibitem[\protect\citeauthoryear{{Arnaud}, {Pointecouteau}  \&
  {Pratt}}{{Arnaud} et~al.}{2007}]{Arnaud2007}
{Arnaud} M.,  {Pointecouteau} E.,   {Pratt} G.~W.,  2007, \mn@doi [\aap]
  {10.1051/0004-6361:20078541}, \href
  {https://ui.adsabs.harvard.edu/abs/2007A&A...474L..37A} {474, L37}

\bibitem[\protect\citeauthoryear{{Arnaud}, {Pratt}, {Piffaretti},
  {B{\"o}hringer}, {Croston}  \& {Pointecouteau}}{{Arnaud}
  et~al.}{2010}]{Arnaud2010}
{Arnaud} M.,  {Pratt} G.~W.,  {Piffaretti} R.,  {B{\"o}hringer} H.,  {Croston}
  J.~H.,   {Pointecouteau} E.,  2010, \mn@doi [\aap]
  {10.1051/0004-6361/200913416}, \href
  {https://ui.adsabs.harvard.edu/abs/2010A&A...517A..92A} {517, A92}

\bibitem[\protect\citeauthoryear{{Avestruz}, {Lau}, {Nagai}  \&
  {Vikhlinin}}{{Avestruz} et~al.}{2014}]{Avestruz2014}
{Avestruz} C.,  {Lau} E.~T.,  {Nagai} D.,   {Vikhlinin} A.,  2014, \mn@doi
  [\apj] {10.1088/0004-637X/791/2/117}, \href
  {https://ui.adsabs.harvard.edu/abs/2014ApJ...791..117A} {791, 117}

\bibitem[\protect\citeauthoryear{{Bah{\'e}} et~al.,}{{Bah{\'e}}
  et~al.}{2017}]{Bahe2017}
{Bah{\'e}} Y.~M.,  et~al., 2017, \mn@doi [\mnras] {10.1093/mnras/stx1403},
  \href {https://ui.adsabs.harvard.edu/abs/2017MNRAS.470.4186B} {470, 4186}

\bibitem[\protect\citeauthoryear{{Barnes}, {Kay}, {Henson}, {McCarthy},
  {Schaye}  \& {Jenkins}}{{Barnes} et~al.}{2017a}]{Barnes2017a}
{Barnes} D.~J.,  {Kay} S.~T.,  {Henson} M.~A.,  {McCarthy} I.~G.,  {Schaye} J.,
    {Jenkins} A.,  2017a, \mn@doi [\mnras] {10.1093/mnras/stw2722}, \href
  {https://ui.adsabs.harvard.edu/abs/2017MNRAS.465..213B} {465, 213}

\bibitem[\protect\citeauthoryear{{Barnes} et~al.,}{{Barnes}
  et~al.}{2017b}]{Barnes2017b}
{Barnes} D.~J.,  et~al., 2017b, \mn@doi [\mnras] {10.1093/mnras/stx1647}, \href
  {https://ui.adsabs.harvard.edu/abs/2017MNRAS.471.1088B} {471, 1088}

\bibitem[\protect\citeauthoryear{{Barnes} et~al.,}{{Barnes}
  et~al.}{2019}]{Barnes2019}
{Barnes} D.~J.,  et~al., 2019, \mn@doi [\mnras] {10.1093/mnras/stz1814}, \href
  {https://ui.adsabs.harvard.edu/abs/2019MNRAS.488.3003B} {488, 3003}

\bibitem[\protect\citeauthoryear{{Barnes}, {Vogelsberger}, {Pearce}, {Pop},
  {Kannan}, {Cao}, {Kay}  \& {Hernquist}}{{Barnes} et~al.}{2021}]{Barnes2021}
{Barnes} D.~J.,  {Vogelsberger} M.,  {Pearce} F.~A.,  {Pop} A.-R.,  {Kannan}
  R.,  {Cao} K.,  {Kay} S.~T.,   {Hernquist} L.,  2021, \mn@doi [\mnras]
  {10.1093/mnras/stab1276}, \href
  {https://ui.adsabs.harvard.edu/abs/2021MNRAS.506.2533B} {506, 2533}

\bibitem[\protect\citeauthoryear{{Battaglia}, {Bond}, {Pfrommer}  \&
  {Sievers}}{{Battaglia} et~al.}{2012}]{Battaglia2012}
{Battaglia} N.,  {Bond} J.~R.,  {Pfrommer} C.,   {Sievers} J.~L.,  2012,
  \mn@doi [\apj] {10.1088/0004-637X/758/2/74}, \href
  {https://ui.adsabs.harvard.edu/abs/2012ApJ...758...74B} {758, 74}

\bibitem[\protect\citeauthoryear{{Battaglia}, {Bond}, {Pfrommer}  \&
  {Sievers}}{{Battaglia} et~al.}{2013}]{Battaglia2013}
{Battaglia} N.,  {Bond} J.~R.,  {Pfrommer} C.,   {Sievers} J.~L.,  2013,
  \mn@doi [\apj] {10.1088/0004-637X/777/2/123}, \href
  {https://ui.adsabs.harvard.edu/abs/2013ApJ...777..123B} {777, 123}

\bibitem[\protect\citeauthoryear{{Bender} et~al.,}{{Bender}
  et~al.}{2018}]{Bender2018}
{Bender} A.~N.,  et~al., 2018, in Millimeter, Submillimeter, and Far-Infrared
  Detectors and Instrumentation for Astronomy IX. p. 1070803 (\mn@eprint
  {arXiv} {1809.00036}), \mn@doi{10.1117/12.2312426}

\bibitem[\protect\citeauthoryear{{Benson} et~al.,}{{Benson}
  et~al.}{2014}]{Benson2014}
{Benson} B.~A.,  et~al., 2014, in Millimeter, Submillimeter, and Far-Infrared
  Detectors and Instrumentation for Astronomy VII. p. 91531P (\mn@eprint
  {arXiv} {1407.2973}), \mn@doi{10.1117/12.2057305}

\bibitem[\protect\citeauthoryear{{Biffi}, {Dolag}, {B{\"o}hringer}  \&
  {Lemson}}{{Biffi} et~al.}{2012}]{Biffi2012}
{Biffi} V.,  {Dolag} K.,  {B{\"o}hringer} H.,   {Lemson} G.,  2012, \mn@doi
  [\mnras] {10.1111/j.1365-2966.2011.20278.x}, \href
  {https://ui.adsabs.harvard.edu/abs/2012MNRAS.420.3545B} {420, 3545}

\bibitem[\protect\citeauthoryear{{Biffi}, {Sembolini}, {De Petris},
  {Valdarnini}, {Yepes}  \& {Gottl{\"o}ber}}{{Biffi} et~al.}{2014}]{Biffi2014}
{Biffi} V.,  {Sembolini} F.,  {De Petris} M.,  {Valdarnini} R.,  {Yepes} G.,
  {Gottl{\"o}ber} S.,  2014, \mn@doi [\mnras] {10.1093/mnras/stu018}, \href
  {https://ui.adsabs.harvard.edu/abs/2014MNRAS.439..588B} {439, 588}

\bibitem[\protect\citeauthoryear{{Biffi} et~al.,}{{Biffi}
  et~al.}{2016}]{Biffi2016}
{Biffi} V.,  et~al., 2016, \mn@doi [\apj] {10.3847/0004-637X/827/2/112}, \href
  {https://ui.adsabs.harvard.edu/abs/2016ApJ...827..112B} {827, 112}

\bibitem[\protect\citeauthoryear{{Bleem} et~al.,}{{Bleem}
  et~al.}{2015}]{Bleem2015}
{Bleem} L.~E.,  et~al., 2015, \mn@doi [\apjs] {10.1088/0067-0049/216/2/27},
  \href {https://ui.adsabs.harvard.edu/abs/2015ApJS..216...27B} {216, 27}

\bibitem[\protect\citeauthoryear{{Bregman}, {Hodges-Kluck}, {Qu}, {Pratt}, {Li}
   \& {Yun}}{{Bregman} et~al.}{2022}]{Bregman2022}
{Bregman} J.~N.,  {Hodges-Kluck} E.,  {Qu} Z.,  {Pratt} C.,  {Li} J.-T.,
  {Yun} Y.,  2022, \mn@doi [\apj] {10.3847/1538-4357/ac51de}, \href
  {https://ui.adsabs.harvard.edu/abs/2022ApJ...928...14B} {928, 14}

\bibitem[\protect\citeauthoryear{{Bulbul} et~al.,}{{Bulbul}
  et~al.}{2021}]{Bulbul2021}
{Bulbul} E.,  et~al., 2021, arXiv e-prints, \href
  {https://ui.adsabs.harvard.edu/abs/2021arXiv211009544B} {p. arXiv:2110.09544}

\bibitem[\protect\citeauthoryear{{Chadayammuri}, {Bogdan}, {Oppenheimer},
  {Kraft}, {Forman}  \& {Jones}}{{Chadayammuri}
  et~al.}{2022}]{Chadayammuri2022}
{Chadayammuri} U.,  {Bogdan} A.,  {Oppenheimer} B.,  {Kraft} R.,  {Forman} W.,
   {Jones} C.,  2022, arXiv e-prints, \href
  {https://ui.adsabs.harvard.edu/abs/2022arXiv220301356C} {p. arXiv:2203.01356}

\bibitem[\protect\citeauthoryear{{Chiu} et~al.,}{{Chiu}
  et~al.}{2018}]{Chiu2018}
{Chiu} I.,  et~al., 2018, \mn@doi [\mnras] {10.1093/mnras/sty1284}, \href
  {https://ui.adsabs.harvard.edu/abs/2018MNRAS.478.3072C} {478, 3072}

\bibitem[\protect\citeauthoryear{{Comparat} et~al.,}{{Comparat}
  et~al.}{2022}]{Comparat2022}
{Comparat} J.,  et~al., 2022, arXiv e-prints, \href
  {https://ui.adsabs.harvard.edu/abs/2022arXiv220105169C} {p. arXiv:2201.05169}

\bibitem[\protect\citeauthoryear{{Dark Energy Survey Collaboration}
  et~al.,}{{Dark Energy Survey Collaboration}
  et~al.}{2016}]{DarkEnergySurveyCollaboration2016}
{Dark Energy Survey Collaboration} et~al., 2016, \mn@doi [\mnras]
  {10.1093/mnras/stw641}, \href
  {https://ui.adsabs.harvard.edu/abs/2016MNRAS.460.1270D} {460, 1270}

\bibitem[\protect\citeauthoryear{{Dolag}, {Borgani}, {Murante}  \&
  {Springel}}{{Dolag} et~al.}{2009}]{Dolag2009}
{Dolag} K.,  {Borgani} S.,  {Murante} G.,   {Springel} V.,  2009, \mn@doi
  [\mnras] {10.1111/j.1365-2966.2009.15034.x}, \href
  {https://ui.adsabs.harvard.edu/abs/2009MNRAS.399..497D} {399, 497}

\bibitem[\protect\citeauthoryear{{Duffy}, {Schaye}, {Kay}  \& {Dalla
  Vecchia}}{{Duffy} et~al.}{2008}]{Duffy2008}
{Duffy} A.~R.,  {Schaye} J.,  {Kay} S.~T.,   {Dalla Vecchia} C.,  2008, \mn@doi
  [\mnras] {10.1111/j.1745-3933.2008.00537.x}, \href
  {https://ui.adsabs.harvard.edu/abs/2008MNRAS.390L..64D} {390, L64}

\bibitem[\protect\citeauthoryear{{Dutton} \& {Macci{\`o}}}{{Dutton} \&
  {Macci{\`o}}}{2014}]{Dutton2014}
{Dutton} A.~A.,  {Macci{\`o}} A.~V.,  2014, \mn@doi [\mnras]
  {10.1093/mnras/stu742}, \href
  {https://ui.adsabs.harvard.edu/abs/2014MNRAS.441.3359D} {441, 3359}

\bibitem[\protect\citeauthoryear{{Eckert}, {Ettori}, {Pointecouteau},
  {Molendi}, {Paltani}  \& {Tchernin}}{{Eckert} et~al.}{2017}]{Eckert2017}
{Eckert} D.,  {Ettori} S.,  {Pointecouteau} E.,  {Molendi} S.,  {Paltani} S.,
  {Tchernin} C.,  2017, \mn@doi [Astronomische Nachrichten]
  {10.1002/asna.201713345}, \href
  {https://ui.adsabs.harvard.edu/abs/2017AN....338..293E} {338, 293}

\bibitem[\protect\citeauthoryear{{Eckert} et~al.,}{{Eckert}
  et~al.}{2019}]{Eckert2019}
{Eckert} D.,  et~al., 2019, \mn@doi [\aap] {10.1051/0004-6361/201833324}, \href
  {https://ui.adsabs.harvard.edu/abs/2019A&A...621A..40E} {621, A40}

\bibitem[\protect\citeauthoryear{{Eckmiller}, {Hudson}  \&
  {Reiprich}}{{Eckmiller} et~al.}{2011}]{Eckmiller2011}
{Eckmiller} H.~J.,  {Hudson} D.~S.,   {Reiprich} T.~H.,  2011, \mn@doi [\aap]
  {10.1051/0004-6361/201116734}, \href
  {https://ui.adsabs.harvard.edu/abs/2011A&A...535A.105E} {535, A105}

\bibitem[\protect\citeauthoryear{{Fabjan}, {Borgani}, {Rasia}, {Bonafede},
  {Dolag}, {Murante}  \& {Tornatore}}{{Fabjan} et~al.}{2011}]{Fabjan2011}
{Fabjan} D.,  {Borgani} S.,  {Rasia} E.,  {Bonafede} A.,  {Dolag} K.,
  {Murante} G.,   {Tornatore} L.,  2011, \mn@doi [\mnras]
  {10.1111/j.1365-2966.2011.18497.x}, \href
  {https://ui.adsabs.harvard.edu/abs/2011MNRAS.416..801F} {416, 801}

\bibitem[\protect\citeauthoryear{{Foster}, {Ji}, {Smith}  \&
  {Brickhouse}}{{Foster} et~al.}{2012}]{Foster2012}
{Foster} A.~R.,  {Ji} L.,  {Smith} R.~K.,   {Brickhouse} N.~S.,  2012, \mn@doi
  [\apj] {10.1088/0004-637X/756/2/128}, \href
  {https://ui.adsabs.harvard.edu/abs/2012ApJ...756..128F} {756, 128}

\bibitem[\protect\citeauthoryear{{Gardini}, {Rasia}, {Mazzotta}, {Tormen}, {De
  Grandi}  \& {Moscardini}}{{Gardini} et~al.}{2004}]{Gardini2004}
{Gardini} A.,  {Rasia} E.,  {Mazzotta} P.,  {Tormen} G.,  {De Grandi} S.,
  {Moscardini} L.,  2004, \mn@doi [\mnras] {10.1111/j.1365-2966.2004.07800.x},
  \href {https://ui.adsabs.harvard.edu/abs/2004MNRAS.351..505G} {351, 505}

\bibitem[\protect\citeauthoryear{{Gaspari}, {Brighenti}, {Temi}  \&
  {Ettori}}{{Gaspari} et~al.}{2014}]{Gaspari2014}
{Gaspari} M.,  {Brighenti} F.,  {Temi} P.,   {Ettori} S.,  2014, \mn@doi
  [\apjl] {10.1088/2041-8205/783/1/L10}, \href
  {https://ui.adsabs.harvard.edu/abs/2014ApJ...783L..10G} {783, L10}

\bibitem[\protect\citeauthoryear{{Genel} et~al.,}{{Genel}
  et~al.}{2014}]{Genel2014}
{Genel} S.,  et~al., 2014, \mn@doi [\mnras] {10.1093/mnras/stu1654}, \href
  {https://ui.adsabs.harvard.edu/abs/2014MNRAS.445..175G} {445, 175}

\bibitem[\protect\citeauthoryear{{Giles} et~al.,}{{Giles}
  et~al.}{2016}]{Giles2016}
{Giles} P.~A.,  et~al., 2016, \mn@doi [\aap] {10.1051/0004-6361/201526886},
  \href {https://ui.adsabs.harvard.edu/abs/2016A&A...592A...3G} {592, A3}

\bibitem[\protect\citeauthoryear{{Giles} et~al.,}{{Giles}
  et~al.}{2017}]{Giles2017}
{Giles} P.~A.,  et~al., 2017, \mn@doi [\mnras] {10.1093/mnras/stw2621}, \href
  {https://ui.adsabs.harvard.edu/abs/2017MNRAS.465..858G} {465, 858}

\bibitem[\protect\citeauthoryear{{Gonzalez}, {Sivanandam}, {Zabludoff}  \&
  {Zaritsky}}{{Gonzalez} et~al.}{2013}]{Gonzalez2013}
{Gonzalez} A.~H.,  {Sivanandam} S.,  {Zabludoff} A.~I.,   {Zaritsky} D.,  2013,
  \mn@doi [\apj] {10.1088/0004-637X/778/1/14}, \href
  {https://ui.adsabs.harvard.edu/abs/2013ApJ...778...14G} {778, 14}

\bibitem[\protect\citeauthoryear{{Harris} et~al.,}{{Harris}
  et~al.}{2020}]{Harris2020}
{Harris} C.~R.,  et~al., 2020, \mn@doi [\nat] {10.1038/s41586-020-2649-2},
  \href {https://ui.adsabs.harvard.edu/abs/2020Natur.585..357H} {585, 357}

\bibitem[\protect\citeauthoryear{{Hasselfield} et~al.,}{{Hasselfield}
  et~al.}{2013}]{Hasselfield2013}
{Hasselfield} M.,  et~al., 2013, \mn@doi [\jcap]
  {10.1088/1475-7516/2013/07/008}, \href
  {https://ui.adsabs.harvard.edu/abs/2013JCAP...07..008H} {2013, 008}

\bibitem[\protect\citeauthoryear{{Heinz} \& {Br{\"u}ggen}}{{Heinz} \&
  {Br{\"u}ggen}}{2009}]{Heinz2009}
{Heinz} S.,  {Br{\"u}ggen} M.,  2009, arXiv e-prints, \href
  {https://ui.adsabs.harvard.edu/abs/2009arXiv0903.0043H} {p. arXiv:0903.0043}

\bibitem[\protect\citeauthoryear{{Henden}, {Puchwein}, {Shen}  \&
  {Sijacki}}{{Henden} et~al.}{2018}]{Henden2018}
{Henden} N.~A.,  {Puchwein} E.,  {Shen} S.,   {Sijacki} D.,  2018, \mn@doi
  [\mnras] {10.1093/mnras/sty1780}, \href
  {https://ui.adsabs.harvard.edu/abs/2018MNRAS.479.5385H} {479, 5385}

\bibitem[\protect\citeauthoryear{{Henden}, {Puchwein}  \& {Sijacki}}{{Henden}
  et~al.}{2019}]{Henden2019}
{Henden} N.~A.,  {Puchwein} E.,   {Sijacki} D.,  2019, \mn@doi [\mnras]
  {10.1093/mnras/stz2301}, \href
  {https://ui.adsabs.harvard.edu/abs/2019MNRAS.489.2439H} {489, 2439}

\bibitem[\protect\citeauthoryear{{Henderson} et~al.,}{{Henderson}
  et~al.}{2016}]{Henderson2016}
{Henderson} S.~W.,  et~al., 2016, \mn@doi [Journal of Low Temperature Physics]
  {10.1007/s10909-016-1575-z}, \href
  {https://ui.adsabs.harvard.edu/abs/2016JLTP..184..772H} {184, 772}

\bibitem[\protect\citeauthoryear{{Hunter}}{{Hunter}}{2007}]{Hunter2007}
{Hunter} J.~D.,  2007, \mn@doi [Computing in Science and Engineering]
  {10.1109/MCSE.2007.55}, \href
  {https://ui.adsabs.harvard.edu/abs/2007CSE.....9...90H} {9, 90}

\bibitem[\protect\citeauthoryear{{Kaiser}}{{Kaiser}}{1986}]{Kaiser1986}
{Kaiser} N.,  1986, \mn@doi [\mnras] {10.1093/mnras/222.2.323}, \href
  {https://ui.adsabs.harvard.edu/abs/1986MNRAS.222..323K} {222, 323}

\bibitem[\protect\citeauthoryear{{Kannan}, {Springel}, {Pakmor}, {Marinacci}
  \& {Vogelsberger}}{{Kannan} et~al.}{2016}]{Kannan2016}
{Kannan} R.,  {Springel} V.,  {Pakmor} R.,  {Marinacci} F.,   {Vogelsberger}
  M.,  2016, \mn@doi [\mnras] {10.1093/mnras/stw294}, \href
  {https://ui.adsabs.harvard.edu/abs/2016MNRAS.458..410K} {458, 410}

\bibitem[\protect\citeauthoryear{{Kannan}, {Vogelsberger}, {Pfrommer},
  {Weinberger}, {Springel}, {Hernquist}, {Puchwein}  \& {Pakmor}}{{Kannan}
  et~al.}{2017}]{Kannan2017}
{Kannan} R.,  {Vogelsberger} M.,  {Pfrommer} C.,  {Weinberger} R.,  {Springel}
  V.,  {Hernquist} L.,  {Puchwein} E.,   {Pakmor} R.,  2017, \mn@doi [\apjl]
  {10.3847/2041-8213/aa624b}, \href
  {https://ui.adsabs.harvard.edu/abs/2017ApJ...837L..18K} {837, L18}

\bibitem[\protect\citeauthoryear{{Kay}, {Peel}, {Short}, {Thomas}, {Young},
  {Battye}, {Liddle}  \& {Pearce}}{{Kay} et~al.}{2012}]{Kay2012}
{Kay} S.~T.,  {Peel} M.~W.,  {Short} C.~J.,  {Thomas} P.~A.,  {Young} O.~E.,
  {Battye} R.~A.,  {Liddle} A.~R.,   {Pearce} F.~R.,  2012, \mn@doi [\mnras]
  {10.1111/j.1365-2966.2012.20623.x}, \href
  {https://ui.adsabs.harvard.edu/abs/2012MNRAS.422.1999K} {422, 1999}

\bibitem[\protect\citeauthoryear{{Kettula} et~al.,}{{Kettula}
  et~al.}{2015}]{Kettula2015}
{Kettula} K.,  et~al., 2015, \mn@doi [\mnras] {10.1093/mnras/stv923}, \href
  {https://ui.adsabs.harvard.edu/abs/2015MNRAS.451.1460K} {451, 1460}

\bibitem[\protect\citeauthoryear{{Khedekar}, {Churazov}, {Kravtsov},
  {Zhuravleva}, {Lau}, {Nagai}  \& {Sunyaev}}{{Khedekar}
  et~al.}{2013}]{Khedekar2013}
{Khedekar} S.,  {Churazov} E.,  {Kravtsov} A.,  {Zhuravleva} I.,  {Lau} E.~T.,
  {Nagai} D.,   {Sunyaev} R.,  2013, \mn@doi [\mnras] {10.1093/mnras/stt224},
  \href {https://ui.adsabs.harvard.edu/abs/2013MNRAS.431..954K} {431, 954}

\bibitem[\protect\citeauthoryear{{Klypin}, {Trujillo-Gomez}  \&
  {Primack}}{{Klypin} et~al.}{2011}]{Klypin2011}
{Klypin} A.~A.,  {Trujillo-Gomez} S.,   {Primack} J.,  2011, \mn@doi [\apj]
  {10.1088/0004-637X/740/2/102}, \href
  {https://ui.adsabs.harvard.edu/abs/2011ApJ...740..102K} {740, 102}

\bibitem[\protect\citeauthoryear{{Klypin}, {Yepes}, {Gottl{\"o}ber}, {Prada}
  \& {He{\ss}}}{{Klypin} et~al.}{2016}]{Klypin2016}
{Klypin} A.,  {Yepes} G.,  {Gottl{\"o}ber} S.,  {Prada} F.,   {He{\ss}} S.,
  2016, \mn@doi [\mnras] {10.1093/mnras/stw248}, \href
  {https://ui.adsabs.harvard.edu/abs/2016MNRAS.457.4340K} {457, 4340}

\bibitem[\protect\citeauthoryear{{Kravtsov}, {Vikhlinin}  \&
  {Nagai}}{{Kravtsov} et~al.}{2006}]{Kravtsov2006}
{Kravtsov} A.~V.,  {Vikhlinin} A.,   {Nagai} D.,  2006, \mn@doi [\apj]
  {10.1086/506319}, \href
  {https://ui.adsabs.harvard.edu/abs/2006ApJ...650..128K} {650, 128}

\bibitem[\protect\citeauthoryear{{LSST Dark Energy Science
  Collaboration}}{{LSST Dark Energy Science
  Collaboration}}{2012}]{LSSTDarkEnergyScienceCollaboration2012}
{LSST Dark Energy Science Collaboration} 2012, arXiv e-prints, \href
  {https://ui.adsabs.harvard.edu/abs/2012arXiv1211.0310L} {p. arXiv:1211.0310}

\bibitem[\protect\citeauthoryear{{Lau}, {Kravtsov}  \& {Nagai}}{{Lau}
  et~al.}{2009}]{Lau2009}
{Lau} E.~T.,  {Kravtsov} A.~V.,   {Nagai} D.,  2009, \mn@doi [\apj]
  {10.1088/0004-637X/705/2/1129}, \href
  {https://ui.adsabs.harvard.edu/abs/2009ApJ...705.1129L} {705, 1129}

\bibitem[\protect\citeauthoryear{{Laureijs} et~al.,}{{Laureijs}
  et~al.}{2011}]{Laureijs2011}
{Laureijs} R.,  et~al., 2011, arXiv e-prints, \href
  {https://ui.adsabs.harvard.edu/abs/2011arXiv1110.3193L} {p. arXiv:1110.3193}

\bibitem[\protect\citeauthoryear{{Le Brun}, {McCarthy}, {Schaye}  \&
  {Ponman}}{{Le Brun} et~al.}{2014}]{LeBrun2014}
{Le Brun} A. M.~C.,  {McCarthy} I.~G.,  {Schaye} J.,   {Ponman} T.~J.,  2014,
  \mn@doi [\mnras] {10.1093/mnras/stu608}, \href
  {https://ui.adsabs.harvard.edu/abs/2014MNRAS.441.1270L} {441, 1270}

\bibitem[\protect\citeauthoryear{{Le Brun}, {McCarthy}, {Schaye}  \&
  {Ponman}}{{Le Brun} et~al.}{2017}]{LeBrun2017}
{Le Brun} A. M.~C.,  {McCarthy} I.~G.,  {Schaye} J.,   {Ponman} T.~J.,  2017,
  \mn@doi [\mnras] {10.1093/mnras/stw3361}, \href
  {https://ui.adsabs.harvard.edu/abs/2017MNRAS.466.4442L} {466, 4442}

\bibitem[\protect\citeauthoryear{{Lieu} et~al.,}{{Lieu}
  et~al.}{2016}]{Lieu2016}
{Lieu} M.,  et~al., 2016, \mn@doi [\aap] {10.1051/0004-6361/201526883}, \href
  {https://ui.adsabs.harvard.edu/abs/2016A&A...592A...4L} {592, A4}

\bibitem[\protect\citeauthoryear{{Lim}, {Barnes}, {Vogelsberger}, {Mo},
  {Nelson}, {Pillepich}, {Dolag}  \& {Marinacci}}{{Lim} et~al.}{2021}]{Lim2021}
{Lim} S.~H.,  {Barnes} D.,  {Vogelsberger} M.,  {Mo} H.~J.,  {Nelson} D.,
  {Pillepich} A.,  {Dolag} K.,   {Marinacci} F.,  2021, \mn@doi [\mnras]
  {10.1093/mnras/stab1172}, \href
  {https://ui.adsabs.harvard.edu/abs/2021MNRAS.504.5131L} {504, 5131}

\bibitem[\protect\citeauthoryear{{Lin}, {Stanford}, {Eisenhardt}, {Vikhlinin},
  {Maughan}  \& {Kravtsov}}{{Lin} et~al.}{2012}]{Lin2012}
{Lin} Y.-T.,  {Stanford} S.~A.,  {Eisenhardt} P. R.~M.,  {Vikhlinin} A.,
  {Maughan} B.~J.,   {Kravtsov} A.,  2012, \mn@doi [\apjl]
  {10.1088/2041-8205/745/1/L3}, \href
  {https://ui.adsabs.harvard.edu/abs/2012ApJ...745L...3L} {745, L3}

\bibitem[\protect\citeauthoryear{{Lovisari}, {Reiprich}  \&
  {Schellenberger}}{{Lovisari} et~al.}{2015}]{Lovisari2015}
{Lovisari} L.,  {Reiprich} T.~H.,   {Schellenberger} G.,  2015, \mn@doi [\aap]
  {10.1051/0004-6361/201423954}, \href
  {https://ui.adsabs.harvard.edu/abs/2015A&A...573A.118L} {573, A118}

\bibitem[\protect\citeauthoryear{{Mahdavi}, {Hoekstra}, {Babul}, {Bildfell},
  {Jeltema}  \& {Henry}}{{Mahdavi} et~al.}{2013}]{Mahdavi2013}
{Mahdavi} A.,  {Hoekstra} H.,  {Babul} A.,  {Bildfell} C.,  {Jeltema} T.,
  {Henry} J.~P.,  2013, \mn@doi [\apj] {10.1088/0004-637X/767/2/116}, \href
  {https://ui.adsabs.harvard.edu/abs/2013ApJ...767..116M} {767, 116}

\bibitem[\protect\citeauthoryear{{Mantz}, {Allen}, {Rapetti}  \&
  {Ebeling}}{{Mantz} et~al.}{2010a}]{Mantz2010a}
{Mantz} A.,  {Allen} S.~W.,  {Rapetti} D.,   {Ebeling} H.,  2010a, \mn@doi
  [\mnras] {10.1111/j.1365-2966.2010.16992.x}, \href
  {https://ui.adsabs.harvard.edu/abs/2010MNRAS.406.1759M} {406, 1759}

\bibitem[\protect\citeauthoryear{{Mantz}, {Allen}, {Ebeling}, {Rapetti}  \&
  {Drlica-Wagner}}{{Mantz} et~al.}{2010b}]{Mantz2010b}
{Mantz} A.,  {Allen} S.~W.,  {Ebeling} H.,  {Rapetti} D.,   {Drlica-Wagner} A.,
   2010b, \mn@doi [\mnras] {10.1111/j.1365-2966.2010.16993.x}, \href
  {https://ui.adsabs.harvard.edu/abs/2010MNRAS.406.1773M} {406, 1773}

\bibitem[\protect\citeauthoryear{{Mantz}, {Allen}, {Morris}  \&
  {Schmidt}}{{Mantz} et~al.}{2016a}]{Mantz2016a}
{Mantz} A.~B.,  {Allen} S.~W.,  {Morris} R.~G.,   {Schmidt} R.~W.,  2016a,
  \mn@doi [\mnras] {10.1093/mnras/stv2899}, \href
  {https://ui.adsabs.harvard.edu/abs/2016MNRAS.456.4020M} {456, 4020}

\bibitem[\protect\citeauthoryear{{Mantz} et~al.,}{{Mantz}
  et~al.}{2016b}]{Mantz2016b}
{Mantz} A.~B.,  et~al., 2016b, \mn@doi [\mnras] {10.1093/mnras/stw2250}, \href
  {https://ui.adsabs.harvard.edu/abs/2016MNRAS.463.3582M} {463, 3582}

\bibitem[\protect\citeauthoryear{Marinacci et~al.,}{Marinacci
  et~al.}{2018}]{Marinacci2018}
Marinacci F.,  et~al., 2018, \mn@doi [Monthly Notices of the Royal Astronomical
  Society] {10.1093/mnras/sty2206}, 480, 5113

\bibitem[\protect\citeauthoryear{{Maughan}}{{Maughan}}{2007}]{Maughan2007}
{Maughan} B.~J.,  2007, \mn@doi [\apj] {10.1086/520831}, \href
  {https://ui.adsabs.harvard.edu/abs/2007ApJ...668..772M} {668, 772}

\bibitem[\protect\citeauthoryear{{Maughan}, {Jones}, {Forman}  \& {Van
  Speybroeck}}{{Maughan} et~al.}{2008}]{Maughan2008}
{Maughan} B.~J.,  {Jones} C.,  {Forman} W.,   {Van Speybroeck} L.,  2008,
  \mn@doi [\apjs] {10.1086/521225}, \href
  {https://ui.adsabs.harvard.edu/abs/2008ApJS..174..117M} {174, 117}

\bibitem[\protect\citeauthoryear{{Mazzotta}, {Rasia}, {Moscardini}  \&
  {Tormen}}{{Mazzotta} et~al.}{2004}]{Mazzotta2004}
{Mazzotta} P.,  {Rasia} E.,  {Moscardini} L.,   {Tormen} G.,  2004, \mn@doi
  [\mnras] {10.1111/j.1365-2966.2004.08167.x}, \href
  {https://ui.adsabs.harvard.edu/abs/2004MNRAS.354...10M} {354, 10}

\bibitem[\protect\citeauthoryear{{McCarthy}, {Schaye}, {Bower}, {Ponman},
  {Booth}, {Dalla Vecchia}  \& {Springel}}{{McCarthy}
  et~al.}{2011}]{McCarthy2011}
{McCarthy} I.~G.,  {Schaye} J.,  {Bower} R.~G.,  {Ponman} T.~J.,  {Booth}
  C.~M.,  {Dalla Vecchia} C.,   {Springel} V.,  2011, \mn@doi [\mnras]
  {10.1111/j.1365-2966.2010.18033.x}, \href
  {https://ui.adsabs.harvard.edu/abs/2011MNRAS.412.1965M} {412, 1965}

\bibitem[\protect\citeauthoryear{{McDonald} et~al.,}{{McDonald}
  et~al.}{2013}]{McDonald2013}
{McDonald} M.,  et~al., 2013, \mn@doi [\apj] {10.1088/0004-637X/774/1/23},
  \href {https://ui.adsabs.harvard.edu/abs/2013ApJ...774...23M} {774, 23}

\bibitem[\protect\citeauthoryear{{McDonald} et~al.,}{{McDonald}
  et~al.}{2017}]{McDonald2017}
{McDonald} M.,  et~al., 2017, \mn@doi [\apj] {10.3847/1538-4357/aa7740}, \href
  {https://ui.adsabs.harvard.edu/abs/2017ApJ...843...28M} {843, 28}

\bibitem[\protect\citeauthoryear{{Mehrtens} et~al.,}{{Mehrtens}
  et~al.}{2012}]{Mehrtens2012}
{Mehrtens} N.,  et~al., 2012, \mn@doi [\mnras]
  {10.1111/j.1365-2966.2012.20931.x}, \href
  {https://ui.adsabs.harvard.edu/abs/2012MNRAS.423.1024M} {423, 1024}

\bibitem[\protect\citeauthoryear{{Merloni} et~al.,}{{Merloni}
  et~al.}{2012}]{Merloni2012}
{Merloni} A.,  et~al., 2012, arXiv e-prints, \href
  {https://ui.adsabs.harvard.edu/abs/2012arXiv1209.3114M} {p. arXiv:1209.3114}

\bibitem[\protect\citeauthoryear{{Motl}, {Hallman}, {Burns}  \&
  {Norman}}{{Motl} et~al.}{2005}]{Motl2005}
{Motl} P.~M.,  {Hallman} E.~J.,  {Burns} J.~O.,   {Norman} M.~L.,  2005,
  \mn@doi [\apjl] {10.1086/430144}, \href
  {https://ui.adsabs.harvard.edu/abs/2005ApJ...623L..63M} {623, L63}

\bibitem[\protect\citeauthoryear{{Nagai}}{{Nagai}}{2006}]{Nagai2006}
{Nagai} D.,  2006, \mn@doi [\apj] {10.1086/506467}, \href
  {https://ui.adsabs.harvard.edu/abs/2006ApJ...650..538N} {650, 538}

\bibitem[\protect\citeauthoryear{{Nagai} \& {Lau}}{{Nagai} \&
  {Lau}}{2011}]{Nagai2011}
{Nagai} D.,  {Lau} E.~T.,  2011, \mn@doi [\apjl] {10.1088/2041-8205/731/1/L10},
  \href {https://ui.adsabs.harvard.edu/abs/2011ApJ...731L..10N} {731, L10}

\bibitem[\protect\citeauthoryear{{Nagai}, {Vikhlinin}  \& {Kravtsov}}{{Nagai}
  et~al.}{2007a}]{Nagai2007a}
{Nagai} D.,  {Vikhlinin} A.,   {Kravtsov} A.~V.,  2007a, \mn@doi [\apj]
  {10.1086/509868}, \href
  {https://ui.adsabs.harvard.edu/abs/2007ApJ...655...98N} {655, 98}

\bibitem[\protect\citeauthoryear{{Nagai}, {Kravtsov}  \& {Vikhlinin}}{{Nagai}
  et~al.}{2007b}]{Nagai2007b}
{Nagai} D.,  {Kravtsov} A.~V.,   {Vikhlinin} A.,  2007b, \mn@doi [\apj]
  {10.1086/521328}, \href
  {https://ui.adsabs.harvard.edu/abs/2007ApJ...668....1N} {668, 1}

\bibitem[\protect\citeauthoryear{{Naiman} et~al.,}{{Naiman}
  et~al.}{2018}]{Naiman2018}
{Naiman} J.~P.,  et~al., 2018, \mn@doi [\mnras] {10.1093/mnras/sty618}, \href
  {https://ui.adsabs.harvard.edu/abs/2018MNRAS.477.1206N} {477, 1206}

\bibitem[\protect\citeauthoryear{{Nandra} et~al.,}{{Nandra}
  et~al.}{2013}]{Nandra2013}
{Nandra} K.,  et~al., 2013, arXiv e-prints, \href
  {https://ui.adsabs.harvard.edu/abs/2013arXiv1306.2307N} {p. arXiv:1306.2307}

\bibitem[\protect\citeauthoryear{{Nelson}, {Rudd}, {Shaw}  \& {Nagai}}{{Nelson}
  et~al.}{2012}]{Nelson2012}
{Nelson} K.,  {Rudd} D.~H.,  {Shaw} L.,   {Nagai} D.,  2012, \mn@doi [\apj]
  {10.1088/0004-637X/751/2/121}, \href
  {https://ui.adsabs.harvard.edu/abs/2012ApJ...751..121N} {751, 121}

\bibitem[\protect\citeauthoryear{{Nelson}, {Lau}, {Nagai}, {Rudd}  \&
  {Yu}}{{Nelson} et~al.}{2014}]{Nelson2014a}
{Nelson} K.,  {Lau} E.~T.,  {Nagai} D.,  {Rudd} D.~H.,   {Yu} L.,  2014,
  \mn@doi [\apj] {10.1088/0004-637X/782/2/107}, \href
  {https://ui.adsabs.harvard.edu/abs/2014ApJ...782..107N} {782, 107}

\bibitem[\protect\citeauthoryear{Nelson et~al.,}{Nelson
  et~al.}{2015}]{Nelson2015}
Nelson D.,  et~al., 2015, \mn@doi [Astronomy and Computing]
  {https://doi.org/10.1016/j.ascom.2015.09.003}, 13, 12

\bibitem[\protect\citeauthoryear{Nelson et~al.,}{Nelson
  et~al.}{2017}]{Nelson2018}
Nelson D.,  et~al., 2017, \mn@doi [Monthly Notices of the Royal Astronomical
  Society] {10.1093/mnras/stx3040}, 475, 624

\bibitem[\protect\citeauthoryear{{Nelson} et~al.,}{{Nelson}
  et~al.}{2019a}]{Nelson2019b}
{Nelson} D.,  et~al., 2019a, \mn@doi [Comput. Astrophy. Cosmol.]
  {10.1186/s40668-019-0028-x}, 6, 2

\bibitem[\protect\citeauthoryear{{Nelson} et~al.,}{{Nelson}
  et~al.}{2019b}]{Nelson2019a}
{Nelson} D.,  et~al., 2019b, \mn@doi [\mnras] {10.1093/mnras/stz2306}, \href
  {https://ui.adsabs.harvard.edu/abs/2019MNRAS.490.3234N} {490, 3234}

\bibitem[\protect\citeauthoryear{{Neto} et~al.,}{{Neto}
  et~al.}{2007}]{Neto2007}
{Neto} A.~F.,  et~al., 2007, \mn@doi [\mnras]
  {10.1111/j.1365-2966.2007.12381.x}, \href
  {https://ui.adsabs.harvard.edu/abs/2007MNRAS.381.1450N} {381, 1450}

\bibitem[\protect\citeauthoryear{{Pakmor} \& {Springel}}{{Pakmor} \&
  {Springel}}{2013}]{Pakmor2013}
{Pakmor} R.,  {Springel} V.,  2013, \mn@doi [\mnras] {10.1093/mnras/stt428},
  \href {https://ui.adsabs.harvard.edu/abs/2013MNRAS.432..176P} {432, 176}

\bibitem[\protect\citeauthoryear{{Pakmor}, {Springel}, {Bauer}, {Mocz},
  {Munoz}, {Ohlmann}, {Schaal}  \& {Zhu}}{{Pakmor} et~al.}{2016}]{Pakmor2016}
{Pakmor} R.,  {Springel} V.,  {Bauer} A.,  {Mocz} P.,  {Munoz} D.~J.,
  {Ohlmann} S.~T.,  {Schaal} K.,   {Zhu} C.,  2016, \mn@doi [\mnras]
  {10.1093/mnras/stv2380}, \href
  {https://ui.adsabs.harvard.edu/abs/2016MNRAS.455.1134P} {455, 1134}

\bibitem[\protect\citeauthoryear{{Pfrommer}, {Pakmor}, {Schaal}, {Simpson}  \&
  {Springel}}{{Pfrommer} et~al.}{2017}]{Pfrommer2017}
{Pfrommer} C.,  {Pakmor} R.,  {Schaal} K.,  {Simpson} C.~M.,   {Springel} V.,
  2017, \mn@doi [\mnras] {10.1093/mnras/stw2941}, \href
  {https://ui.adsabs.harvard.edu/abs/2017MNRAS.465.4500P} {465, 4500}

\bibitem[\protect\citeauthoryear{{Pike}, {Kay}, {Newton}, {Thomas}  \&
  {Jenkins}}{{Pike} et~al.}{2014}]{Pike2014}
{Pike} S.~R.,  {Kay} S.~T.,  {Newton} R. D.~A.,  {Thomas} P.~A.,   {Jenkins}
  A.,  2014, \mn@doi [\mnras] {10.1093/mnras/stu1788}, \href
  {https://ui.adsabs.harvard.edu/abs/2014MNRAS.445.1774P} {445, 1774}

\bibitem[\protect\citeauthoryear{{Pillepich}, {Porciani}  \&
  {Reiprich}}{{Pillepich} et~al.}{2012}]{Pillepich2012}
{Pillepich} A.,  {Porciani} C.,   {Reiprich} T.~H.,  2012, \mn@doi [\mnras]
  {10.1111/j.1365-2966.2012.20443.x}, \href
  {https://ui.adsabs.harvard.edu/abs/2012MNRAS.422...44P} {422, 44}

\bibitem[\protect\citeauthoryear{{Pillepich} et~al.,}{{Pillepich}
  et~al.}{2018a}]{Pillepich2018b}
{Pillepich} A.,  et~al., 2018a, \mn@doi [\mnras] {10.1093/mnras/stx2656}, \href
  {https://ui.adsabs.harvard.edu/abs/2018MNRAS.473.4077P} {473, 4077}

\bibitem[\protect\citeauthoryear{{Pillepich} et~al.,}{{Pillepich}
  et~al.}{2018b}]{Pillepich2018a}
{Pillepich} A.,  et~al., 2018b, \mn@doi [\mnras] {10.1093/mnras/stx3112}, \href
  {https://ui.adsabs.harvard.edu/abs/2018MNRAS.475..648P} {475, 648}

\bibitem[\protect\citeauthoryear{{Pillepich}, {Reiprich}, {Porciani}, {Borm}
  \& {Merloni}}{{Pillepich} et~al.}{2018c}]{Pillepich2018}
{Pillepich} A.,  {Reiprich} T.~H.,  {Porciani} C.,  {Borm} K.,   {Merloni} A.,
  2018c, \mn@doi [\mnras] {10.1093/mnras/sty2240}, \href
  {https://ui.adsabs.harvard.edu/abs/2018MNRAS.481..613P} {481, 613}

\bibitem[\protect\citeauthoryear{{Pillepich} et~al.,}{{Pillepich}
  et~al.}{2019}]{Pillepich2019}
{Pillepich} A.,  et~al., 2019, \mn@doi [\mnras] {10.1093/mnras/stz2338}, \href
  {https://ui.adsabs.harvard.edu/abs/2019MNRAS.490.3196P} {490, 3196}

\bibitem[\protect\citeauthoryear{{Planck Collaboration} et~al.,}{{Planck
  Collaboration} et~al.}{2011a}]{Planck2011}
{Planck Collaboration} et~al., 2011a, \mn@doi [\aap]
  {10.1051/0004-6361/201116457}, \href
  {https://ui.adsabs.harvard.edu/abs/2011A&A...536A..10P} {536, A10}

\bibitem[\protect\citeauthoryear{{Planck Collaboration} et~al.,}{{Planck
  Collaboration} et~al.}{2011b}]{PlanckCollaboration2011}
{Planck Collaboration} et~al., 2011b, \mn@doi [\aap]
  {10.1051/0004-6361/201116458}, \href
  {https://ui.adsabs.harvard.edu/abs/2011A&A...536A..11P} {536, A11}

\bibitem[\protect\citeauthoryear{{Planck Collaboration} et~al.,}{{Planck
  Collaboration} et~al.}{2014}]{PlanckCollaboration2014b}
{Planck Collaboration} et~al., 2014, \mn@doi [\aap]
  {10.1051/0004-6361/201321521}, \href
  {https://ui.adsabs.harvard.edu/abs/2014A&A...571A..20P} {571, A20}

\bibitem[\protect\citeauthoryear{{Planck Collaboration} et~al.,}{{Planck
  Collaboration} et~al.}{2016}]{PlanckCollaboration2016b}
{Planck Collaboration} et~al., 2016, \mn@doi [\aap]
  {10.1051/0004-6361/201525833}, \href
  {https://ui.adsabs.harvard.edu/abs/2016A&A...594A..24P} {594, A24}

\bibitem[\protect\citeauthoryear{{Planelles}, {Borgani}, {Dolag}, {Ettori},
  {Fabjan}, {Murante}  \& {Tornatore}}{{Planelles}
  et~al.}{2013}]{Planelles2013}
{Planelles} S.,  {Borgani} S.,  {Dolag} K.,  {Ettori} S.,  {Fabjan} D.,
  {Murante} G.,   {Tornatore} L.,  2013, \mn@doi [\mnras]
  {10.1093/mnras/stt265}, \href
  {https://ui.adsabs.harvard.edu/abs/2013MNRAS.431.1487P} {431, 1487}

\bibitem[\protect\citeauthoryear{{Planelles}, {Borgani}, {Fabjan}, {Killedar},
  {Murante}, {Granato}, {Ragone-Figueroa}  \& {Dolag}}{{Planelles}
  et~al.}{2014}]{Planelles2014}
{Planelles} S.,  {Borgani} S.,  {Fabjan} D.,  {Killedar} M.,  {Murante} G.,
  {Granato} G.~L.,  {Ragone-Figueroa} C.,   {Dolag} K.,  2014, \mn@doi [\mnras]
  {10.1093/mnras/stt2141}, \href
  {https://ui.adsabs.harvard.edu/abs/2014MNRAS.438..195P} {438, 195}

\bibitem[\protect\citeauthoryear{{Pratt}, {Croston}, {Arnaud}  \&
  {B{\"o}hringer}}{{Pratt} et~al.}{2009}]{Pratt2009}
{Pratt} G.~W.,  {Croston} J.~H.,  {Arnaud} M.,   {B{\"o}hringer} H.,  2009,
  \mn@doi [\aap] {10.1051/0004-6361/200810994}, \href
  {https://ui.adsabs.harvard.edu/abs/2009A&A...498..361P} {498, 361}

\bibitem[\protect\citeauthoryear{{Pratt}, {Arnaud}, {Biviano}, {Eckert},
  {Ettori}, {Nagai}, {Okabe}  \& {Reiprich}}{{Pratt} et~al.}{2019}]{Pratt2019}
{Pratt} G.~W.,  {Arnaud} M.,  {Biviano} A.,  {Eckert} D.,  {Ettori} S.,
  {Nagai} D.,  {Okabe} N.,   {Reiprich} T.~H.,  2019, \mn@doi [\ssr]
  {10.1007/s11214-019-0591-0}, \href
  {https://ui.adsabs.harvard.edu/abs/2019SSRv..215...25P} {215, 25}

\bibitem[\protect\citeauthoryear{{Puchwein}, {Sijacki}  \&
  {Springel}}{{Puchwein} et~al.}{2008}]{Puchwein2008}
{Puchwein} E.,  {Sijacki} D.,   {Springel} V.,  2008, \mn@doi [\apjl]
  {10.1086/593352}, \href
  {https://ui.adsabs.harvard.edu/abs/2008ApJ...687L..53P} {687, L53}

\bibitem[\protect\citeauthoryear{{Raghunathan} et~al.,}{{Raghunathan}
  et~al.}{2022}]{Raghunathan2022}
{Raghunathan} S.,  et~al., 2022, \mn@doi [\apj] {10.3847/1538-4357/ac4712},
  \href {https://ui.adsabs.harvard.edu/abs/2022ApJ...926..172R} {926, 172}

\bibitem[\protect\citeauthoryear{{Rasia}, {Mazzotta}, {Borgani}, {Moscardini},
  {Dolag}, {Tormen}, {Diaferio}  \& {Murante}}{{Rasia}
  et~al.}{2005}]{Rasia2005}
{Rasia} E.,  {Mazzotta} P.,  {Borgani} S.,  {Moscardini} L.,  {Dolag} K.,
  {Tormen} G.,  {Diaferio} A.,   {Murante} G.,  2005, \mn@doi [\apjl]
  {10.1086/427554}, \href
  {https://ui.adsabs.harvard.edu/abs/2005ApJ...618L...1R} {618, L1}

\bibitem[\protect\citeauthoryear{{Rasia} et~al.,}{{Rasia}
  et~al.}{2006}]{Rasia2006}
{Rasia} E.,  et~al., 2006, \mn@doi [\mnras] {10.1111/j.1365-2966.2006.10466.x},
  \href {https://ui.adsabs.harvard.edu/abs/2006MNRAS.369.2013R} {369, 2013}

\bibitem[\protect\citeauthoryear{{Rasia}, {Mazzotta}, {Bourdin}, {Borgani},
  {Tornatore}, {Ettori}, {Dolag}  \& {Moscardini}}{{Rasia}
  et~al.}{2008}]{Rasia2008}
{Rasia} E.,  {Mazzotta} P.,  {Bourdin} H.,  {Borgani} S.,  {Tornatore} L.,
  {Ettori} S.,  {Dolag} K.,   {Moscardini} L.,  2008, \mn@doi [\apj]
  {10.1086/524345}, \href
  {https://ui.adsabs.harvard.edu/abs/2008ApJ...674..728R} {674, 728}

\bibitem[\protect\citeauthoryear{{Rasia} et~al.,}{{Rasia}
  et~al.}{2014}]{Rasia2014}
{Rasia} E.,  et~al., 2014, \mn@doi [\apj] {10.1088/0004-637X/791/2/96}, \href
  {https://ui.adsabs.harvard.edu/abs/2014ApJ...791...96R} {791, 96}

\bibitem[\protect\citeauthoryear{{Reichert}, {B{\"o}hringer}, {Fassbender}  \&
  {M{\"u}hlegger}}{{Reichert} et~al.}{2011}]{Reichert2011}
{Reichert} A.,  {B{\"o}hringer} H.,  {Fassbender} R.,   {M{\"u}hlegger} M.,
  2011, \mn@doi [\aap] {10.1051/0004-6361/201116861}, \href
  {https://ui.adsabs.harvard.edu/abs/2011A&A...535A...4R} {535, A4}

\bibitem[\protect\citeauthoryear{{Sanderson}, {Ponman}, {Finoguenov},
  {Lloyd-Davies}  \& {Markevitch}}{{Sanderson} et~al.}{2003}]{Sanderson2003}
{Sanderson} A.~J.~R.,  {Ponman} T.~J.,  {Finoguenov} A.,  {Lloyd-Davies} E.~J.,
    {Markevitch} M.,  2003, \mn@doi [\mnras]
  {10.1046/j.1365-8711.2003.06401.x}, \href
  {https://ui.adsabs.harvard.edu/abs/2003MNRAS.340..989S} {340, 989}

\bibitem[\protect\citeauthoryear{{Sarazin}}{{Sarazin}}{1986}]{Sarazin1986}
{Sarazin} C.~L.,  1986, \mn@doi [Reviews of Modern Physics]
  {10.1103/RevModPhys.58.1}, \href
  {https://ui.adsabs.harvard.edu/abs/1986RvMP...58....1S} {58, 1}

\bibitem[\protect\citeauthoryear{{Sehgal} et~al.,}{{Sehgal}
  et~al.}{2019}]{CMB-HD}
{Sehgal} N.,  et~al., 2019, in Bulletin of the American Astronomical Society.
  p.~6 (\mn@eprint {arXiv} {1906.10134})

\bibitem[\protect\citeauthoryear{{Short}, {Thomas}, {Young}, {Pearce},
  {Jenkins}  \& {Muanwong}}{{Short} et~al.}{2010}]{Short2010}
{Short} C.~J.,  {Thomas} P.~A.,  {Young} O.~E.,  {Pearce} F.~R.,  {Jenkins} A.,
    {Muanwong} O.,  2010, \mn@doi [\mnras] {10.1111/j.1365-2966.2010.17267.x},
  \href {https://ui.adsabs.harvard.edu/abs/2010MNRAS.408.2213S} {408, 2213}

\bibitem[\protect\citeauthoryear{{Smith}, {Brickhouse}, {Liedahl}  \&
  {Raymond}}{{Smith} et~al.}{2001}]{Smith2001}
{Smith} R.~K.,  {Brickhouse} N.~S.,  {Liedahl} D.~A.,   {Raymond} J.~C.,  2001,
  \mn@doi [\apjl] {10.1086/322992}, \href
  {https://ui.adsabs.harvard.edu/abs/2001ApJ...556L..91S} {556, L91}

\bibitem[\protect\citeauthoryear{{Springel}}{{Springel}}{2010}]{Springel2010}
{Springel} V.,  2010, \mn@doi [\mnras] {10.1111/j.1365-2966.2009.15715.x},
  \href {https://ui.adsabs.harvard.edu/abs/2010MNRAS.401..791S} {401, 791}

\bibitem[\protect\citeauthoryear{{Springel}, {Yoshida}  \& {White}}{{Springel}
  et~al.}{2001}]{Springel2001}
{Springel} V.,  {Yoshida} N.,   {White} S. D.~M.,  2001, \mn@doi [\na]
  {10.1016/S1384-1076(01)00042-2}, \href
  {https://ui.adsabs.harvard.edu/abs/2001NewA....6...79S} {6, 79}

\bibitem[\protect\citeauthoryear{{Springel} et~al.,}{{Springel}
  et~al.}{2018}]{Springel2018}
{Springel} V.,  et~al., 2018, \mn@doi [\mnras] {10.1093/mnras/stx3304}, \href
  {https://ui.adsabs.harvard.edu/abs/2018MNRAS.475..676S} {475, 676}

\bibitem[\protect\citeauthoryear{{Stanek}, {Rasia}, {Evrard}, {Pearce}  \&
  {Gazzola}}{{Stanek} et~al.}{2010}]{Stanek2010}
{Stanek} R.,  {Rasia} E.,  {Evrard} A.~E.,  {Pearce} F.,   {Gazzola} L.,  2010,
  \mn@doi [\apj] {10.1088/0004-637X/715/2/1508}, \href
  {https://ui.adsabs.harvard.edu/abs/2010ApJ...715.1508S} {715, 1508}

\bibitem[\protect\citeauthoryear{{Sun}}{{Sun}}{2012}]{Sun2012}
{Sun} M.,  2012, \mn@doi [New Journal of Physics]
  {10.1088/1367-2630/14/4/045004}, \href
  {https://ui.adsabs.harvard.edu/abs/2012NJPh...14d5004S} {14, 045004}

\bibitem[\protect\citeauthoryear{{Sun}, {Voit}, {Donahue}, {Jones}, {Forman}
  \& {Vikhlinin}}{{Sun} et~al.}{2009}]{Sun2009}
{Sun} M.,  {Voit} G.~M.,  {Donahue} M.,  {Jones} C.,  {Forman} W.,
  {Vikhlinin} A.,  2009, \mn@doi [\apj] {10.1088/0004-637X/693/2/1142}, \href
  {https://ui.adsabs.harvard.edu/abs/2009ApJ...693.1142S} {693, 1142}

\bibitem[\protect\citeauthoryear{{Truong} et~al.,}{{Truong}
  et~al.}{2018}]{Truong2018}
{Truong} N.,  et~al., 2018, \mn@doi [\mnras] {10.1093/mnras/stx2927}, \href
  {https://ui.adsabs.harvard.edu/abs/2018MNRAS.474.4089T} {474, 4089}

\bibitem[\protect\citeauthoryear{{Vazza}, {Eckert}, {Simionescu}, {Br{\"u}ggen}
   \& {Ettori}}{{Vazza} et~al.}{2013}]{Vazza2013}
{Vazza} F.,  {Eckert} D.,  {Simionescu} A.,  {Br{\"u}ggen} M.,   {Ettori} S.,
  2013, \mn@doi [\mnras] {10.1093/mnras/sts375}, \href
  {https://ui.adsabs.harvard.edu/abs/2013MNRAS.429..799V} {429, 799}

\bibitem[\protect\citeauthoryear{{Vikhlinin}, {Kravtsov}, {Forman}, {Jones},
  {Markevitch}, {Murray}  \& {Van Speybroeck}}{{Vikhlinin}
  et~al.}{2006}]{Vikhlinin2006}
{Vikhlinin} A.,  {Kravtsov} A.,  {Forman} W.,  {Jones} C.,  {Markevitch} M.,
  {Murray} S.~S.,   {Van Speybroeck} L.,  2006, \mn@doi [\apj]
  {10.1086/500288}, \href
  {https://ui.adsabs.harvard.edu/abs/2006ApJ...640..691V} {640, 691}

\bibitem[\protect\citeauthoryear{{Vikhlinin} et~al.,}{{Vikhlinin}
  et~al.}{2009a}]{Vikhlinin2009a}
{Vikhlinin} A.,  et~al., 2009a, \mn@doi [\apj] {10.1088/0004-637X/692/2/1033},
  \href {https://ui.adsabs.harvard.edu/abs/2009ApJ...692.1033V} {692, 1033}

\bibitem[\protect\citeauthoryear{{Vikhlinin} et~al.,}{{Vikhlinin}
  et~al.}{2009b}]{Vikhlinin2009b}
{Vikhlinin} A.,  et~al., 2009b, \mn@doi [\apj] {10.1088/0004-637X/692/2/1060},
  \href {https://ui.adsabs.harvard.edu/abs/2009ApJ...692.1060V} {692, 1060}

\bibitem[\protect\citeauthoryear{{Virtanen} et~al.,}{{Virtanen}
  et~al.}{2020}]{Virtanen2020}
{Virtanen} P.,  et~al., 2020, \mn@doi [Nature Methods]
  {10.1038/s41592-019-0686-2}, \href
  {https://ui.adsabs.harvard.edu/abs/2020NatMe..17..261V} {17, 261}

\bibitem[\protect\citeauthoryear{{Vogelsberger} et~al.,}{{Vogelsberger}
  et~al.}{2014a}]{Vogelsberger2014b}
{Vogelsberger} M.,  et~al., 2014a, \mn@doi [\mnras] {10.1093/mnras/stu1536},
  \href {https://ui.adsabs.harvard.edu/abs/2014MNRAS.444.1518V} {444, 1518}

\bibitem[\protect\citeauthoryear{{Vogelsberger} et~al.,}{{Vogelsberger}
  et~al.}{2014b}]{Vogelsberger2014a}
{Vogelsberger} M.,  et~al., 2014b, \mn@doi [\nat] {10.1038/nature13316}, \href
  {https://ui.adsabs.harvard.edu/abs/2014Natur.509..177V} {509, 177}

\bibitem[\protect\citeauthoryear{{Weinberger} et~al.,}{{Weinberger}
  et~al.}{2017a}]{Weinberger2017b}
{Weinberger} R.,  et~al., 2017a, \mn@doi [\mnras] {10.1093/mnras/stw2944},
  \href {https://ui.adsabs.harvard.edu/abs/2017MNRAS.465.3291W} {465, 3291}

\bibitem[\protect\citeauthoryear{{Weinberger}, {Ehlert}, {Pfrommer}, {Pakmor}
  \& {Springel}}{{Weinberger} et~al.}{2017b}]{Weinberger2017}
{Weinberger} R.,  {Ehlert} K.,  {Pfrommer} C.,  {Pakmor} R.,   {Springel} V.,
  2017b, \mn@doi [\mnras] {10.1093/mnras/stx1409}, \href
  {https://ui.adsabs.harvard.edu/abs/2017MNRAS.470.4530W} {470, 4530}

\bibitem[\protect\citeauthoryear{{Weinberger} et~al.,}{{Weinberger}
  et~al.}{2018}]{Weinberger2018}
{Weinberger} R.,  et~al., 2018, \mn@doi [\mnras] {10.1093/mnras/sty1733}, \href
  {https://ui.adsabs.harvard.edu/abs/2018MNRAS.479.4056W} {479, 4056}

\bibitem[\protect\citeauthoryear{{Yang}, {Cai}, {Cui}, {Dav{\'e}}, {Peacock}
  \& {Sorini}}{{Yang} et~al.}{2022}]{Yang2022}
{Yang} T.,  {Cai} Y.-C.,  {Cui} W.,  {Dav{\'e}} R.,  {Peacock} J.~A.,
  {Sorini} D.,  2022, arXiv e-prints, \href
  {https://ui.adsabs.harvard.edu/abs/2022arXiv220211430Y} {p. arXiv:2202.11430}

\bibitem[\protect\citeauthoryear{{Zhuravleva}, {Churazov}, {Kravtsov}, {Lau},
  {Nagai}  \& {Sunyaev}}{{Zhuravleva} et~al.}{2013}]{Zhuravleva2013}
{Zhuravleva} I.,  {Churazov} E.,  {Kravtsov} A.,  {Lau} E.~T.,  {Nagai} D.,
  {Sunyaev} R.,  2013, \mn@doi [\mnras] {10.1093/mnras/sts275}, \href
  {https://ui.adsabs.harvard.edu/abs/2013MNRAS.428.3274Z} {428, 3274}

\bibitem[\protect\citeauthoryear{{ZuHone}, {Biffi}, {Hallman}, {Randall},
  {Foster}  \& {Schmid}}{{ZuHone} et~al.}{2014}]{ZuHone2014}
{ZuHone} J.~A.,  {Biffi} V.,  {Hallman} E.~J.,  {Randall} S.~W.,  {Foster}
  A.~R.,   {Schmid} C.,  2014, arXiv e-prints, \href
  {https://ui.adsabs.harvard.edu/abs/2014arXiv1407.1783Z} {p. arXiv:1407.1783}

\makeatother
\end{thebibliography}

\appendix
\section{Core-Excised Scaling Relations}

\begin{figure*}
\centering
\includegraphics[width=0.9\textwidth]{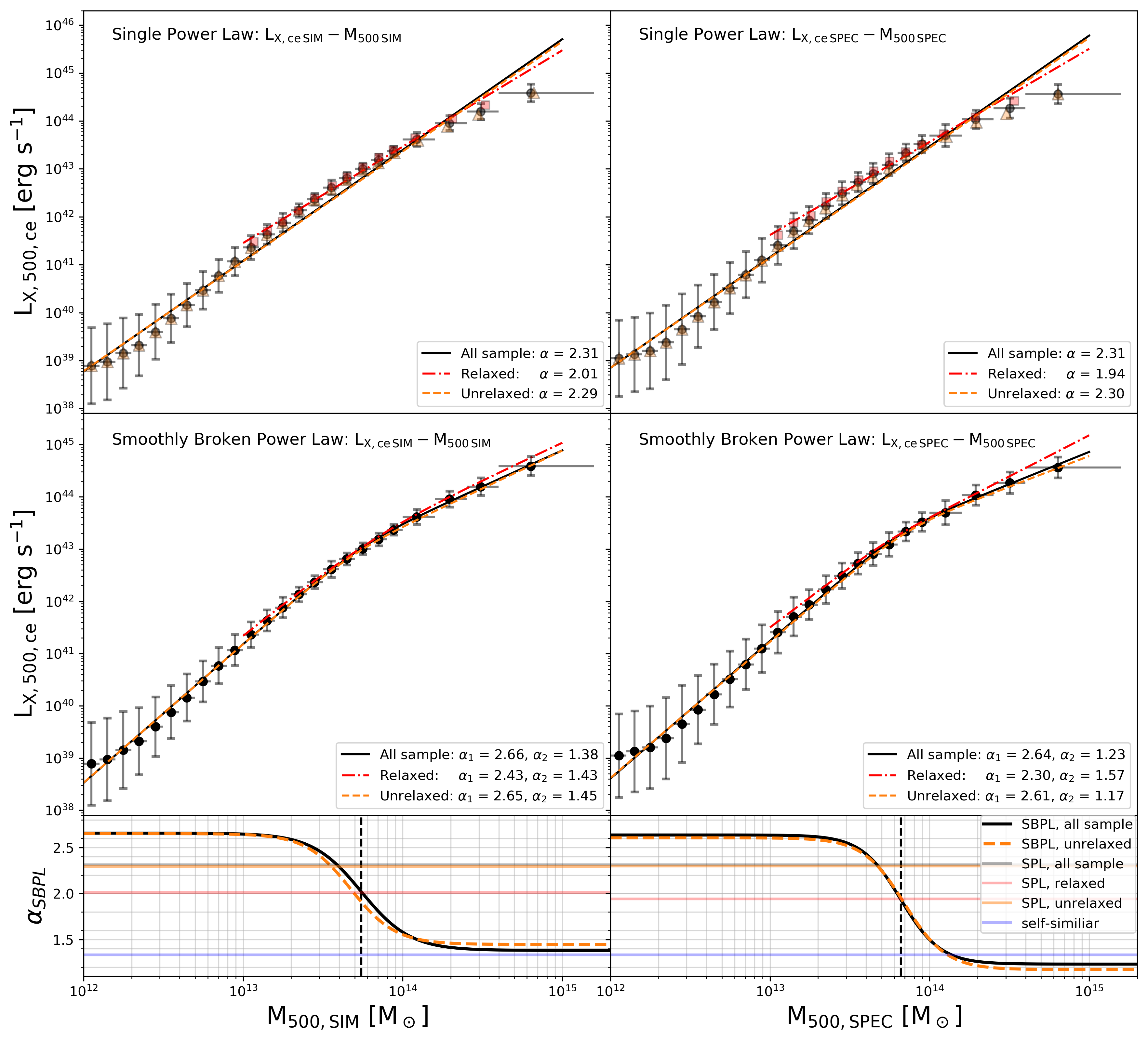}
\caption{Best-fits for the $L_{\rm X, \, ce}$ -- \mtot\, scaling relation at $z=0$ in TNG300. See Figure~\ref{fig:Mgas4panel} for the description of the panels and lines, noting that $\alpha = 4/3$ for the self-similar prediction of this relation. 
}
\label{fig:Lxce4panel}
\end{figure*}

\begin{figure*}
\centering
\includegraphics[width=0.9\textwidth]{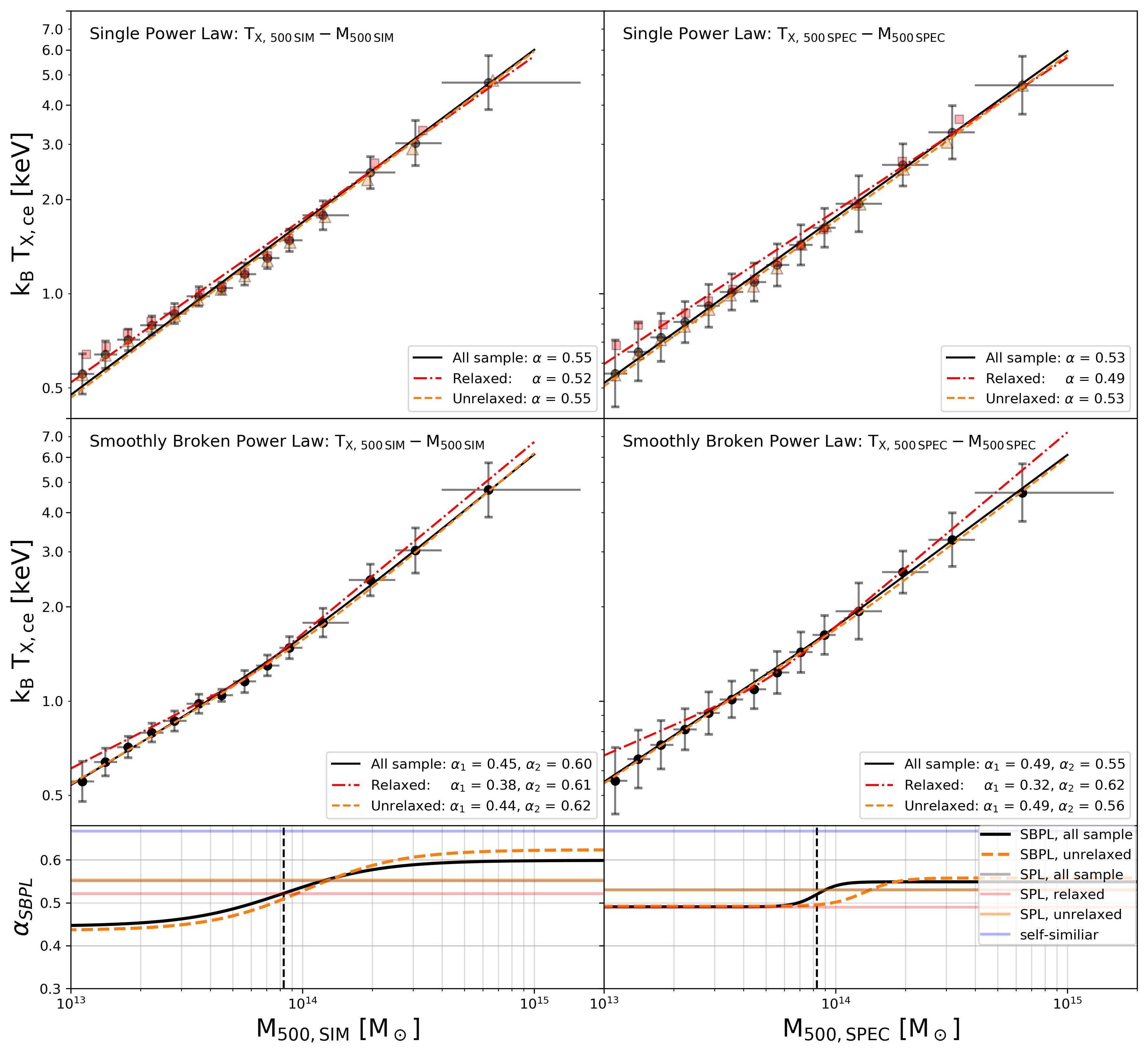}
\caption{Best-fits for the $T_{\rm X, \, ce}$ -- \mtot\, scaling relation at $z=0$ in TNG300. See Figure~\ref{fig:Mgas4panel} for the description of the panels and lines, noting that $\alpha = 2/3$ for the self-similar prediction of this relation. }
\label{fig:Txce4panel}
\end{figure*}

Figure~\ref{fig:Lxce4panel} presents results from fitting the SPL and SBPL models to the core-excised X-ray luminosity as a function of halo mass. Similarly, the best-fitting models for core-excised spectroscopic temperatures as a function of halo mass are shown in Figure~\ref{fig:Txce4panel}. Detailed discussions of the best-fitting models for these two core-excised observables can be found in Section~\ref{subsec:xraymass} and Section~\ref{subsec:scalingstx}, respectively.

\section{Best-fitting parameters}

Throughout this work we have explored the best-fit models for X-ray and SZ scaling relations, with quantities measured inside the true \rsim\, aperture from the simulation, as well as the spectroscopic aperture \rspec.
Tables (\ref{tab:tabmg} -- \ref{tab:tabysz}) include summaries of the best-fitting parameters for a simple power law (SPL, eqn.~\ref{eqn:eqnSPL}) and our new model for a smoothly broken power law (SBPL, eqn.~\ref{eqn:eqnSBPL}). Table~\ref{tab:tablxall} also includes a more detailed comparison of the best-fit values for a broken power law with a pivot fixed at \mcrit $=10^{14}$\ms\, (BPL fixed, eqn.~\ref{eqn:eqnBPLfixed}) and a broken power law with free pivot (BPL free, eqn.~\ref{eqn:eqnBPLfree}). We include results from fits performed for the full sample, as well as just for relaxed and unrelaxed haloes.

\begin{table*}
\def\arraystretch{2}
\caption{\label{tab:tabmg} Summary of the best-fitting parameters for the \mgm\, scaling relation. 
The top half of the table summarizes results for both gas and total masses 
measured within \rsim\,, while the bottom half of the table presents results for masses measured within \rspec.
SPL rows present the best-fit slope ($\alpha$)  and normalization ($A$) parameters from fitting the simple power law model from equation~(\ref{eqn:eqnSPL}) to the geometric means in each mass bin. 
SBPL rows show the best-fit slopes ($\alpha_1$ for small mass scales, $\alpha_2$ for massive clusters), normalization ($A$), transition width ($\delta$), and pivot location ($X_{\rm pivot}$) from fitting the smoothly broken power law model from equation~(\ref{eqn:eqnSBPL}) to the geometric means in each mass bin. 
Errors for the best-fit parameters correspond to 16\textsuperscript{th} and 84\textsuperscript{th} percentiles, calculated by bootstrap resampling the data 10,000 times. The normalization point is set to $X_{\rm norm} = 10^{14}$\,\ms\, for all fits.}
\resizebox{\textwidth}{!}{%
\begin{tabular}{|l|l|l|l|l|l|l|l|l|}
\hline
\multicolumn{1}{|c|}{\multirow{2}{*}{Scaling}} & 
\multicolumn{1}{c|}{\multirow{2}{*}{Aperture}} & 
\multicolumn{1}{c|}{\multirow{2}{*}{Fit Type}} & 
\multicolumn{1}{c|}{\multirow{2}{*}{Sample}} & 
\multicolumn{2}{c|}{$\alpha$} & 
\multicolumn{1}{c|}{\multirow{2}{*}{$A$}} & 
\multicolumn{1}{c|}{\multirow{2}{*}{$\delta$}} & 
\multicolumn{1}{c|}{\multirow{2}{*}{$X_{{\rm pivot}}$ [ $\times 10^{14} \, {{\rm M}}_\odot$]}} \\ \cline{5-6}
\multicolumn{1}{|c|}{}                         & 
\multicolumn{1}{c|}{}                          & 
\multicolumn{1}{c|}{}                          & 
\multicolumn{1}{c|}{}                        & 
\multicolumn{1}{c|}{$\alpha_1$}    & 
\multicolumn{1}{c|}{$\alpha_2$}    & 
\multicolumn{1}{c|}{}                   & 
\multicolumn{1}{c|}{}                          & 
\multicolumn{1}{c|}{}                                     \\ \hline

\multirow{12}{*}{${M}_{{\rm 500 \, gas}}$  - ${ M}_{{\rm 500 \, crit}}$}  & \multirow{6}{*}{$R_{\rm 500,\, SIM}$} & \multirow{3}{*}{SPL}          & All Sample    & \multicolumn{2}{c|}{$1.534^{+0.004}_{-0.004}$}                         & $13.062^{+0.004}_{-0.004}$                    & \multicolumn{1}{c|}{---}      & \multicolumn{1}{c|}{---}          \\ \cline{4-9} 
                                                    &                                          &                                       & Relaxed       & \multicolumn{2}{c|}{$1.358^{+0.031}_{-0.009}$}       & $13.077^{+0.022}_{-0.003}$        & \multicolumn{1}{c|}{---}      & \multicolumn{1}{c|}{---}          \\ \cline{4-9} 
                                               &                                                &                                      & Unrelaxed     & \multicolumn{2}{c|}{$1.528^{+0.032}_{-0.004}$} & $13.053^{+0.039}_{-0.005}$  & \multicolumn{1}{c|}{---}      & \multicolumn{1}{c|}{---} \\ \cline{3-9} 
                        &   & \multirow{3}{*}{SBPL}   & All Sample    & $1.671^{+0.002}_{-0.002}$                    & $1.086^{+0.014}_{-0.013}$                    & $13.093^{+0.003}_{-0.003}$                    & $0.19^{+0.02}_{-0.02}$                      & \multicolumn{1}{c|}{$0.57^{+0.02}_{-0.02}$}      \\ \cline{4-9} 
                        &   &                               & Relaxed       & $1.559^{+0.046}_{-0.034}$        & $1.100^{+0.021}_{-0.027}$        & $13.106^{+0.003}_{-0.003}$        & $0.19^{+0.07}_{-0.05}$          & \multicolumn{1}{c|}{$0.62^{+0.06}_{-0.05}$}                                                                                                                      \\ \cline{4-9} 
                        &   &                               & Unrelaxed     & $1.665^{+0.003}_{-0.003}$  & $1.078^{+0.033}_{-0.014}$  & $13.084^{+0.004}_{-0.006}$  & $0.25^{+0.03}_{-0.05}$    & \multicolumn{1}{c|}{$0.60^{+0.01}_{-0.06}$}                                                                                                                      \\ \cline{2-9} 
                        
                                                & \multirow{6}{*}{$R_{\rm 500,\, SPEC}$} & \multirow{3}{*}{SPL}     & All Sample    & \multicolumn{2}{c|}{$1.493^{+0.030}_{-0.008}$}        & $13.044^{+0.036}_{-0.010}$                    & \multicolumn{1}{c|}{---}      & \multicolumn{1}{c|}{---}          \\ \cline{4-9} 
                                                        &                                          &                                       & Relaxed       & \multicolumn{2}{c|}{$1.347^{+0.044}_{-0.016}$}       & $13.068^{+0.027}_{-0.004}$        & \multicolumn{1}{c|}{---}      & \multicolumn{1}{c|}{---}          \\ \cline{4-9} 
                                                   &                                                &                                      & Unrelaxed     & \multicolumn{2}{c|}{$1.491^{+0.027}_{-0.010}$} & $13.038^{+0.033}_{-0.012}$  & \multicolumn{1}{c|}{---}      & \multicolumn{1}{c|}{---}          \\ \cline{3-9} 
                                                                    
                        &   & \multirow{3}{*}{SBPL}   & All Sample    & $1.597^{+0.004}_{-0.004}$                    & $1.051^{+0.037}_{-0.042}$                    & $13.097^{+0.009}_{-0.008}$                    & $0.08^{+0.04}_{-0.03}$                      & \multicolumn{1}{c|}{$0.72^{+0.05}_{-0.04}$}      \\ \cline{4-9} 
                        &   &                               & Relaxed       & $1.605^{+0.138}_{-0.065}$        & $1.091^{+0.026}_{-0.040}$        & $13.091^{+0.006}_{-0.006}$        & $0.20^{+0.14}_{-0.07}$          & \multicolumn{1}{c|}{$0.51^{+0.07}_{-0.09}$}                                                                                                                      \\ \cline{4-9} 
                        &   &                               & Unrelaxed     & $1.584^{+0.005}_{-0.005}$  & $1.028^{+0.051}_{-0.059}$  & $13.101^{+0.013}_{-0.014}$  & $0.05^{+0.06}_{-0.00}$    & \multicolumn{1}{c|}{$0.80^{+0.07}_{-0.07}$}                                                                                                                      \\ \cline{1-9} 
                        
\end{tabular}
}%
\end{table*}

\begin{table*}
\def\arraystretch{2}
\caption{\label{tab:tablxall} Summary of the best-fit values of the single power law (SPL, eqn.~\ref{eqn:eqnSPL}), the broken power law with fixed pivot (BPL fixed, eqn.~\ref{eqn:eqnBPLfixed}) and free pivot (BPL free, eqn.~\ref{eqn:eqnBPLfree}), as well as the smoothly-broken power law (SBPL, eqn.~\ref{eqn:eqnSBPL}) models applied to the \lxm\, scaling relation at $z=0$ in TNG300.  
The normalization factor is set to $X_{\rm norm} = 10^{14}$\,\ms\, for all models. See the caption of Table~\ref{tab:tabmg} for more details.} 
\resizebox{\textwidth}{!}{%
\begin{tabular}{|l|l|l|l|l|l|l|l|l|}
\hline
\multicolumn{1}{|c|}{\multirow{2}{*}{Scaling}} & 
\multicolumn{1}{c|}{\multirow{2}{*}{Aperture}} & 
\multicolumn{1}{c|}{\multirow{2}{*}{Fit Type}} & 
\multicolumn{1}{c|}{\multirow{2}{*}{Sample}} & 
\multicolumn{2}{c|}{$\alpha$} & 
\multicolumn{1}{c|}{\multirow{2}{*}{$A$}} & 
\multicolumn{1}{c|}{\multirow{2}{*}{$\delta$}} & 
\multicolumn{1}{c|}{\multirow{2}{*}{${X}_{{\rm pivot}}$ [ $\times 10^{14} \, {{\rm M}}_\odot$]}} \\ \cline{5-6}
\multicolumn{1}{|c|}{}                         & 
\multicolumn{1}{c|}{}                          & 
\multicolumn{1}{c|}{}                          & 
\multicolumn{1}{c|}{}                        & 
\multicolumn{1}{c|}{$\alpha_1$}    & 
\multicolumn{1}{c|}{$\alpha_2$}    & 
\multicolumn{1}{c|}{}                   & 
\multicolumn{1}{c|}{}                          & 
\multicolumn{1}{c|}{}                                     \\ \hline
\multirow{24}{*}{${L}_{{\rm X}}$  - ${ M}_{{\rm 500 \, crit}}$}  & \multirow{12}{*}{$ R_{\rm 500,\, SIM}$} & \multirow{3}{*}{SPL}    & All Sample    & \multicolumn{2}{c|}{$2.264^{+0.016}_{-0.013}$}                        & $43.463^{+0.019}_{-0.014}$                       & \multicolumn{1}{c|}{---}      & \multicolumn{1}{c|}{---}          \\ \cline{4-9} 
                                                    &                                          &                                       & Relaxed       & \multicolumn{2}{c|}{$2.069^{+0.033}_{-0.031}$}       & $43.547^{+0.018}_{-0.015}$        & \multicolumn{1}{c|}{---}      & \multicolumn{1}{c|}{---}          \\ \cline{4-9} 
                                               &                                                &                                      & Unrelaxed     & \multicolumn{2}{c|}{$2.237^{+0.054}_{-0.013}$} & $43.424^{+0.061}_{-0.015}$  & \multicolumn{1}{c|}{---}      & \multicolumn{1}{c|}{---}          \\ \cline{3-9} 
                                                                        
                        &   & \multirow{3}{*}{BPL fixed}   & All Sample    & $2.424^{+0.005}_{-0.005}$                    & $1.377^{+0.076}_{-0.068}$                    & $43.656^{+0.006}_{-0.006}$               & \multicolumn{1}{c|}{---}    & \multicolumn{1}{c|}{$1$}    \\ \cline{4-9} 
                        &   &                               & Relaxed       & $2.156^{+0.026}_{-0.025}$        & $1.804^{+0.113}_{-0.105}$        & $43.598^{+0.011}_{-0.011}$        & \multicolumn{1}{c|}{---}    & \multicolumn{1}{c|}{$1$}    \\ \cline{4-9} 
                        &   &                               & Unrelaxed     & $2.127^{+0.016}_{-0.016}$  & $1.505^{+0.073}_{-0.064}$  & $43.522^{+0.011}_{-0.011}$  & \multicolumn{1}{c|}{---}    & \multicolumn{1}{c|}{$1$}    \\ \cline{3-9} 
                                                                       
                        &   & \multirow{3}{*}{BPL free}   & All Sample    & $2.498^{+0.008}_{-0.010}$                    & $1.679^{+0.084}_{-0.076}$                    & $43.495^{+0.012}_{-0.008}$                    & \multicolumn{1}{c|}{---}     & \multicolumn{1}{c|}{$0.47^{+0.03}_{-0.06}$}                               \\ \cline{4-9} 
                        &   &                                   & Relaxed       & $2.233^{+0.059}_{-0.047}$        & $1.905^{+0.079}_{-0.093}$        & $43.562^{+0.011}_{-0.010}$        & \multicolumn{1}{c|}{---}     & \multicolumn{1}{c|}{$0.51^{+0.14}_{-0.11}$}                  \\ \cline{4-9} 
                        &   &                                   & Unrelaxed     & $2.496^{+0.009}_{-0.013}$  & $1.679^{+0.067}_{-0.066}$  & $43.440^{+0.011}_{-0.011}$  & \multicolumn{1}{c|}{---}     & \multicolumn{1}{c|}{$0.40^{+0.06}_{-0.02}$}             \\ \cline{3-9} 
                                                                       
                        &   & \multirow{3}{*}{SBPL}   & All Sample    & $2.509^{+0.006}_{-0.006}$                    & $1.610^{+0.125}_{-0.163}$                    & $43.514^{+0.022}_{-0.016}$                    & $0.24^{+0.11}_{-0.09}$                      & \multicolumn{1}{c|}{$0.52^{+0.14}_{-0.09}$}      \\ \cline{4-9} 
                        &   &                               & Relaxed       & $2.259^{+0.079}_{-0.057}$        & $1.901^{+0.083}_{-0.259}$        & $43.565^{+0.012}_{-0.011} $       & $0.08^{+0.38}_{-0.08}$          & \multicolumn{1}{c|}{$0.50^{+0.50}_{-0.11}$}                                                                                                                      \\ \cline{4-9} 
                        &   &                               & Unrelaxed     & $2.503^{+0.006}_{-0.006}$  & $1.631^{+0.101}_{-0.090}$  & $43.453^{+0.017}_{-0.013}$  & $0.20^{+0.07}_{-0.09}$    & \multicolumn{1}{c|}{$0.44^{+0.06}_{-0.06}$}                                                                                                                      \\ \cline{2-9} 
                        
                                                            & \multirow{12}{*}{$ R_{\rm 500,\, SPEC}$} & \multirow{3}{*}{SPL}    & All Sample    & \multicolumn{2}{c|}{$2.266^{+0.072}_{-0.022}$}                     & $43.541^{+0.083}_{-0.026}$                    & \multicolumn{1}{c|}{---}      & \multicolumn{1}{c|}{---}          \\ \cline{4-9} 
                                                        &                                          &                                       & Relaxed       & \multicolumn{2}{c|}{$2.005^{+0.041}_{-0.038}$}       & $43.650^{+0.018}_{-0.018}$        & \multicolumn{1}{c|}{---}      & \multicolumn{1}{c|}{---}          \\ \cline{4-9} 
                                                   &                                                &                                      & Unrelaxed     & \multicolumn{2}{c|}{$2.244^{+0.058}_{-0.029}$} & $43.504^{+0.066}_{-0.035}$  & \multicolumn{1}{c|}{---}      & \multicolumn{1}{c|}{---}          \\ \cline{3-9} 
                                                                    
                        &   & \multirow{3}{*}{BPL fixed}   & All Sample    & $2.446^{+0.009}_{-0.009}$                    & $1.250^{+0.114}_{-0.116}$                    & $43.760^{+0.011}_{-0.011}$               & \multicolumn{1}{c|}{---}    & \multicolumn{1}{c|}{$1$}    \\ \cline{4-9} 
                        &   &                               & Relaxed       & $2.076^{+0.048}_{-0.048}$        & $1.771^{+0.132}_{-0.133}$        & $43.690^{+0.019}_{-0.019}$        & \multicolumn{1}{c|}{---}    & \multicolumn{1}{c|}{$1$}    \\ \cline{4-9} 
                        &   &                               & Unrelaxed     & $2.229^{+0.037}_{-0.037}$  & $1.311^{+0.144}_{-0.156}$  & $43.657^{+0.025}_{-0.025}$  & \multicolumn{1}{c|}{---}    & \multicolumn{1}{c|}{$1$}    \\ \cline{3-9} 
                                                                       
                        &   & \multirow{3}{*}{BPL free}   & All Sample    & $2.484^{+0.013}_{-0.014}$                    & $1.486^{+0.170}_{-0.168}$                    & $43.633^{+0.040}_{-0.024} $               & \multicolumn{1}{c|}{---}     & \multicolumn{1}{c|}{$0.64^{+0.13}_{-0.07}$}                               \\ \cline{4-9} 
                        &   &                               & Relaxed       & $2.163^{+0.115}_{-0.083}$        & $1.850^{+0.078}_{-0.094}$        & $43.659^{+0.014}_{-0.014} $       & \multicolumn{1}{c|}{---}     & \multicolumn{1}{c|}{$0.51^{+0.15}_{-0.13}$}                  \\ \cline{4-9} 
                        &   &                               & Unrelaxed     & $2.448^{+0.014}_{-0.014}$  & $1.403^{+0.159}_{-0.168}$  & $43.608^{+0.042}_{-0.038} $ & \multicolumn{1}{c|}{---}     & \multicolumn{1}{c|}{$0.73^{+0.09}_{-0.11}$}             \\ \cline{3-9} 
                                                                       
                        &   & \multirow{3}{*}{SBPL}   & All Sample    & $2.503^{+0.009}_{-0.009}$                    & $1.379^{+0.271}_{-0.272}$                    & $43.641^{+0.023}_{-0.028}$                    & $0.25^{+0.11}_{-0.11}$                      & \multicolumn{1}{c|}{$0.72^{+0.28}_{-0.17}$}      \\ \cline{4-9} 
                        &   &                               & Relaxed       & $2.440^{+0.001}_{-0.000}$        & $1.793^{+0.106}_{-0.255}$        & $43.662^{+0.013}_{-0.013}$        & $0.76^{+0.24}_{-0.48}$          & \multicolumn{1}{c|}{$0.30^{+0.22}_{-0.15}$}                                                                                                                      \\ \cline{4-9} 
                        &   &                               & Unrelaxed     & $2.472^{+0.011}_{-0.013}$  & $1.356^{+0.189}_{-0.239}$  & $43.596^{+0.030}_{-0.031}$  & $0.22^{+0.10}_{-0.08}$    & \multicolumn{1}{c|}{$0.72^{+0.28}_{-0.13}$}                                                                                                                      \\ \cline{1-9} 
                        
\end{tabular}
}%
\end{table*}

\newpage

\begin{table*}
\def\arraystretch{2}
\caption{\label{tab:tablxce} Summary of the best-fitting parameters for the \Lxce\, -- \mcrit\, scaling relation. This table follows the same format introduced in the caption of Table~\ref{tab:tabmg}.
}
\resizebox{\textwidth}{!}{%
\begin{tabular}{|l|l|l|l|l|l|l|l|l|l|}
\hline
\multicolumn{1}{|c|}{\multirow{2}{*}{Scaling}} & 
\multicolumn{1}{c|}{\multirow{2}{*}{Aperture}} & 
\multicolumn{1}{c|}{\multirow{2}{*}{Fit Type}} & 
\multicolumn{1}{c|}{\multirow{2}{*}{Sample}} & 
\multicolumn{2}{c|}{$\alpha$} & 
\multicolumn{1}{c|}{\multirow{2}{*}{$A$}} & 
\multicolumn{1}{c|}{\multirow{2}{*}{$\delta$}} & 
\multicolumn{1}{c|}{\multirow{2}{*}{${X}_{{\rm pivot}}$ [ $\times 10^{14} \, {{\rm M}}_\odot$]}}  \\ \cline{5-6}
\multicolumn{1}{|c|}{}                         & 
\multicolumn{1}{c|}{}                          & 
\multicolumn{1}{c|}{}                          & 
\multicolumn{1}{c|}{}                        & 
\multicolumn{1}{c|}{$\alpha_1$}    & 
\multicolumn{1}{c|}{$\alpha_2$}    & 
\multicolumn{1}{c|}{}                   & 
\multicolumn{1}{c|}{}                          & 
\multicolumn{1}{c|}{}                                     \\ \hline

\multirow{12}{*}{${ L}_{{\rm X , \, ce}}$  - ${M}_{{\rm 500 \, crit}}$}  & \multirow{6}{*}{$ R_{\rm 500,\, SIM}$} & \multirow{3}{*}{SPL}          & All Sample    & \multicolumn{2}{c|}{$2.313^{+0.013}_{-0.011}$}                         & $43.395^{+0.014}_{-0.011}$                    & \multicolumn{1}{c|}{---}      & \multicolumn{1}{c|}{---}          \\ \cline{4-9} 
                                                    &                                          &                                       & Relaxed       & \multicolumn{2}{c|}{$2.012^{+0.057}_{-0.020}$}       & $43.465^{+0.035}_{-0.009}$        & \multicolumn{1}{c|}{---}      & \multicolumn{1}{c|}{---}          \\ \cline{4-9} 
                                               &                                                &                                      & Unrelaxed     & \multicolumn{2}{c|}{$2.294^{+0.077}_{-0.011}$} & $43.366^{+0.089}_{-0.012}$  & \multicolumn{1}{c|}{---}      & \multicolumn{1}{c|}{---}          \\ \cline{3-9} 
                                                                        
                        &   & \multirow{3}{*}{SBPL}   & All Sample    & $2.657^{+0.006}_{-0.006}$                    & $1.381^{+0.084}_{-0.069}$                    & $43.474^{+0.011}_{-0.013} $                   & $0.31^{+0.04}_{-0.05}$                      & \multicolumn{1}{c|}{$0.55^{+0.05}_{-0.05}$}      \\ \cline{4-9} 
                        &   &                               & Relaxed       & $2.432^{+0.248}_{-0.113}$        & $1.434^{+0.262}_{-0.234}$        & $43.516^{+0.010}_{-0.012}$        & $0.45^{+0.23}_{-0.27}$          & \multicolumn{1}{c|}{$0.67^{+0.28}_{-0.18}$}                                                                                                                      \\ \cline{4-9} 
                        &   &                               & Unrelaxed     & $2.651^{+0.007}_{-0.006}$  & $1.447^{+0.092}_{-0.056}$  & $43.421^{+0.013}_{-0.014}$  & $0.28^{+0.04}_{-0.06}$    & \multicolumn{1}{c|}{$0.47^{+0.04}_{-0.05}$}                                                                                                                      \\ \cline{2-9} 
                        
                                                & \multirow{6}{*}{$ R_{\rm 500,\, SPEC}$} & \multirow{3}{*}{SPL}     & All Sample    & \multicolumn{2}{c|}{$2.313^{+0.090}_{-0.025}$}        & $43.470^{+0.105}_{-0.028}$                    & \multicolumn{1}{c|}{---}      & \multicolumn{1}{c|}{---}          \\ \cline{4-9} 
                                                        &                                          &                                       & Relaxed       & \multicolumn{2}{c|}{$1.943^{+0.068}_{-0.038}$}       & $43.563^{+0.040}_{-0.013}$        & \multicolumn{1}{c|}{---}      & \multicolumn{1}{c|}{---}          \\ \cline{4-9} 
                                                   &                                                &                                      & Unrelaxed     & \multicolumn{2}{c|}{$2.297^{+0.081}_{-0.029}$} & $43.441^{+0.093}_{-0.034}$  & \multicolumn{1}{c|}{---}      & \multicolumn{1}{c|}{---}          \\ \cline{3-9} 
                                                                    
                        &   & \multirow{3}{*}{SBPL}   & All Sample    & $2.637^{+0.009}_{-0.009}$                    & $1.233^{+0.204}_{-0.265}$                    & $43.590^{+0.024}_{-0.025}$                    & $0.25^{+0.09}_{-0.07}$                      & \multicolumn{1}{c|}{$0.66^{+0.17}_{-0.11}$}      \\ \cline{4-9} 
                        &   &                               & Relaxed       & $2.301^{+0.349}_{-0.140}$        & $1.571^{+0.101}_{-0.178}$        & $43.600^{+0.016}_{-0.016}$        & $0.20^{+0.27}_{-0.15}$          & \multicolumn{1}{c|}{$0.54^{+0.17}_{-0.14}$}                                                                                                                      \\ \cline{4-9} 
                        &   &                               & Unrelaxed     & $2.606^{+0.011}_{-0.012}$  & $1.173^{+0.177}_{-0.235}$  & $43.558^{+0.026}_{-0.028}$  & $0.26^{+0.07}_{-0.06}$    & \multicolumn{1}{c|}{$0.70^{+0.16}_{-0.10}$}                                                                                                                      \\ \cline{1-9} 
                        
\end{tabular}
}%
\end{table*}

\newpage

\begin{table*}
\def\arraystretch{2}
\caption{\label{tab:tabtx} Summary of the best-fitting parameters for the \tx\, -- \mcrit\, scaling relation. This table follows the same format introduced in the caption of Table~\ref{tab:tabmg}. 
}
\resizebox{\textwidth}{!}{%
\begin{tabular}{|l|l|l|l|l|l|l|l|l|}
\hline
\multicolumn{1}{|c|}{\multirow{2}{*}{Scaling}} & 
\multicolumn{1}{c|}{\multirow{2}{*}{Aperture}} & 
\multicolumn{1}{c|}{\multirow{2}{*}{Fit Type}} & 
\multicolumn{1}{c|}{\multirow{2}{*}{Sample}} & 
\multicolumn{2}{c|}{$\alpha$} & 
\multicolumn{1}{c|}{\multirow{2}{*}{$A$}} & 
\multicolumn{1}{c|}{\multirow{2}{*}{$\delta$}} & 
\multicolumn{1}{c|}{\multirow{2}{*}{${ X}_{{\rm pivot}}$ [ $\times 10^{14} \, {{\rm M}}_\odot$]}} \\ \cline{5-6}
\multicolumn{1}{|c|}{}                         & 
\multicolumn{1}{c|}{}                          & 
\multicolumn{1}{c|}{}                          & 
\multicolumn{1}{c|}{}                        & 
\multicolumn{1}{c|}{$\alpha_1$}    & 
\multicolumn{1}{c|}{$\alpha_2$}    & 
\multicolumn{1}{c|}{}                   & 
\multicolumn{1}{c|}{}                          & 
\multicolumn{1}{c|}{}                                     \\ \hline

\multirow{12}{*}{${ T}_{{\rm X}}$  - ${ M}_{{\rm 500 \, crit}}$}  & \multirow{6}{*}{$R_{\rm 500,\, SIM}$} & \multirow{3}{*}{SPL}          & All Sample    & \multicolumn{2}{c|}{$0.537^{+0.024}_{-0.037}$}                         & $0.248^{+0.016}_{-0.012}$                    & \multicolumn{1}{c|}{---}      & \multicolumn{1}{c|}{---}          \\ \cline{4-9} 
                                                    &                                          &                                       & Relaxed       & \multicolumn{2}{c|}{$0.549^{+0.012}_{-0.014}$}       & $0.260^{+0.004}_{-0.006}$        & \multicolumn{1}{c|}{---}      & \multicolumn{1}{c|}{---}          \\ \cline{4-9} 
                                               &                                                &                                      & Unrelaxed     & \multicolumn{2}{c|}{$0.541^{+0.027}_{-0.030}$} & $0.236^{+0.013}_{-0.013}$  & \multicolumn{1}{c|}{---}      & \multicolumn{1}{c|}{---}          \\ \cline{3-9} 
                                                                        
                        &   & \multirow{3}{*}{SBPL}   & All Sample    & $0.432^{+0.028}_{-0.018}$                    & $0.556^{+0.043}_{-0.042}$                    & $0.238^{+0.023}_{-0.017}$                    & $0.01^{+2.98}_{-0.01}$                      & \multicolumn{1}{c|}{$0.38^{+0.15}_{-0.25}$}      \\ \cline{4-9} 
                        &   &                               & Relaxed       & $0.341^{+0.011}_{-0.013}$        & $0.642^{+0.058}_{-0.019}$        & $0.234^{+0.003}_{-0.003}$        & $0.02^{+0.10}_{-0.02}$          & \multicolumn{1}{c|}{$0.51^{+0.04}_{-0.03}$}                                                                                                                      \\ \cline{4-9} 
                        &   &                               & Unrelaxed     & $0.418^{+0.041}_{-0.015}$  & $0.568^{+0.059}_{-0.044}$  & $0.218^{+0.023}_{-0.016}$  & $0.01^{+0.49}_{-0.01}$    & \multicolumn{1}{c|}{$0.44^{+0.21}_{-0.23}$}                                                                                                                      \\ \cline{2-9} 
                        
                                                & \multirow{6}{*}{$R_{\rm 500,\, SPEC}$} & \multirow{3}{*}{SPL}     & All Sample    & \multicolumn{2}{c|}{$0.500^{+0.040}_{-0.043}$}        & $0.254^{+0.012}_{-0.014}$                    & \multicolumn{1}{c|}{---}      & \multicolumn{1}{c|}{---}          \\ \cline{4-9} 
                                                        &                                          &                                       & Relaxed       & \multicolumn{2}{c|}{$0.523^{+0.033}_{-0.027}$}       & $0.289^{+0.014}_{-0.010}$        & \multicolumn{1}{c|}{---}      & \multicolumn{1}{c|}{---}          \\ \cline{4-9} 
                                                   &                                                &                                      & Unrelaxed     & \multicolumn{2}{c|}{$0.480^{+0.059}_{-0.065}$} & $0.232^{+0.016}_{-0.029}$  & \multicolumn{1}{c|}{---}      & \multicolumn{1}{c|}{---}          \\ \cline{3-9} 
                                                                    
                        &   & \multirow{3}{*}{SBPL}   & All Sample    & $0.395^{+0.079}_{-0.175}$                    & $0.511^{+0.080}_{-0.031}$                    & $0.249^{+0.011}_{-0.016}$                    & $0.24^{+1.26}_{-0.23}$                      & \multicolumn{1}{c|}{$0.46^{+0.14}_{-0.35}$}      \\ \cline{4-9} 
                        &   &                               & Relaxed       & $0.220^{+0.062}_{-0.000}$        & $0.648^{+0.108}_{-0.044}$        & $0.262^{+0.005}_{-0.006}$        & $0.17^{+0.16}_{-0.09}$          & \multicolumn{1}{c|}{$0.42^{+0.10}_{-0.05}$}                                                                                                                      \\ \cline{4-9} 
                        &   &                               & Unrelaxed     & $0.330^{+0.100}_{-0.110}$  & $0.490^{+0.116}_{-0.010}$  & $0.227^{+0.017}_{-0.017}$  & $1.50^{+0.00}_{-1.46}$    & \multicolumn{1}{c|}{$0.60^{+0.00}_{-0.40}$}                                                                                                                      \\ \cline{1-9} 
                        
\end{tabular}
}%
\end{table*}

\begin{table*}
\def\arraystretch{2}
\caption{\label{tab:tabtxce} Summary of the best-fitting parameters for the \txce\, -- \mcrit\, scaling relation. This table follows the same format introduced in the caption of Table~\ref{tab:tabmg}. 
}
\resizebox{\textwidth}{!}{%
\begin{tabular}{|l|l|l|l|l|l|l|l|l|}
\hline
\multicolumn{1}{|c|}{\multirow{2}{*}{Scaling}} & 
\multicolumn{1}{c|}{\multirow{2}{*}{Aperture}} & 
\multicolumn{1}{c|}{\multirow{2}{*}{Fit Type}} & 
\multicolumn{1}{c|}{\multirow{2}{*}{Sample}} & 
\multicolumn{2}{c|}{$\alpha$} & 
\multicolumn{1}{c|}{\multirow{2}{*}{$A$}} & 
\multicolumn{1}{c|}{\multirow{2}{*}{$\delta$}} & 
\multicolumn{1}{c|}{\multirow{2}{*}{${X}_{{\rm pivot}}$ [ $\times 10^{14} \, {{\rm M}}_\odot$]}}  \\ \cline{5-6}
\multicolumn{1}{|c|}{}                         & 
\multicolumn{1}{c|}{}                          & 
\multicolumn{1}{c|}{}                          & 
\multicolumn{1}{c|}{}                        & 
\multicolumn{1}{c|}{$\alpha_1$}    & 
\multicolumn{1}{c|}{$\alpha_2$}    & 
\multicolumn{1}{c|}{}                   & 
\multicolumn{1}{c|}{}                          & 
\multicolumn{1}{c|}{}                                     \\ \hline

\multirow{12}{*}{${T}_{{\rm X , \, ce}}$  - ${ M}_{{\rm 500 \, crit}}$}  & \multirow{6}{*}{$ R_{\rm 500,\, SIM}$} & \multirow{3}{*}{SPL}          & All Sample    & \multicolumn{2}{c|}{$0.551^{+0.013}_{-0.017}$}                         & $0.228^{+0.004}_{-0.005}$                    & \multicolumn{1}{c|}{---}      & \multicolumn{1}{c|}{---}          \\ \cline{4-9} 
                                                    &                                          &                                       & Relaxed       & \multicolumn{2}{c|}{$0.521^{+0.013}_{-0.017}$}       & $0.237^{+0.005}_{-0.007}$        & \multicolumn{1}{c|}{---}      & \multicolumn{1}{c|}{---}          \\ \cline{4-9} 
                                               &                                                &                                      & Unrelaxed     & \multicolumn{2}{c|}{$0.553^{+0.015}_{-0.029}$} & $0.221^{+0.005}_{-0.007}$  & \multicolumn{1}{c|}{---}      & \multicolumn{1}{c|}{---}          \\ \cline{3-9} 
                                                                        
                        &   & \multirow{3}{*}{SBPL}   & All Sample    & $0.445^{+0.019}_{-0.075}$                    & $0.599^{+0.273}_{-0.032}$                    & $0.206^{+0.008}_{-0.004}$                    & $0.43^{+0.73}_{-0.42}$                      & \multicolumn{1}{c|}{$0.83^{+4.17}_{-0.24}$}      \\ \cline{4-9} 
                        &   &                               & Relaxed       & $0.380^{+0.016}_{-0.029}$        & $0.613^{+0.387}_{-0.023}$        & $0.215^{+0.004}_{-0.003}$        & $0.06^{+1.11}_{-0.05}$          & \multicolumn{1}{c|}{$0.60^{+2.67}_{-0.05}$}                                                                                                                      \\ \cline{4-9} 
                        &   &                               & Unrelaxed     & $0.436^{+0.018}_{-0.051}$  & $0.624^{+0.274}_{-0.048}$  & $0.195^{+0.007}_{-0.005}$  & $0.41^{+0.65}_{-0.40}$    & \multicolumn{1}{c|}{$1.02^{+3.97}_{-0.37}$}                                                                                                                      \\ \cline{2-9} 
                        
                                                & \multirow{6}{*}{$ R_{\rm 500,\, SPEC}$} & \multirow{3}{*}{SPL}     & All Sample    & \multicolumn{2}{c|}{$0.530^{+0.026}_{-0.031}$}        & $0.244^{+0.008}_{-0.009}$                    & \multicolumn{1}{c|}{---}      & \multicolumn{1}{c|}{---}          \\ \cline{4-9} 
                                                        &                                          &                                       & Relaxed       & \multicolumn{2}{c|}{$0.489^{+0.021}_{-0.030}$}       & $0.264^{+0.009}_{-0.015}$        & \multicolumn{1}{c|}{---}      & \multicolumn{1}{c|}{---}          \\ \cline{4-9} 
                                                   &                                                &                                      & Unrelaxed     & \multicolumn{2}{c|}{$0.529^{+0.036}_{-0.035}$} & $0.234^{+0.011}_{-0.013}$  & \multicolumn{1}{c|}{---}      & \multicolumn{1}{c|}{---}          \\ \cline{3-9} 
                                                                    
                        &   & \multirow{3}{*}{SBPL}   & All Sample    & $0.491^{+0.054}_{-0.020}$                    & $0.549^{+0.092}_{-0.106}$                    & $0.238^{+0.014}_{-0.011}$                    & $0.10^{+0.37}_{-0.05}$                      & \multicolumn{1}{c|}{$0.83^{+1.17}_{-0.33}$}      \\ \cline{4-9} 
                        &   &                               & Relaxed       & $0.321^{+0.042}_{-0.055}$        & $0.621^{+0.090}_{-0.028}$        & $0.238^{+0.005}_{-0.005}$        & $0.18^{+0.19}_{-0.10}$          & \multicolumn{1}{c|}{$0.50^{+0.15}_{-0.00}$}                                                                                                                      \\ \cline{4-9} 
                        &   &                               & Unrelaxed     & $0.492^{+0.050}_{-0.042}$  & $0.558^{+0.145}_{-0.120}$  & $0.229^{+0.013}_{-0.015}$  & $0.14^{+0.50}_{-0.09}$    & \multicolumn{1}{c|}{$1.34^{+0.66}_{-0.84}$}                                                                                                                      \\ \cline{1-9} 
                        
\end{tabular}
}%
\end{table*}

\begin{table*}
\def\arraystretch{2}
\caption{\label{tab:tabyx} Summary of the best-fitting parameters for the \yxm\, scaling relation. 
This table follows the same format introduced in the caption of Table~\ref{tab:tabmg}. 
}
\resizebox{\textwidth}{!}{%
\begin{tabular}{|l|l|l|l|l|l|l|l|l|}
\hline
\multicolumn{1}{|c|}{\multirow{2}{*}{Scaling}} & 
\multicolumn{1}{c|}{\multirow{2}{*}{Aperture}} & 
\multicolumn{1}{c|}{\multirow{2}{*}{Fit Type}} & 
\multicolumn{1}{c|}{\multirow{2}{*}{Sample}} & 
\multicolumn{2}{c|}{$\alpha$} & 
\multicolumn{1}{c|}{\multirow{2}{*}{$A$}} & 
\multicolumn{1}{c|}{\multirow{2}{*}{$\delta$}} & 
\multicolumn{1}{c|}{\multirow{2}{*}{${ X}_{{\rm pivot}}$ [ $\times 10^{14} \, {{\rm M}}_\odot$]}}  \\ \cline{5-6}
\multicolumn{1}{|c|}{}                         & 
\multicolumn{1}{c|}{}                          & 
\multicolumn{1}{c|}{}                          & 
\multicolumn{1}{c|}{}                        & 
\multicolumn{1}{c|}{$\alpha_1$}    & 
\multicolumn{1}{c|}{$\alpha_2$}    & 
\multicolumn{1}{c|}{}                   & 
\multicolumn{1}{c|}{}                          & 
\multicolumn{1}{c|}{}                                     \\ \hline

\multirow{12}{*}{${ Y}_{{\rm X}}$  - ${ M}_{{\rm 500 \, crit}}$}  & \multirow{6}{*}{$ R_{\rm 500,\, SIM}$} & \multirow{3}{*}{SPL}          & All Sample    & \multicolumn{2}{c|}{$2.098^{+0.004}_{-0.004}$}                         & $13.296^{+0.005}_{-0.004}$                    & \multicolumn{1}{c|}{---}      & \multicolumn{1}{c|}{---}         \\ \cline{4-9} 
                                                    &                                          &                                       & Relaxed       & \multicolumn{2}{c|}{$1.847^{+0.016}_{-0.011}$}       & $13.305^{+0.009}_{-0.003}$        & \multicolumn{1}{c|}{---}      & \multicolumn{1}{c|}{---}          \\ \cline{4-9} 
                                               &                                                &                                      & Unrelaxed     & \multicolumn{2}{c|}{$2.088^{+0.028}_{-0.004}$} & $13.279^{+0.035}_{-0.005}$  & \multicolumn{1}{c|}{---}      & \multicolumn{1}{c|}{---}          \\ \cline{3-9} 
                                                                        
                        &   & \multirow{3}{*}{SBPL}   & All Sample    & $2.260^{+0.003}_{-0.003}$                    & $1.710^{+0.021}_{-0.018}$                    & $13.283^{+0.004}_{-0.003}$                    & $0.13^{+0.03}_{-0.04}$                      & \multicolumn{1}{c|}{$0.38^{+0.02}_{-0.02}$}      \\ \cline{4-9} 
                        &   &                               & Relaxed       & $1.949^{+0.319}_{-0.053}$        & $1.581^{+0.160}_{-0.581}$        & $13.320^{+0.004}_{-0.005}$        & $0.49^{+0.73}_{-0.29}$          & \multicolumn{1}{c|}{$1.10^{+3.15}_{-0.73}$}                                                                                                                      \\ \cline{4-9} 
                        &   &                               & Unrelaxed     & $2.253^{+0.003}_{-0.003}$  & $1.731^{+0.020}_{-0.019}$  & $13.260^{+0.004}_{-0.004}$  & $0.11^{+0.03}_{-0.03}$    & \multicolumn{1}{c|}{$0.34^{+0.02}_{-0.02}$}                                                                                                                     \\ \cline{2-9} 
                        
                                                & \multirow{6}{*}{$R_{\rm 500,\, SPEC}$} & \multirow{3}{*}{SPL}     & All Sample    & \multicolumn{2}{c|}{$2.052^{+0.028}_{-0.012}$}        & $13.293^{+0.035}_{-0.015}$                    & \multicolumn{1}{c|}{---}      & \multicolumn{1}{c|}{---}          \\ \cline{4-9} 
                                                        &                                          &                                       & Relaxed       & \multicolumn{2}{c|}{$1.823^{+0.024}_{-0.020}$}       & $13.322^{+0.011}_{-0.007}$        & \multicolumn{1}{c|}{---}      & \multicolumn{1}{c|}{---}          \\ \cline{4-9} 
                                                   &                                                &                                      & Unrelaxed     & \multicolumn{2}{c|}{$2.046^{+0.024}_{-0.015}$} & $13.280^{+0.030}_{-0.019}$  & \multicolumn{1}{c|}{---}      & \multicolumn{1}{c|}{---}          \\ \cline{3-9} 
                                                                    
                        &   & \multirow{3}{*}{SBPL}   & All Sample    & $2.153^{+0.005}_{-0.005}$                    & $1.584^{+0.081}_{-0.132}$                    & $13.338^{+0.015}_{-0.014}$                    & $0.14^{+0.09}_{-0.07}$                      & \multicolumn{1}{c|}{$0.72^{+0.18}_{-0.09}$}      \\ \cline{4-9} 
                        &   &                               & Relaxed       & $2.130^{+0.034}_{-0.000}$        & $1.425^{+0.237}_{-0.109}$        & $13.332^{+0.007}_{-0.007}$        & $1.46^{+1.10}_{-0.88}$          & \multicolumn{1}{c|}{$0.79^{+0.61}_{-0.49}$}                                                                                                                      \\ \cline{4-9} 
                        &   &                               & Unrelaxed     & $2.134^{+0.007}_{-0.004}$  & $1.541^{+0.097}_{-0.148}$  & $13.338^{+0.020}_{-0.021}$  & $0.05^{+0.16}_{-0.00}$    & \multicolumn{1}{c|}{$0.82^{+0.18}_{-0.11}$}                                                                                                                      \\ \cline{1-9} 
                        
\end{tabular}
}%
\end{table*}

\begin{table*}
\def\arraystretch{2}
\caption{\label{tab:tabysz} Summary of the best-fitting parameters for the \yszm\, scaling relation. 
This table follows the same format introduced in the caption of Table~\ref{tab:tabmg}. 
}
\resizebox{\textwidth}{!}{%
\begin{tabular}{|l|l|l|l|l|l|l|l|l|}
\hline
\multicolumn{1}{|c|}{\multirow{2}{*}{Scaling}} & 
\multicolumn{1}{c|}{\multirow{2}{*}{Aperture}} & 
\multicolumn{1}{c|}{\multirow{2}{*}{Fit Type}} & 
\multicolumn{1}{c|}{\multirow{2}{*}{Sample}} & 
\multicolumn{2}{c|}{$\alpha$} & 
\multicolumn{1}{c|}{\multirow{2}{*}{$A$}} & 
\multicolumn{1}{c|}{\multirow{2}{*}{$\delta$}} & 
\multicolumn{1}{c|}{\multirow{2}{*}{${X}_{{\rm pivot}}$ [ $\times 10^{14} \, {{\rm M}}_\odot$]}}  \\ \cline{5-6}
\multicolumn{1}{|c|}{}                         & 
\multicolumn{1}{c|}{}                          & 
\multicolumn{1}{c|}{}                          & 
\multicolumn{1}{c|}{}                        & 
\multicolumn{1}{c|}{$\alpha_1$}    & 
\multicolumn{1}{c|}{$\alpha_2$}    & 
\multicolumn{1}{c|}{}                   & 
\multicolumn{1}{c|}{}                          & 
\multicolumn{1}{c|}{}   \\ \hline

\multirow{12}{*}{${Y}_{{\rm SZ}}$  - ${ M}_{{\rm 500 \, crit}}$}  & \multirow{6}{*}{$ R_{\rm 500,\, SIM}$} & \multirow{3}{*}{SPL}          & All Sample    & \multicolumn{2}{c|}{$2.036^{+0.021}_{-0.016}$}                         & $-3.786^{+0.017}_{-0.013}$                   
& \multicolumn{1}{c|}{---}      
& \multicolumn{1}{c|}{---}          \\ \cline{4-9} 
                                                    &                                          &                                       & Relaxed       & \multicolumn{2}{c|}{$1.851^{+0.032}_{-0.008}$}       & $-3.696^{+0.016}_{-0.003}$        & \multicolumn{1}{c|}{---}      & \multicolumn{1}{c|}{---}         \\ \cline{4-9} 
                                               &                                                &                                      & Unrelaxed     & \multicolumn{2}{c|}{$2.033^{+0.082}_{-0.017}$} & $-3.794^{+0.071}_{-0.013}$  & \multicolumn{1}{c|}{---}      & \multicolumn{1}{c|}{---}         \\ \cline{3-9} 
                                                                        
                        &   & \multirow{3}{*}{SBPL}   & All Sample    & $2.418^{+0.004}_{-0.003}$                    & $1.687^{+0.015}_{-0.015}$                    & $-3.682^{+0.003}_{-0.003}$                    & $0.25^{+0.02}_{-0.02}$                      & \multicolumn{1}{c|}{$0.37^{+0.01}_{-0.01}$}     \\ \cline{4-9} 
                        &   &                               & Relaxed       & $2.068^{+0.058}_{-0.042}$        & $1.696^{+0.026}_{-0.052}$        & $-3.665^{+0.005}_{-0.004}$        & $0.20^{+0.14}_{-0.07}$          & \multicolumn{1}{c|}{$0.55^{+0.10}_{-0.07}$}                                                                                                                     \\ \cline{4-9} 
                        &   &                               & Unrelaxed     & $2.414^{+0.004}_{-0.004}$  & $1.699^{+0.018}_{-0.021}$  & $-3.695^{+0.004}_{-0.004}$  & $0.25^{+0.03}_{-0.03}$    & \multicolumn{1}{c|}{$0.35^{+0.02}_{-0.01}$}                                                                                                                     \\ \cline{2-9} 
                        
                                                & \multirow{6}{*}{$R_{\rm 500,\, SPEC}$} & \multirow{3}{*}{SPL}     & All Sample    & \multicolumn{2}{c|}{$2.037^{+0.084}_{-0.038}$}        & $-3.716^{+0.071}_{-0.031}$                    & \multicolumn{1}{c|}{---}      & \multicolumn{1}{c|}{---}          \\ \cline{4-9} 
                                                        &                                          &                                       & Relaxed       & \multicolumn{2}{c|}{$1.806^{+0.031}_{-0.020}$}       & $-3.613^{+0.015}_{-0.007}$        & \multicolumn{1}{c|}{---}      & \multicolumn{1}{c|}{---}          \\ \cline{4-9} 
                                                   &                                                &                                      & Unrelaxed     & \multicolumn{2}{c|}{$2.039^{+0.079}_{-0.040}$} & $-3.724^{+0.066}_{-0.033}$  & \multicolumn{1}{c|}{---}      & \multicolumn{1}{c|}{---}          \\ \cline{3-9} 
                                                                    
                        &   & \multirow{3}{*}{SBPL}   & All Sample    & $2.386^{+0.007}_{-0.007}$                    & $1.556^{+0.100}_{-0.056}$                    & $-3.597^{+0.013}_{-0.014}$                    & $0.18^{+0.05}_{-0.05}$                      & \multicolumn{1}{c|}{$0.52^{+0.05}_{-0.06}$}      \\ \cline{4-9} 
                        &   &                               & Relaxed       & $2.019^{+0.209}_{-0.077}$        & $1.667^{+0.030}_{-0.040}$        & $-3.593^{+0.008}_{-0.007}$        & $0.09^{+0.16}_{-0.08}$          & \multicolumn{1}{c|}{$0.47^{+0.11}_{-0.13}$}                                                                                                                      \\ \cline{4-9} 
                        &   &                               & Unrelaxed     & $2.349^{+0.013}_{-0.017}$  & $1.507^{+0.129}_{-0.007}$  & $-3.603^{+0.022}_{-0.023}$  & $0.16^{+0.07}_{-0.09}$    & \multicolumn{1}{c|}{$0.59^{+0.08}_{-0.07}$}                                                                                                                     \\ \cline{1-9} 
                        
\end{tabular}
}%
\end{table*}

\bsp	
\label{lastpage}
\end{document}